\documentclass[1p,preprint]{elsarticle}

\usepackage{graphicx}
\usepackage{subfigure}
\usepackage{cancel}
\usepackage{color}
\usepackage{colortbl,xcolor}
\usepackage{amsfonts}
\usepackage{amsmath,amssymb}
\usepackage{pifont}
\usepackage{booktabs}
\usepackage{rotating}
\usepackage[pdfpagemode=UseNone,pdfstartview=FitH]{hyperref}
\usepackage{algorithm}
\usepackage{algorithmicx}
\usepackage{algpseudocode}
\usepackage{multirow}
\usepackage{bbm}
\usepackage{calc}

\newcommand{\bsol}{\begin{proof}[Solution]}
\newcommand{\esol}{\end{proof}}
\newcommand{\bq}{\begin{equation}}
\newcommand{\eq}{\end{equation}}

\newcommand{\R}{\mathbb{R}}

\newcommand{\dt}{\Delta t}
\newcommand{\Laplacian}{\nabla^2}
\newcommand{\diffop}[1]{\mathbbmss{#1}} %\mathbb{}, \mathbbm{}
\newcommand{\ctsop}[1]{\textbf{#1}}
\newcommand{\mycode}[1]{\texttt{#1}}
\newcommand{\bs}[1]{\boldsymbol{#1}}
\newcommand{\half}{\frac{1}{2}}
\newcommand{\brac}[1]{\left(#1 \right)}

\newcommand{\pd}[2]{\frac{\partial #1}{\partial #2}}
\newcommand{\pdd}[2]{\frac{\partial^2 #1}{\partial #2^2}}

\newcommand{\assign}[1]{\textcolor{red}{#1}}
  \renewcommand{\assign}[1]{#1}
\newcommand{\leavethisout}[1]{}
\newcommand{\domwidth}{H}
\newcommand{\eqmstrain}{L}
\newcommand{\dx}{h}
\newcommand{\ds}{h_s}

\newcommand{\EE}{\text{e}}
\newcommand{\order}[1]{{\mathcal O}\left(#1\right)}
\newcommand{\myerror}[2]{{\mathcal E}\left[#1;#2\right]}
\newcommand{\myrate}[2]{{\mathcal R}\left[#1;#2\right]}
\newcommand{\mydot}{\mbox{\raisebox{1.5pt}{\scriptsize $\bullet$}}}

\newcommand{\mystar}{\ast}
\renewcommand{\mystar}{\text{\large $\ast$}}
\newcommand{\mydstar}{\mystar\mystar}
\newcommand{\MAC}{E}% MAC - edges
\newcommand{\Reynolds}{\mbox{\itshape Re}}
\newcommand{\changed}[1]{{\color{red}{#1}}}
\renewcommand{\changed}[1]{#1}

\journal{Journal of Computational Physics}

\begin{document}

\begin{frontmatter}

\title{An Efficient Parallel Immersed Boundary Algorithm using a
  Pseudo-Compressible Fluid Solver\tnoteref{t4}}  

\author[sfu]{Jeffrey K. Wiens\corref{cor1}}
\ead{jwiens@sfu.ca}
\ead[url]{http://www.jkwiens.com/}
\author[sfu]{John M. Stockie}
\ead{jstockie@sfu.ca}
\ead[url]{http://www.math.sfu.ca/~stockie}

\address[sfu]{Department of Mathematics, Simon Fraser University, 8888
  University Drive, Burnaby, BC, Canada, V5A 1S6}
\cortext[cor1]{Corresponding author.}

\begin{abstract}
  We propose an efficient algorithm for the immersed boundary method on
  distributed-memory architectures that has the computational complexity of
  a completely explicit method and also has excellent parallel scaling.  The
  algorithm utilizes the pseudo-compressibility method recently proposed
  by Guermond and Minev that uses a directional splitting strategy to
  discretize the incompressible Navier-Stokes equations, thereby
  reducing the linear systems to a series of one-dimensional tridiagonal
  systems.  We perform numerical simulations of several fluid-structure
  interaction problems in two and three dimensions and study the
  accuracy and convergence rates of the proposed algorithm. We also
  compare the proposed algorithm with other second-order 
  projection-based fluid solvers.  Lastly, the \changed{execution time
    and scaling} properties of the proposed algorithm are investigated
  \changed{and compared to alternate approaches}.
\end{abstract}

\tnotetext[t4]{We acknowledge support from the Natural Sciences and
  Engineering Research Council of Canada (NSERC) through a Postgraduate
  Scholarship (JKW) and a Discovery Grant (JMS).  The numerical
  simulations in this paper were performed using computing resources
  provided by WestGrid and Compute Canada.}


\end{frontmatter}

%\linenumbers

\section{Introduction}
\label{sec:Introduction}

The immersed boundary (IB) method is a mathematical framework for
studying fluid-structure interaction that was originally developed by
Peskin to simulate the flow of blood through a heart
valve~\cite{Peskin1972}.  The IB method has been used
in a wide variety of biofluids applications including blood flow through
heart valves~\cite{Griffith2009,Peskin1972},
aerodynamics of the vocal cords~\cite{Duncan2006}, sperm
motility~\cite{Dillon2007}, insect flight~\cite{Miller2000},
and jellyfish feeding dynamics~\cite{Hamlet2011}.  The method is also
increasingly being applied in non-biological
applications~\cite{MittalIaccarino2005}.

The immersed boundary equations capture the dynamics of both fluid and
immersed elastic structure using a mixture of Eulerian and Lagrangian
variables: the fluid is represented using Eulerian coordinates that are
fixed in space, and the immersed boundary is described by a set of
moving Lagrangian coordinates.  An essential component of the model is
the Dirac delta function that mediates interactions between fluid and IB
quantities in two ways.  First of all, the immersed boundary exerts an
elastic force (possibly singular) on the fluid through an external forcing
term in the Navier-Stokes equations that is calculated using the current
IB configuration.  Secondly, the immersed boundary is constrained to
move at the same velocity as the surrounding fluid, which is just the
no-slip condition.  The greatest advantage of this approach is that when
the governing equations are discretized, no boundary-fitted coordinates
are required to handle the solid structure and the influence of the
immersed boundary on the fluid is captured solely through an external
body force.

When devising a numerical method for solving the IB equations, a common
approach is to use a fractional-step scheme in which the fluid is
decoupled from the immersed boundary, thereby reducing the overall
complexity of the method.  Typically, these fractional-step schemes
employ some permutation of the following steps:
\begin{itemize}
\item \emph{Velocity interpolation:} the fluid
  velocity is interpolated onto the immersed boundary.
\item \emph{IB evolution:} the immersed boundary is evolved in time
  using the interpolated velocity field.
\item \emph{Force spreading:} calculate the force exerted by the
  immersed boundary and spreads it onto the nearby fluid grid points,
  with the resulting force appearing as an external forcing
  term in the Navier-Stokes equations. 
\item \emph{Fluid solve:} evolve the fluid variables in time
  using the external force calculated in the force spreading step.
\end{itemize}
Algorithms that fall into this category include Peskin's original
method~\cite{Peskin1972} as well as algorithms developed by Lai and
Peskin~\cite{Lai2000}, Griffith and Peskin~\cite{Griffith2005}, and many
others. 

A popular recent implementation of fractional-step type is the IBAMR
code~\cite{ibamr} that supports distributed-memory parallelism and
adaptive mesh refinement.  This project grew out of Griffith's doctoral
thesis~\cite{GriffithThesis2005} and was outlined in the
papers~\cite{Griffith2007,Griffith2005}.  In the original IBAMR
algorithm, the incompressible Navier-Stokes equations are solved using a
second-order accurate projection scheme in which the viscous term is
handled with an L-stable
discretization~\cite{McCorquodale2001,Twizell1996} while an explicit
second-order Godunov scheme~\cite{Colella1990,Minion1996} is applied to
the nonlinear advection terms.  The IB evolution equation is then
integrated in time using a strong stability-preserving Runge-Kutta
method~\cite{Gottlieb2001}. Since IBAMR's release, drastic
improvements have been made that increase both the accuracy and
generality of the software~\cite{Griffith2012,Griffith2009}.

Fractional-step schemes often suffer from a severe time step restriction
due to numerical stiffness that arises from an explicit treatment of the
immersed boundary in the most commonly used splitting
approaches~\cite{Stockie1999}.  Because of this limitation, many
researchers have proposed new algorithms that couple the fluid and
immersed boundary together in an implicit fashion, for
example~\cite{Ceniceros2009,Hou2008,Le2009,Mori2008,Newren2007}.  These
methods alleviate the severe time step restriction, but do so at the
expense of solving large nonlinear systems of algebraic equations in
each time step. Although these implicit schemes have been shown in some
cases to be competitive with their explicit
counterparts~\cite{Newren2008}, there is not yet sufficient evidence to
prefer one approach over the other, especially when considering parallel
implementations.

Projection methods are a common class of fractional-step schemes for solving the incompressible
Navier-Stokes equations, and are divided
into two steps. First, the discretized momentum equations are integrated
in time to obtain an intermediate velocity field that in general is not
divergence-free.  In the second step, the intermediate velocity is
projected onto the space of divergence-free fields using the Hodge
decomposition.  The projection step typically requires the solution of
large linear systems in each time step that are computationally costly
\changed{and form a significant bottleneck in CFD codes}.  This cost is increased even more
when a small time step is required for explicit implementations.  Note
that even though some researchers make use of unsplit discretizations of
the Navier-Stokes equations~\cite{Griffith2012,Newren2008}, there is
significant benefit to be had by using a split-step
projection method as a preconditioner~\cite{Griffith2009-2}. Therefore,
any improvements made to a multi-step fluid solver can reasonably be
incorporated into unsplit schemes as well.

In this paper, we develop a fractional-step IB method that has the
computational complexity of a completely explicit method and 
exhibits excellent parallel scaling 
on distributed-memory architectures.  This is achieved
by abandoning the projection method paradigm and instead adopting the
pseudo-compressible fluid solver developed by Guermond and
Minev~\cite{Guermond2010,Guermond2011}.  Pseudo-compressibility methods
relax the incompressibility constraint by perturbing it in an
appropriate manner, such as in Temam's penalty
method~\cite{Temam1968}, the artificial compressibility
method~\cite{Chorin1967}, and Chorin's projection
method~\cite{Chorin1968,Rannacher1992}.  Guermond and Minev's algorithm
differentiates itself by employing a directional-splitting strategy,
thereby permitting the linear systems of size $N^d\times N^d$ typically
arising in projection methods (where $d=2$ or 3 is the problem
dimension) to be replaced with a set of one-dimensional tridiagonal
systems of size $N \times N$.  These tridiagonal systems can be solved
efficiently on distributed-memory computing architectures by combining
Thomas's algorithm with a Schur-complement technique. This allows the
proposed IB algorithm to \changed{efficiently utilize parallel resources~\cite{Ganzha2011}}. 
The only serious limitation of the IB algorithm is that it is
restricted to simple geometries and boundary conditions
due to the directional-splitting strategy adopted by Guermond and
Minev. However, since IB practitioners often use a rectangular fluid
domain with periodic boundary conditions, this is not a serious
limitation.  Instead, the IB method provides a natural setting to
leverage the strengths of Guermond and Minev's algorithm allowing
complex geometries to be incorporated into the domain through an
immersed boundary. This is a simple alternative to the fictitious domain
procedure proposed by Angot et al.~\cite{Angot2012}.

In section~\ref{sec:Equations}, we begin by stating the governing
equations for the immersed boundary method.  We continue by describing
our proposed numerical scheme in section~\ref{sec:Algorithm} where we
incorporate the higher-order rotational form of Guermond and Minev's
algorithm that discretizes an $\order{\Delta t^2}$ perturbation of the
Navier-Stokes equations to yield a formally $\order{\Delta t^{3/2}}$
accurate method.  As a result, the proposed method has convergence
properties similar to a fully second-order projection method, while
maintaining the computational complexity of a completely explicit
method.  \changed{In section~\ref{sec:Implementation}, we discuss
  implementation details and highlight the novel aspects of our
  algorithm.}  Finally, in section~\ref{sec:Results}
\changed{and~\ref{sec:Performance}}, we \changed{demonstrate} the
accuracy, efficiency and parallel \changed{performance} of our method by
means of several test problems in 2D and 3D.

\section{Immersed Boundary Equations}
\label{sec:Equations}

In this paper, we consider a $d$-dimensional Newtonian, incompressible
fluid that fills a periodic box $\Omega = [0,\domwidth]^d$ having side
length $\domwidth$ and dimension $d=2$ or 3.  The fluid is specified
using Eulerian coordinates, $\bs{x}=(x,y)$ in 2D or $(x,y,z)$ in 3D.
Immersed within the fluid is a neutrally-buoyant, elastic structure
$\Gamma \subset \Omega$ that we assume is either a single
one-dimensional elastic fiber, or else is constructed from a collection
of such fibers.  In other words, $\Gamma$ can be a curve, surface or
region.  The immersed boundary can therefore be described using a
fiber-based Lagrangian parameterization, in which the position along any
fiber is described by a single parameter $s$.  If there are multiple
fibers making up $\Gamma$ (for example, for a ``thick'' elastic region
in 2D, or a surface in 3D) then a second parameter $r$ is introduced to
identify individual fibers.  The Lagrangian parameters are assumed to be
dimensionless and lie in the interval $s,r\in[0,1]$.

In the following derivation, we state the governing equations for a
single elastic fiber in dimension $d=2$, and the extension to the
three-dimensional case or for multiple fibers is straightforward.  The
fluid velocity $\bs{u}(\bs{x},t)=(u(\bs{x},t),v(\bs{x},t))$ and pressure
$p(\bs{x},t)$ at location $\bs{x}$ and time $t$ are governed by the
incompressible Navier-Stokes equations
\begin{gather}
  \label{eq:NSE}
  \rho \brac{\pd{\bs{u}}{t} + \bs{u}\cdot\nabla\bs{u}} + \nabla p = \mu 
  \Laplacian \bs{u} + \bs{f}, 
  \\
  \label{eq:incompressible}
  \nabla \cdot \bs{u} = 0,
\end{gather}
where $\rho$ is the fluid density and $\mu$ is the dynamic viscosity
(both constants).  The term $\bs{f}$ appearing on the right hand side of
\eqref{eq:NSE} is an elastic force arising from the immersed boundary
that is given by
\begin{gather}
  \label{eq:force}
  \bs{f}(\bs{x},t) = \int\limits_\Gamma \bs{F}(s,t) \, \delta(\bs{x} -
  \bs{X}(s,t)) \,ds, 
\end{gather}
where $\bs{x}=\bs{X}(s,t)=(X(s,t), Y(s,t))$ represents the IB
configuration and $\bs{F}(s,t)$ is the elastic force density.  The delta
function $\delta(\bs{x}) = \delta(x)\delta(y)$ is a Cartesian product of
1D Dirac delta functions, and acts to ``spread'' the IB
force from $\Gamma$ onto adjacent fluid particles.  In general, the
force density $\bs{F}$ is a functional of the current IB configuration
\begin{gather}
  \label{eq:forceDensity}
  \bs{F}(s,t) = \bs{\mathcal{F}} \left[\bs{X}(s,t)\right].
\end{gather}
For example, the force density 
\begin{gather}
  \label{eq:forceDensityDefinition}
  \bs{\mathcal{F}}[\bs{X}(s,t)] = \sigma \pd{ }{s}\brac{\pd{\bs{X}}{s}
    \brac{ 1 - \frac{\eqmstrain}{|\pd{\bs{X}}{s}|} }} 
\end{gather}
corresponds to a single elastic fiber having ``spring constant''
$\sigma$ and an equilibrium state in which the elastic strain $|\partial
\bs{X} / \partial s| \equiv \eqmstrain$.

The final equation needed to close the system is an evolution equation
for the immersed boundary, which comes from the simple requirement that
$\Gamma$ must move at the local fluid velocity:
\begin{gather}
  \label{eq:membrane}
  \pd{\bs{X}(s,t)}{t} = \bs{u}(\bs{X}(s,t),t) = \int\limits_\Omega
  \bs{u}(\bs{x},t) \, \delta(\bs{x}-\bs{X}(s,t)) \, d\bs{x}. 
\end{gather}
This last equation is simply the no-slip condition, with the
rightmost equality corresponding to the delta function convolution form being more convenient for
numerical computations because of its resemblance to the IB forcing
term~\eqref{eq:force}.  Periodic boundary conditions are imposed on both
the fluid and the immersed structure and appropriate initial values are
prescribed for the fluid velocity $\bs{u}(\bs{x},0)$ and IB position
$\bs{X}(s,0)$.  \changed{Note that our assumption of periodicity in the
  fluid and immersed boundary is a choice made for purposes of
  convenience only, and is not a necessary restriction on either the
  mathematical model or the algorithms developed in
  section~\ref{sec:Algorithm}.}  Further details on the mathematical
formulation of the immersed boundary problem and its extension to three
dimensions can be found in \cite{Peskin2002}.

\section{Algorithm}
\label{sec:Algorithm}

We now provide a detailed description of our algorithm for
solving the immersed boundary problem.  The novelty in our approach
derives first and foremost from the use of a pseudo-compressibility method for
solving the incompressible Navier-Stokes equations, which is new in the
IB context and is described in this section.  The second novel aspect of our
algorithm is in the \changed{implementation}, which is detailed in
section~\ref{sec:Implementation}.

\subsection{Pseudo-Compressibility Methods}

Pseudo-compressibility methods~\cite{Rannacher1992,Shen1997} belong to a
general class of numerical schemes for approximating the incompressible
Navier-Stokes equations by appropriately relaxing the incompressibility
constraint.  An $\order{\epsilon}$ perturbation of the governing
equations is introduced in the following manner
\begin{gather}
  \label{eq:PerturbNSE}
  \rho \brac{\pd{\bs{u}_\epsilon}{t} +
    \bs{u}_\epsilon\cdot\nabla\bs{u}_\epsilon} + \nabla p_\epsilon = \mu
  \Laplacian \bs{u}_\epsilon + \bs{f}, 
  \\
  \label{eq:PerturbIncompressible}
  \frac{\epsilon}{\rho} \ctsop{A} p_\epsilon + \nabla \cdot \bs{u}_\epsilon = 0, 
\end{gather}
where various choices of the generic operator $\ctsop{A}$ lead to a
number of familiar numerical schemes. For example, choosing
$\ctsop{A}=\ctsop{1}$ (the identity) corresponds to the penalty method
of Temam~\cite{Temam1968}, $\ctsop{A}=\partial_t$ yields the artificial
compressibility method~\cite{Chorin1967}, $\ctsop{A}=-\Laplacian$ is
equivalent to Chorin's projection scheme~\cite{Chorin1968,Rannacher1992}
(as long as the perturbation parameter is set equal to the time step,
$\epsilon=\dt$), and $\ctsop{A}=-\Laplacian \partial_t$ yields Shen's
method~\cite{Shen1996} (when $\epsilon=\beta\rho(\dt)^2$ for some positive
constant $\beta$).

Recently, Guermond and Minev~\cite{Guermond2010,Guermond2011} proposed a
new pseudo-compressibility method \changed{with excellent parallel
scaling properties}.  The first-order version of their method can be cast
in the form of an $\order{\epsilon}$-perturbation such as in
equations \eqref{eq:PerturbNSE}--\eqref{eq:PerturbIncompressible} with 
$\epsilon = \dt$ and 
\begin{align*}
  \ctsop{A} =
  \begin{cases}
    (1-\partial_{xx})(1-\partial_{yy}) & \text{in 2D},\\
    (1-\partial_{xx})(1-\partial_{yy})(1-\partial_{zz}) & \text{in 3D}.
  \end{cases}
\end{align*}
They also proposed an $\order{\epsilon^2}$ (second-order in time) variant
that corresponds to the three-stage scheme
\begin{gather}
  \label{eq:PerturbNSE2}
  \rho \brac{\pd{\bs{u}_\epsilon}{t} +
    \bs{u}_\epsilon\cdot\nabla\bs{u}_\epsilon} + \nabla p_\epsilon = \mu
  \Laplacian \bs{u}_\epsilon + \bs{f}, 
  \\
  \label{eq:PerturbIncompressible2}
  \frac{\epsilon}{\rho} \ctsop{A} \psi_\epsilon + \nabla \cdot \bs{u}_\epsilon =
  0, \\
  \label{eq:PerturbCorrection2}
  \epsilon \pd{p_\epsilon}{t} = \psi_\epsilon - \chi \mu \nabla \cdot
  \bs{u}_\epsilon ,
\end{gather}
where $\psi_\epsilon$ is an intermediate variable and $\chi\in
(0,1]$ is an adjustable parameter. 

For both variants of the method, corresponding to either
\eqref{eq:PerturbNSE2}--\eqref{eq:PerturbIncompressible2}
or~\eqref{eq:PerturbNSE2}--\eqref{eq:PerturbCorrection2}, the momentum
equation is discretized in time using a Crank-Nicolson step and the
viscous term is directionally-split using the technique proposed by
Douglas~\cite{Douglas1962}. The perturbed incompressibility constraint
is solved using a straightforward discretization of the direction-split
factors in the operator $\ctsop{A}$ that reduces to a set of
one-dimensional tridiagonal systems.  These simple linear systems can be
solved very efficiently on a distributed-memory machine by combining
Thomas's algorithm with a Schur-complement technique.  This is achieved
by expressing each tridiagonal system using block matrices and
manipulating the original system into a set of block-structured systems
and a Schur complement system. By solving these block-structured systems
in parallel, the domain decomposition can be effectively parallelized.

It is important to note that Guermond and Minev's fluid solver cannot be
recast as a pressure projection algorithm; nevertheless, it has been
demonstrated both analytically~\cite{Guermond2012} and
computationally~\cite{Guermond2011-2} to have comparable convergence
properties to related projection methods. More precisely, the
higher-order algorithm we apply here yields a formally $\order{\Delta
  t^{3/2}}$ accurate method for 2D flows, although in practice higher
convergence rates are observed in both 2D and 3D computations.

The main disadvantage of the algorithm is that it is limited to simple
(rectangular) geometries because of the use of directional-splitting.
However, this is not a real disadvantage in the immersed boundary
context because complex solid boundaries can be introduced by using
immersed boundary points (attached to fixed ``tether points'') that are
embedded within a regular computational domain.  In this way, the IB
method provides a simple and efficient alternative to the fictitious
domain approach~\cite{Angot2012} and related methods that could be used
to incorporate complex geometries into Guermond and Minev's fluid solver.

\subsection{Discretization of Fluid and IB Domains}

\begin{figure}[htbp]
  \begin{center}
    \includegraphics[width=0.6\textwidth]{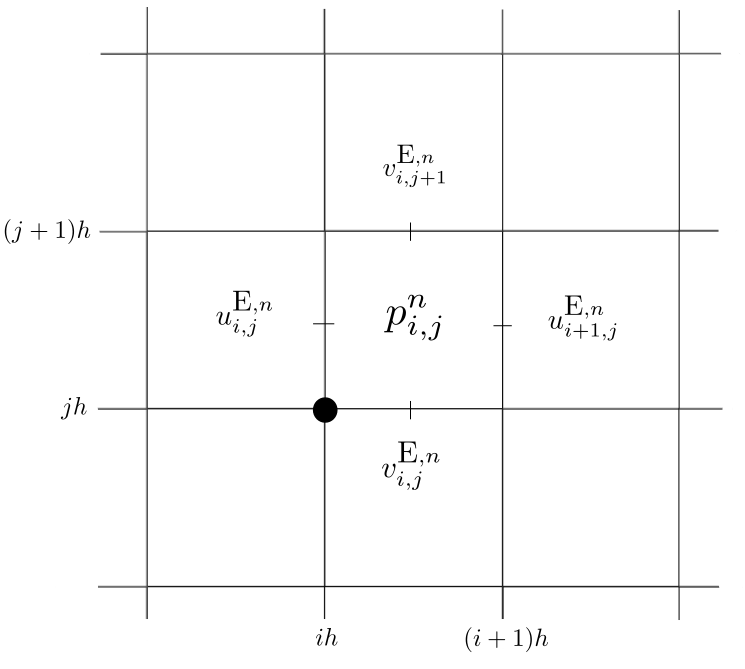}
    \caption{Location of fluid velocity and pressure variables on the
      staggered marker-and-cell (MAC) grid.}
  \label{fig:Grid}
  \end{center}
\end{figure}

When discretizing the governing equations, we require two separate
computational grids, one each for the Eulerian and Lagrangian
variables. For simplicity, we state our discrete scheme for a
two-dimensional fluid ($d=2$) and a fiber consisting of a single
one-dimensional closed curve.  The immersed structure is discretized
using $N_s$ uniformly-spaced points $s_k=k\ds$ in the interval $[0,1]$,
with mesh spacing $\ds=1/N_s$ and $k=0, 1, \ldots, N_s-1$.  As a
short-hand, we denote discrete approximations of the IB position at time
$t_n=n\dt$ by
\begin{gather*}
  \bs{X}_{k}^n \approx \brac{X(k \ds, t_n),\; Y(k \ds, t_n)} ,
\end{gather*}
where $n=0,1,2,\ldots$.  Similarly, the fluid domain
$\Omega=[0,\domwidth]^2$ is divided into an $N \times N$, uniform,
rectangular mesh in which each cell has side length $\dx=\domwidth/N$.
We employ a \emph{marker-and-cell} (MAC)
discretization~\cite{Harlow1965} as illustrated in
Figure~\ref{fig:Grid}, in which the pressure 
\begin{gather*}
  p_{i,j}^n \approx p( \bs{x}_{i,j}, t_n) 
\end{gather*}
is approximated at the cell center points 
\begin{gather*}
  \bs{x}_{i,j} = \brac{(i+{1}/{2}) \dx,~(j+{1}/{2}) \dx},
\end{gather*}
for $i,j = 0,1,\ldots,N-1$.  The velocities on the other
hand are approximated at the edges of cells
\begin{gather*}
  \bs{u}^{\text{\MAC},n}_{i,j} = \brac{ u^{\text{\MAC},n}_{i,j},~
    v^{\text{\MAC},n}_{i,j} }, \\
  \intertext{where}
  u^{\text{\MAC},n}_{i,j} \approx u( i \dx, (j+{1}/{2}) \dx , t_n)
  \qquad \text{and} \qquad 
  v^{\text{\MAC},n}_{i,j} \approx v( (i+{1}/{2}) \dx, j \dx , t_n).
\end{gather*}
The $x$-component of the fluid velocity is defined on the east and west
cell edges, while the $y$-component is located on the north and south
edges.

\subsection{Spatial Finite Difference Operators}

Next, we introduce the discrete difference operators that are used for
approximating spatial derivatives.  The second derivatives of a scalar
Eulerian variable are replaced using the second-order centered
difference stencils
\begin{align*}
  \diffop{D}_{xx} p_{i,j} &= \frac{p_{i+1,j} - 2 p_{i,j} + p_{i-1,j}}{\dx^2}\\
  \mbox{and}\qquad
  \diffop{D}_{yy} p_{i,j} &= \frac{p_{i,j+1} -2 p_{i,j} + p_{i,j-1}}{\dx^2}.
\end{align*}
The same operators may be applied to the vector velocity, so that for example
\begin{gather*}
  \diffop{D}_{xx} \bs{u}^{\text{\MAC}}_{i,j} = 
  \left[ 
    \begin{array}{c}
      \diffop{D}_{xx} u^{\text{\MAC}}_{i,j}\\[0.2cm]
      \diffop{D}_{xx} v^{\text{\MAC}}_{i,j}
    \end{array}
  \right].
\end{gather*}
Since the fluid pressure and velocity variables are defined at different
locations (i.e., cell centers and edges respectively), we also require
difference operators whose input and output are at different locations,
and for this purpose we indicate explicitly the locations of the input
and output using a superscript of the form
${}^\text{\emph{Input}$\to$\emph{Output}}$.  For example, an operator
with the superscript ${}^\text{C$\to$\text{\MAC}}$ takes a cell-centered
input variable (denoted ``C'') and returns an output value located on a
cell edge (denoted ``E'').  Using this notation, we may then define the
discrete gradient operator $\diffop{G}^{\text{C}\to\text{\MAC}}$ as
\begin{gather*}
  \diffop{G}^{\text{C} \to \text{\MAC}} p_{i,j} = \brac{
    \frac{p_{i,j}-p_{i-1,j}}{\dx},~ \frac{p_{i,j}-p_{i,j-1}}{\dx}}, 
\end{gather*}
which acts on the cell-centered pressure variable and returns a
vector-valued quantity on the edges of a cell.  Likewise, the discrete
divergence of the edge-valued velocity
\begin{gather*}
  \diffop{D}^{\text{\MAC} \to \text{C}} \cdot
  \bs{u}_{i,j}^{\text{\MAC}} = \frac{u_{i+1,j} - u_{i,j}}{\dx} +
  \frac{v_{i,j+1} - v_{i,j}}{\dx},
\end{gather*}
which returns a cell-centered value.

Difference formulas are also required for Lagrangian variables such as
$\bs{X}_{k}$, for which we use the first-order one-sided difference
approximations:
\begin{align*}
  \diffop{D}_{s}^+ \bs{X}_{k} &= \frac{\bs{X}_{k+1} - \bs{X}_{k}}{\ds} 
  \\
  \mbox{and}~~~ \diffop{D}_{s}^- \bs{X}_{k} &= \frac{\bs{X}_{k} -
    \bs{X}_{k-1}}{\ds}.  
\end{align*}
Finally, when discretizing the integrals in \eqref{eq:force}
and \eqref{eq:membrane}, we require a discrete approximation to the Dirac
delta function.  Here, we make use of the following approximation
\begin{gather*}
  \delta_h(\bs{x}) = \frac{1}{\dx^2} \phi \brac{ \frac{x}{\dx} } \phi
  \brac{ \frac{y}{\dx} }
\end{gather*}
where
\begin{gather}
  \label{eq:discretedelta}
  \phi(r) = 
    \begin{cases}
      \frac{1}{8}(3-2|r| + \sqrt{1+4|r|-4r^2}) & 
      \text{if $0 \leq |r| < 1$}, \\
      \frac{1}{8}(5-2|r| - \sqrt{-7+12|r|-4r^2}) & 
      \text{if $1 \leq |r| < 2$}, \\
      0 & \text{if $2 \leq |r|$}.
    \end{cases}
\end{gather}
\changed{Peskin~\cite{Peskin2002} derives this and other regularized
  delta function kernels by imposing various desirable smoothness and
  interpolation properties.  We have chosen the form of $\delta_h$ in
  equation \eqref{eq:discretedelta} because numerical simulations have
  shown that it offers a good balance between accuracy and
  cost~\cite{BringleyPeskin2008,Griffith2005,Stockie1997}, not to
  mention that it is currently the approximate delta function that is
  most commonly applied in other IB simulations.}

\subsection{IB-GM Algorithm}
\label{sec:algorithm}

We are now prepared to describe our algorithm for the IB problem based
on the fluid solver of Guermond and Minev~\cite{Guermond2011}, which we
abbreviate ``IB-GM''.  The fluid is evolved in time in two main stages,
both of which reduce to solving one-dimensional tridiagonal linear
systems.  In the first stage, the diffusion terms in the momentum
equations are integrated in time using the directional-splitting
technique proposed by Douglas~\cite{Douglas1962}.  The nonlinear
advection term on the other hand is dealt with explicitly using the
second-order Adams-Bashforth extrapolation
\begin{gather}
  \label{eq:nonlinear}
  N^{n+1/2} = \frac{3}{2}
  \diffop{N}(\bs{u}^{\text{\MAC},n}_{i,j}) \changed{-} \half
  \diffop{N}(\bs{u}^{\text{\MAC},n-1}_{i,j}),
\end{gather}
where $\diffop{N}(\mydot)$ is an approximation of the advection term
$\bs{u}\cdot\nabla\bs{u}$.  In this paper, we write the advection
term in skew-symmetric form
\begin{gather}
  \diffop{N}(\bs{u}) \approx  \half \bs{u}\cdot\nabla\bs{u} + \half
  \nabla \cdot (\bs{u}\bs{u}), 
\end{gather}
and then discretize the resulting expression using the second-order
centered difference scheme studied by Morinishi et
al.~\cite{Morinishi1998}.

In the second stage, the correction term $\psi$ is calculated using
Guermond and Minev's splitting operator~\cite{Guermond2011}, and the
actual pressure variable is updated using the higher-order variant of
their algorithm corresponding
to~\eqref{eq:PerturbNSE2}--\eqref{eq:PerturbCorrection2}.  \changed{For
  all simulations, we use the same parameter values $\chi=0.6$ and
  $\epsilon = \Delta t$.}

For the remaining force spreading and velocity interpolation steps, we
apply standard techniques.  The integrals appearing in equations
\eqref{eq:force} and \eqref{eq:membrane} are approximated to second
order using the trapezoidal quadrature rule and the fiber evolution
equation \eqref{eq:membrane} is integrated using the second-order
Adams-Bashforth extrapolation.

Assuming that the state variables are known at the $(n-1)$th and $n$th
time steps, the IB-GM algorithm proceeds as follows.
\begin{description}
\item[Step 1.] Evolve the IB position to time $t_{n+1/2}=(n+1/2)\dt$:

  \begin{enumerate}
    \renewcommand{\theenumi}{1\alph{enumi}}
  \item Interpolate the fluid velocity onto immersed boundary
    points:
    \begin{gather*}
      \assign{\bs{U}^n_{k}} = \sum_{i,j} \bs{u}^{\text{\MAC},n}_{i,j}
      \delta_h(\bs{x}^{\text{\MAC}}_{i,j} - \bs{X}^n_{k}) \, \dx^2.
    \end{gather*}

  \item Evolve \label{step:1b} the IB position to time $t_{n+1}$ using
    an Adams-Bashforth discretization of \eqref{eq:membrane}:
    \begin{gather*}
      \frac{\assign{\bs{X}^{n+1}_{k}} - \bs{X}^{n}_{k}}{\dt} =
      \frac{3}{2} \bs{U}^n_{k} - \half \bs{U}^{n-1}_{k}.
    \end{gather*}

  \item Approximate the IB position at time $t_{n+1/2}$ using the
    arithmetic average:
    \begin{gather*}
      \assign{\bs{X}^{n+1/2}_{k}} = \half \brac{\bs{X}^{n+1}_{k} +
        \bs{X}^{n}_{k}}.
    \end{gather*}
  \end{enumerate}
  
\item[Step 2.] Calculate the fluid forcing term:

  \begin{enumerate}
    \renewcommand{\theenumi}{2\alph{enumi}}
  \item Approximate the IB force density at time $t_{n+1/2}$ using
    \eqref{eq:forceDensityDefinition}:  
    \begin{gather*}
      \assign{\bs{F}^{n+1/2}_{k}} = \sigma \diffop{D}_{s}^- \brac{
        \diffop{D}_{s}^+ \bs{X}_{k}^{n+1/2} \brac{\diffop{1} -
          \frac{\eqmstrain}{\left|\diffop{D}_{s}^+
              \bs{X}_{k}^{n+1/2}\right|}}}. 
    \end{gather*}

  \item Spread the IB force density onto fluid grid points:
    \begin{gather*}
      \assign{\bs{f}^{\text{\MAC},n+1/2}_{i,j}} = \sum_{k}
      \bs{F}^{n+1/2}_{k} \, \delta_h(\bs{x}^{\text{\MAC}}_{i,j} -
      \bs{X}^{n+1/2}_k) \, \ds . 
    \end{gather*}
  \end{enumerate}
  
\item[Step 3.] Solve the incompressible Navier--Stokes equations:
  \begin{enumerate}
    \renewcommand{\theenumi}{3\alph{enumi}}

  \item Predict \label{step:3a} the fluid pressure at time $t_{n+1/2}$:
    \begin{gather*}
      \assign{p^{\mystar,n+1/2}_{i,j}} = p^{n-1/2}_{i,j} +
      \psi^{n-1/2}_{i,j}. 
    \end{gather*}
    
  \item Compute \label{step:3b} the first intermediate velocity field
    $\bs{u}^{\text{\MAC},\mystar}$ by integrating the momentum equations 
    explicitly:
    \begin{multline*} 
      \rho \brac{ \frac{ \assign{\bs{u}^{\text{\MAC},\mystar}_{i,j}} -
          \bs{u}^{\text{\MAC},n}_{i,j} }{\dt} + {N}^{n+1/2} } = \\
      \mu \brac{ \diffop{D}_{xx} + \diffop{D}_{yy}}
      \bs{u}^{\text{\MAC},n}_{i,j} - \diffop{G}^{\text{C}\to
        \text{\MAC}} p^{\mystar,n+1/2}_{i,j} +
      \bs{f}^{\text{\MAC},n+1/2}_{i,j}. 
    \end{multline*}

  \item Determine the second intermediate velocity
    $\bs{u}^{\text{\MAC},\mydstar}$ by solving the tridiagonal systems
    corresponding to the $x$-derivative in the directional-split
    Laplacian: 
    \begin{gather*}
      \rho \brac{ \frac{ \assign{\bs{u}^{\text{\MAC},\mydstar}_{i,j}} -
          \bs{u}^{\text{\MAC},\mystar}_{i,j} }{\dt} } = \frac{\mu}{2}
      \diffop{D}_{xx} \brac{ \assign{\bs{u}^{\text{\MAC},\mydstar}_{i,j}} -
        \bs{u}^{\text{\MAC},n}_{i,j}}.
    \end{gather*}
    
  \item Obtain the final velocity approximation at time $t_{n+1}$ by
    solving the following tridiagonal systems corresponding to the 
    $y$-derivative piece of the directional-split Laplacian for
    $\assign{\bs{u}^{\text{\MAC},n+1}_{i,j}}$: 
    \begin{gather*}
      \rho \brac{ \frac{ \assign{\bs{u}^{\text{\MAC},n+1}_{i,j}} -
          \bs{u}^{\text{\MAC},\mydstar}_{i,j} }{\dt} } = \frac{\mu}{2}
      \diffop{D}_{yy} \brac{ \assign{\bs{u}^{\text{\MAC},n+1}_{i,j}} -
        \bs{u}^{\text{\MAC},n}_{i,j}}.
    \end{gather*}
    
  \item Determine the pressure correction term $\psi_{i,j}^{n+1/2}$ by
    solving  
    \begin{gather*}
      \brac{\diffop{1}-\diffop{D}_{xx}}
      \brac{\diffop{1}-\diffop{D}_{yy}} \assign{\psi_{i,j}^{n+1/2}} =
      - \frac{\rho}{\dt} \diffop{D}^{\text{\MAC}\to \text{C}}
      \cdot \bs{u}^{\text{\MAC},n+1}_{i,j}.
    \end{gather*}
    
  \item Calculate the pressure at time $t_{n+1/2}$ using
    \begin{gather*}
      \assign{p^{n+1/2}_{i,j}} = p^{n-1/2}_{i,j} + \psi^{n+1/2}_{i,j} -
      \chi \mu \diffop{D}^{\text{\MAC}\to \text{C}} \cdot \left(
        \half(\bs{u}^{\text{\MAC},n+1}_{i,j} +
        \bs{u}^{\text{\MAC},n}_{i,j})\right).
    \end{gather*}
  \end{enumerate}
\end{description}
Note that in the first step of the algorithm with $n=0$, we do not yet
have an approximation of the solution at the previous time step,
and therefore we make the following replacements:
\begin{itemize}
\item In Step \ref{step:1b}, approximate the fiber evolution equation
  using a first-order forward Euler approximation $\bs{X}^{1}_{k} =
  \bs{X}^{0}_{k} + \dt \bs{U}^0_{k}$.
\item In Step \ref{step:3a}, set $p^{\mystar,1/2}_{i,j} = 0$.
\item In Step \ref{step:3b}, the nonlinear term from equation
  \eqref{eq:nonlinear} is replaced with $N^{1/2} =
  \diffop{N}(\bs{u}^{\text{\MAC},0}_{i,j})$. 
\end{itemize}

\section{Parallel Implementation}
\label{sec:Implementation}

Here we outline the details of the algorithm that relate specifically to
the parallel implementation.  Since a primary feature of our algorithm
is its parallel scaling properties, it is important to discuss our
implementation in order to understand the parallel characteristics of
the method.

\subsection{Partitioning of the Eulerian and Lagrangian Grids}
\label{sec:Partitioning}

\begin{figure}[htbp]
  \begin{center}
    \subfigure[]{\includegraphics[width=0.44\textwidth]{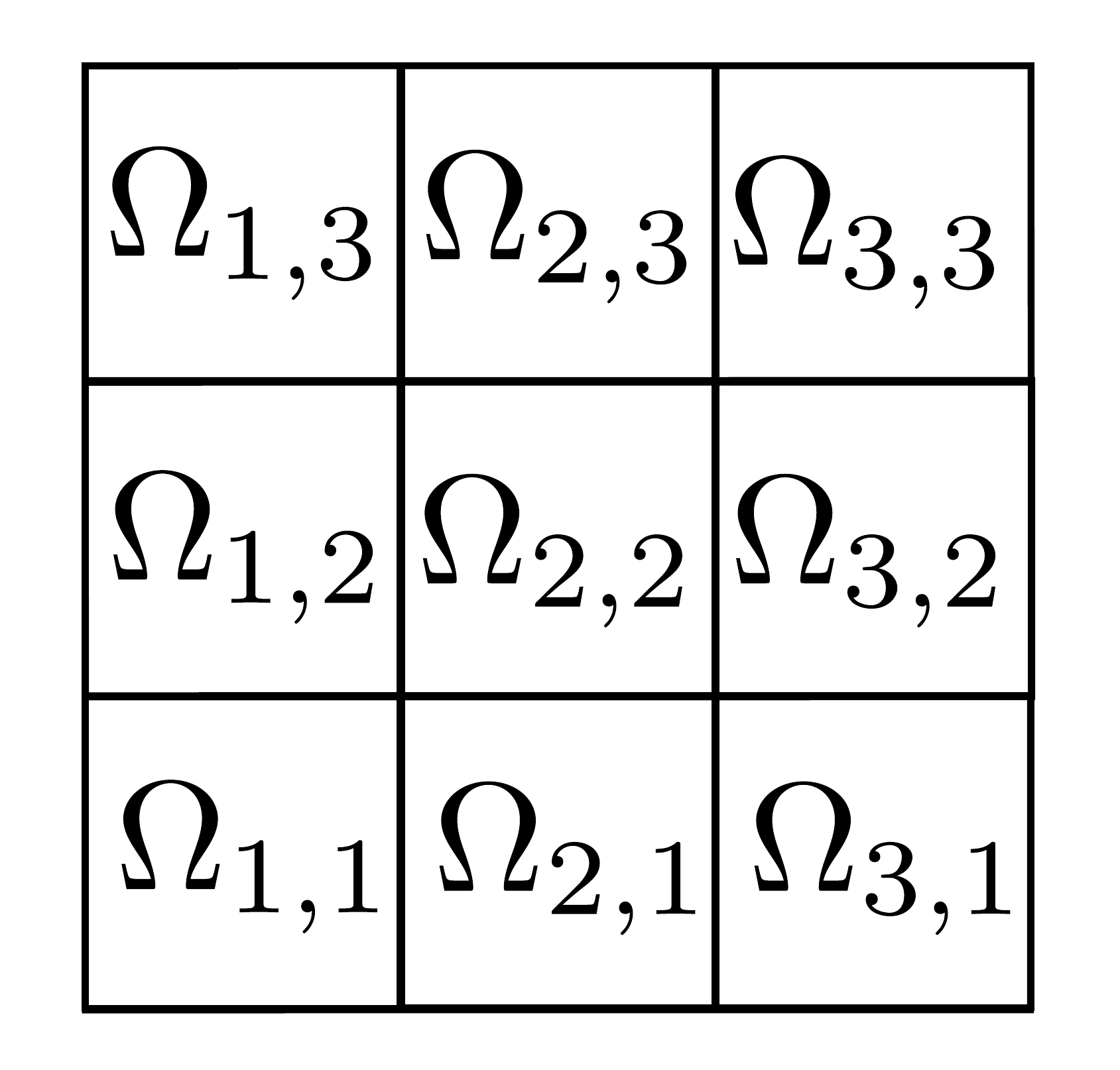} 
      \label{fig:DomainDecomp}}
    \qquad 
    \subfigure[]{\includegraphics[width=0.44\textwidth]{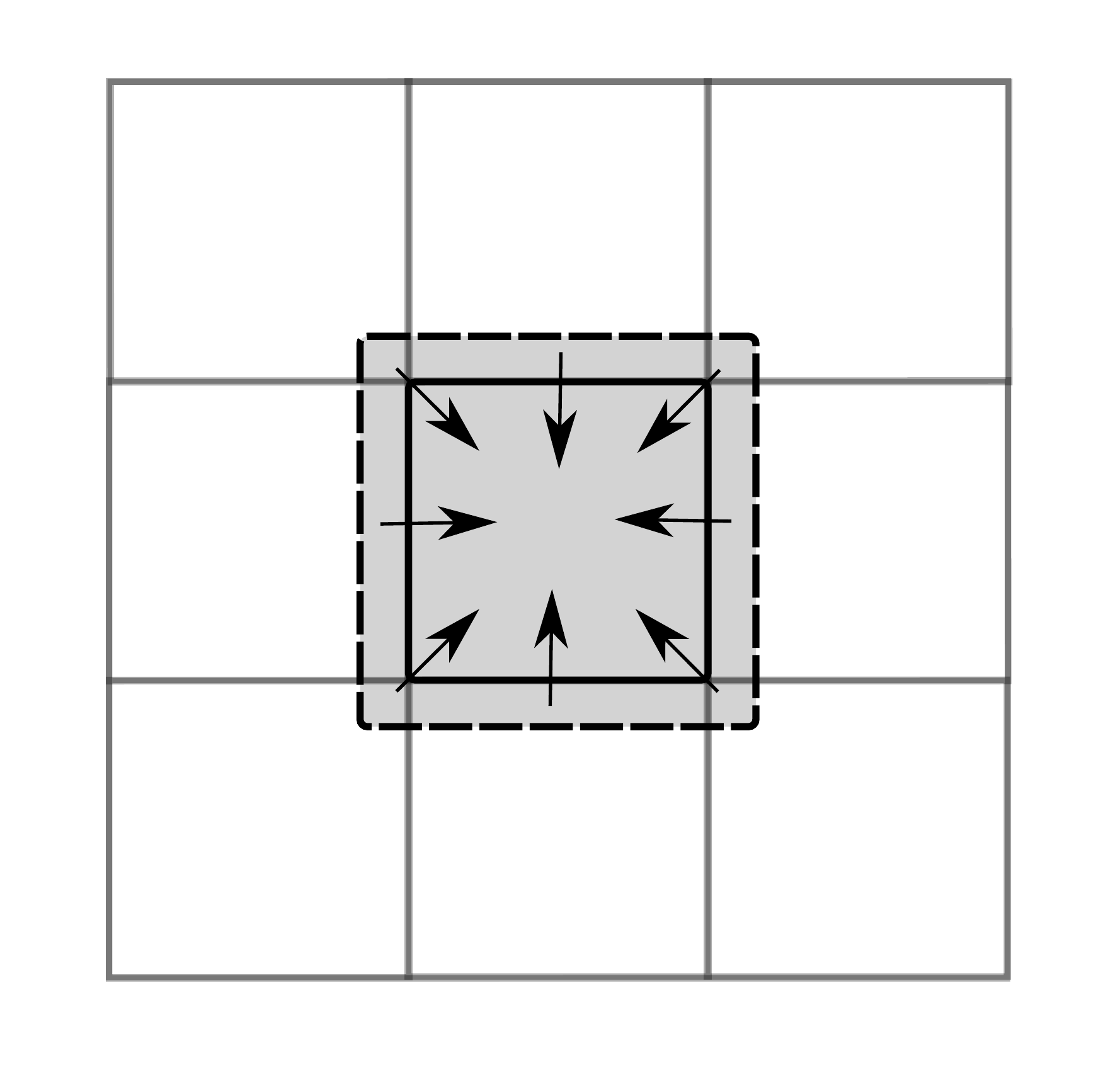} 
      \label{fig:EulerianGhostRegion}}
    \caption{\subref{fig:DomainDecomp} Parallel domain decomposition
      with $P_x=P_y=3$. \subref{fig:EulerianGhostRegion} Communication
      required to update ghost cell regions for the subdomain
      $\Omega_{2,2}$.}
    \label{fig:EulerianPartitioning} 
  \end{center}
\end{figure}

Suppose that the algorithm in section~\ref{sec:algorithm} is implemented
on a distributed-memory computing machine with $P=P_x \cdot P_y$
processing nodes.  The parallelization is performed by subdividing the
rectangular domain $\Omega$ into equally-sized rectangular partitions
$\{\Omega_{\ell,m}\}$, with $\ell=1,2,\ldots, P_x$ and $m=1,2,\ldots,
P_y$, where $P_x$ and $P_y$ refer to the number of subdivisions in the
$x$-- and $y$--directions respectively.  Each node is allocated a single
domain partition $\Omega_{\ell,m}$, along with the values of the Eulerian
and Lagrangian variables contained within it.  For example, the
$(\ell,m)$ node would contain in its memory the fluid variables
$\bs{u}_{i,j}^{\text{\MAC}}$ and $p_{i,j}$ for all $\bs{x}_{i,j} \in
\Omega_{\ell,m}$, along with all immersed boundary data $\bs{X}_k$ and
$\bs{F}_k$ such that $\bs{X}_k \in \Omega_{\ell,m}$.  This partitioning
is illustrated for a simple $3\times 3$ subdivision in
Figure~\ref{fig:EulerianPartitioning}\subref{fig:DomainDecomp}.

In this way, the computational work required in each time step is
effectively divided between processing nodes by requiring that each node
update only those state variables located within its assigned subdomain.
\changed{This approach is similar to that taken by
  Uhlmann~\cite{Uhlmann2004} and Griffith~\cite{Griffith2010}, who
  partitioned the Lagrangian data so that IB points belong to the
  same processor as the surrounding fluid.  An alternate approach would
  be to independently partition the Eulerian and Lagrangian data as done
  by Givelberg~\cite{Givelberg2006}.}

\changed{ The primary novelty in our algorithm derives from that way
  that it introduces parallelism naturally into the discretization.
  This differentiates our method from prior approaches (in
  Griffith~\cite{Griffith2005} or Givelberg~\cite{Givelberg2006}) that
  rely heavily on black-box parallel solvers such as
  Hypre~\cite{Hypre}.  In particular, we use a fractional-step scheme
  that permits the immersed boundary and fluid to be treated
  independently.  The IB component of the algorithm is discretized in
  the same manner as Griffith~\cite{Griffith2012} who used an
  Adams-Bashforth discretization to reduce the number of floating-point
  operations.  Since this is an explicit discretization, the discrete
  immersed boundary can be viewed as a simple particle system -- a
  collection of IB material points connected by force-generating
  connections -- which is a well-established class of problems in the
  parallel computing community~\cite{Asanovic2006,Asanovic2009}.
  Therefore, the major differences in parallel implementation come from
  the discrete delta function whose support allows particles to interact
  over multiple subdomains in the velocity interpolation and force
  spreading steps.

  For the fluid portion of the algorithm, we use the GM fluid solver as
  described in~\cite{Guermond2011} with the following minor
  modifications:
  \begin{itemize}
  \item the advection term is discretized in skew-symmetric form;
  \item periodic boundary conditions are imposed on the fluid domain;
  \item the directional-splitting order is rotated in each time step
    to reduce the possibility of a directional bias; and
  \item minor alterations are required to the parallel tridiagonal solver.   
  \end{itemize}
  By using the directional split strategy of Guermond and Minev, the
  discretized fluid equations deflate into a sequence of one-dimensional
  problems, which is where parallelism is introduced directly into the
  discretization.  The most significant departure from other common IB
  schemes is that the GM solver is a pseudo-compressibility method that
  only approximately satisfies the incompressibility constraint.  It is
  yet to be seen how such a fluid solver will handle the near-singular
  body forces that occur naturally in IB problems.  Therefore, a
  comprehensive numerical study is required to test the accuracy,
  convergence, and volume conservation of the method.

  Next, we describe our approach for implementing data partitioning and
  communication, which makes use of infrastructure provided by
  MPI~\cite{OpenMPI} and PETSc~\cite{PETSc-2012}. } Since the fluid and
immersed boundary are discretized on two separate grids, the data
partitioning between nodes must be handled differently in each case.
The partitioning of Eulerian variables is much simpler because the
spatial locations are fixed in time and remain associated with the same
node for the entire computation.  In contrast, Lagrangian variables are
free to move throughout the fluid domain and so a given IB point may
move between two adjacent subdomains during the course of a single time
step.  As a result, the data structure and communication patterns for
the Lagrangian variables are more complex.

Consider the communication required for the update of fluid variables in
each time step, for which the algorithm in section~\ref{sec:algorithm}
requires the explicit computation of several discrete difference
operators.  For points located inside a subdomain $\Omega_{\ell,m}$,
the discrete operators are easily computed; however for points on the
edge of a domain partition, a difference operator may require data that
is not contained in the current node's local memory. For example, when
calculating the discrete Laplacian (using the 5-point stencil), data at
points adjacent to the given state variable are required.  As a result,
when an adjacent variable does not reside in $\Omega_{\ell,m}$,
communication is required with a neighbouring node to obtain the
required value. This communication is aggregated together using
\emph{ghost cells} that lie inside a strip surrounding 
each $\Omega_{\ell,m}$ as illustrated in Figure
\ref{fig:EulerianPartitioning}\subref{fig:EulerianGhostRegion}.  The
width of the ghost region is set equal to the support of the discrete
delta function used in the velocity interpolation and force spreading
steps; that is, two grid points in the case of the
delta-approximation~\eqref{eq:discretedelta}.  When a difference
operator is applied to a state variable stored in the $(\ell,m)$ node,
the neighbouring nodes communicate the data contained in the ghost cells
adjacent to $\Omega_{\ell,m}$.  After the ghost region is filled, the
discrete difference operators may then be calculated for all points in
$\Omega_{\ell,m}$.  When combined with the parallel linear solver
discussed later in section~\ref{sec:linearsolver}, this parallel
communication technique permits the fluid variables to be evolved in
time.

\begin{figure}[htbp]
  \begin{center}
    \subfigure[]{\includegraphics[width=0.9\textwidth]{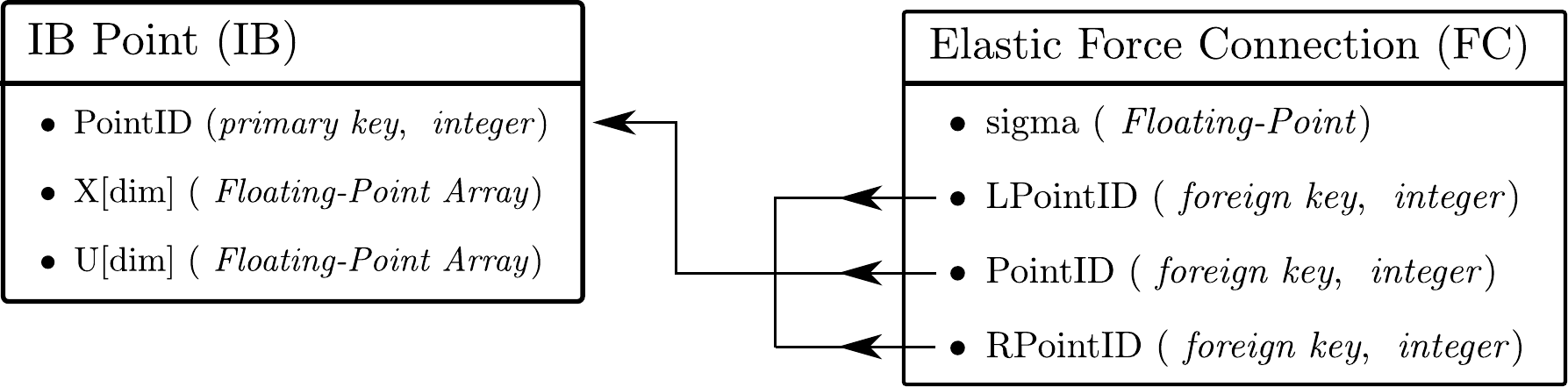} 
      \label{fig:IBRelation}}  
    \qquad 
    \subfigure[]{\includegraphics[width=0.6\textwidth]{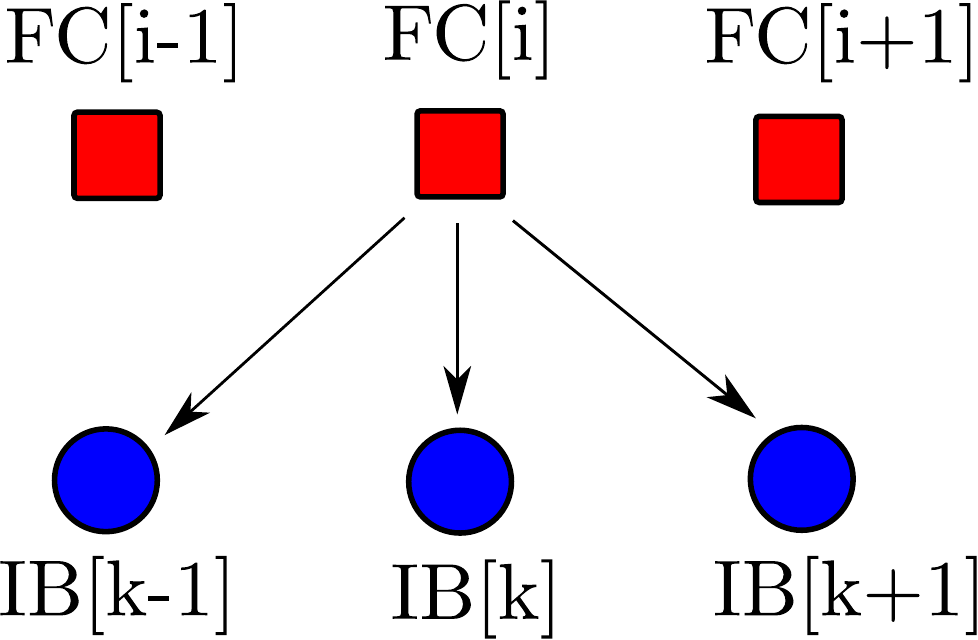} 
      \label{fig:ForceConnectionRef}}
    \caption{\subref{fig:IBRelation} Relationship between the data
      structures for the IB points (\mycode{IB}) and elastic force
      connections (\mycode{FC}). \subref{fig:ForceConnectionRef}
      References from a chosen force connection to the corresponding IB
      points.}
    \label{fig:IBDataStructure} 
  \end{center}
\end{figure}

As the IB points move through the fluid, the number of IB points
residing in any particular subdomain can vary from one time step to the
next.  Therefore, the memory required to store the local IB data
structure changes with time, as does the communication load. These
complications are dealt with by splitting the data structure defining
the immersed boundary into two separate components corresponding to IB
points and force connections.  The IB point (\mycode{IB}) data structure
contains the position and velocity of all IB points resident in a given
subdomain, whereas the force connection (\mycode{FC}) data structure
keeps track of all force-generating connections between these
points. The force density calculations depend on spatial information and
so the \mycode{IB} data structure requires a globally unique index
(which we call the ``primary key'') that is referenced by the
\mycode{FC} data structure (the ``foreign key'').  This relationship is
illustrated in Figure~\ref{fig:IBDataStructure}, where the force
connections shown are consistent with the elastic force
function~\eqref{eq:forceDensityDefinition}.  If the \mycode{IB} data
structure is represented as an associative array using \mycode{PointID}
as the key (and referenced as $\mycode{IB[PointID]}$) and
$\mycode{FC[i]}$ represents a specific element of the force connection
array, then the force density calculation may be written as
\begin{multline*}
  \mycode{FC[i].Fdens} = \frac{\mycode{FC[i].sigma}}{\ds^2} \, \big(\,
  \mycode{IB[\,FC[i].LPointID\,].X} \;+\; \mycode{IB[\,FC[i].RPointID\,].X} \\
  -\; \mycode{2 * IB[\,FC[i].PointID\,].X} \, \big) ,
\end{multline*}
where we have assumed here that the force parameter $\eqmstrain=0$. 
\changed{The \mycode{IB} and \mycode{FC} data structures 
are stored in a hash table and vector (respectively) using the standard 
STL containers in C++.}

\begin{figure}[htbp]
  \begin{center}
    \includegraphics[width=0.9\textwidth]{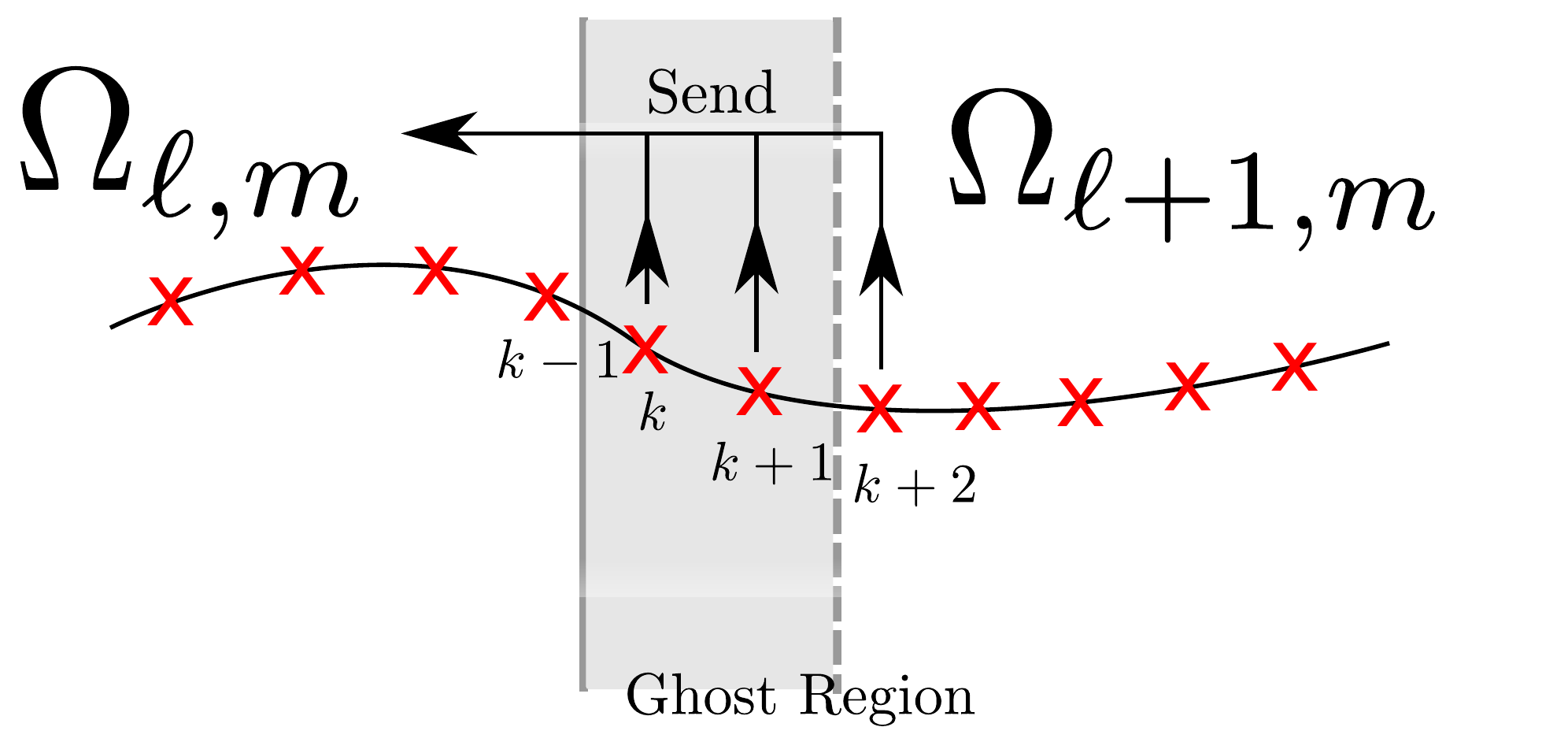} 
    \caption{IB points inside the ghost region surrounding
      $\Omega_{\ell,m}$ are communicated from $\Omega_{\ell+1,m}$.}
    \label{fig:IBSend} 
  \end{center}
\end{figure}

We are now prepared to summarize the complete parallel procedure that is
used to evolve the fluid and immersed boundary.  Keep in mind that at
the beginning of each time step, a processing node contains only those
IB points and force connections that reside in the corresponding
subdomain.  The individual solution steps are:
\begin{itemize}
\item \emph{Velocity interpolation:} Interpolate the fluid velocity
  onto the IB points and store the result in \mycode{IB[\mydot].U}.
  This step requires fluid velocity data from the ghost region. 
  
\item \emph{Immersed boundary evolution:} Evolve the IB points in time
  by updating $\mycode{IB[\mydot].X} = \bs{X}^{n+1}_{\mydot}$.  Note
  that the IB point position at the half time step
  ($\mycode{IB[\mydot].Xh} = \bs{X}^{n+1/2}_{\mydot}$) must also be
  stored for the force spreading step.
  
\item \emph{Immersed boundary communication:} Send the data from IB
  points lying within the ghost region to the neighbouring processing
  nodes.  Figure~\ref{fig:IBSend} illustrates how the IB points residing
  in the ghost region corresponding to $\Omega_{\ell,m}$ are copied from
  $\Omega_{\ell+1,m}$ (for both the full time step $n+1$ and the
  half-step $n+1/2$).  In this example, three IB points (corresponding
  to $\mycode{PointID}=k, k+1, k+2$) and two force connections (with
  $\mycode{FC[i].PointID}=k,k+1$) are communicated to
  $\Omega_{\ell,m}$. The additional IB point is required to calculate
  the force density for $\mycode{FC[i].PointID}=k+1$.  Because the IB
  point $k-1$ already resides in $\Omega_{\ell,m}$, the force density
  can be computed for $\mycode{FC[i].PointID}=k$ without any additional
  communication.
       
\item \emph{Force spreading:} Calculate the force density for all IB
  points in $\Omega_{\ell,m}$ and the surrounding ghost region at the
  time step $n+1/2$.  Then spread the force density onto the Eulerian
  grid points residing in $\Omega_{\ell,m}$.
                             
\item \emph{Immersed boundary cleanup:} Remove all IB points and
  corresponding force connections that do not reside in
  $\Omega_{\ell,m}$ at time step $n+1$.

\item \emph{Evolve fluid:} Evolve the fluid variables in time using the
  parallel techniques discussed above. This requires communication with
  the neighbouring processing nodes to update the ghost cell region, and
  further communication is needed while solving the linear systems.
\end{itemize}

Using the approach outlined above, each processing node only needs to
communicate with its neighbouring nodes, with the only exception being
the linear solver which we address in the next section. Since
communication \changed{often hinders} the performance of a
parallel algorithm, this is the property that allows our method to scale
so well.  For example, if the problem size and number of processing
nodes are doubled, then we would ideally want the execution time per
time step to remain unchanged. 

\subsection{Linear Solver}
\label{sec:linearsolver}

A key remaining component of the algorithm outlined in
section~\ref{sec:algorithm} is the solution of the tridiagonal linear
systems arising in the fluid solver.  When solving the
momentum equations the following linear systems arise:
\begin{gather}
  \brac{\diffop{1} - \frac{\mu\dt}{2\rho} \diffop{D}_{xx}}
  \bs{u}^{\text{\MAC},\mydstar}_{i,j} = 
  \bs{u}^{\text{\MAC},\mystar}_{i,j} - \frac{\mu\dt}{2\rho} \diffop{D}_{xx}
  \bs{u}^{\text{\MAC},n}_{i,j},  
  \label{eq:umacstep1}
  \\
  \brac{\diffop{1} - \frac{\mu\dt}{2\rho} \diffop{D}_{yy}}
  \bs{u}^{\text{\MAC},n+1}_{i,j} = 
  \bs{u}^{\text{\MAC},\mydstar}_{i,j} - \frac{\mu\dt}{2\rho} \diffop{D}_{yy}
  \bs{u}^{\text{\MAC},n}_{i,j},  
  \label{eq:umacstep2}
\end{gather}
while the pressure update step requires solving 
\begin{gather*}
  \brac{\diffop{1} - \diffop{D}_{xx}} \brac{\diffop{1} -
    \diffop{D}_{yy}} \psi_{i,j}^{n+1/2} = - \frac{\rho}{\dt}
  \diffop{D}^{\text{\MAC}\to \text{C}} \cdot
  \bs{u}^{\text{\MAC},n+1}_{i,j} .
\end{gather*}
This last equation can be split into two steps
\begin{align}
  \label{eq:PenaltyStepPart1} 
  \brac{\diffop{1} - \diffop{D}_{xx}} \psi_{i,j}^{\mystar,n+1/2} &= 
  - \frac{\rho}{\dt} \diffop{D}^{\text{\MAC}\to \text{C}} \cdot
  \bs{u}^{\text{\MAC},n+1}_{i,j},
  \\
  \label{eq:PenaltyStepPart2} 
  \mbox{and} \qquad \brac{\diffop{1} - \diffop{D}_{yy}}
  \psi_{i,j}^{n+1/2} &= \psi_{i,j}^{\mystar,n+1/2},
\end{align}
where $\psi_{i,j}^{\mystar}$ is an intermediate variable.  Note that
each linear system in \eqref{eq:umacstep1}--\eqref{eq:PenaltyStepPart2}
involves a difference operator that acts in one spatial dimension only
and decouples into a set of one-dimensional periodic (or cyclic)
tridiagonal systems.  For example, equations \eqref{eq:umacstep1} and
\eqref{eq:PenaltyStepPart1} consist of $N$ tridiagonal systems of size
$N\times N$ having the general form
\begin{gather}
  \label{eq:TriDiagX} 
  \diffop{A}^{(j)} \Psi_{i,j} = b_{i,j} ,
\end{gather}
for each $j=0,1,\ldots, N-1$.

\begin{figure}[htbp]
  \begin{center}
    \subfigure[]{\includegraphics[width=0.44\textwidth]{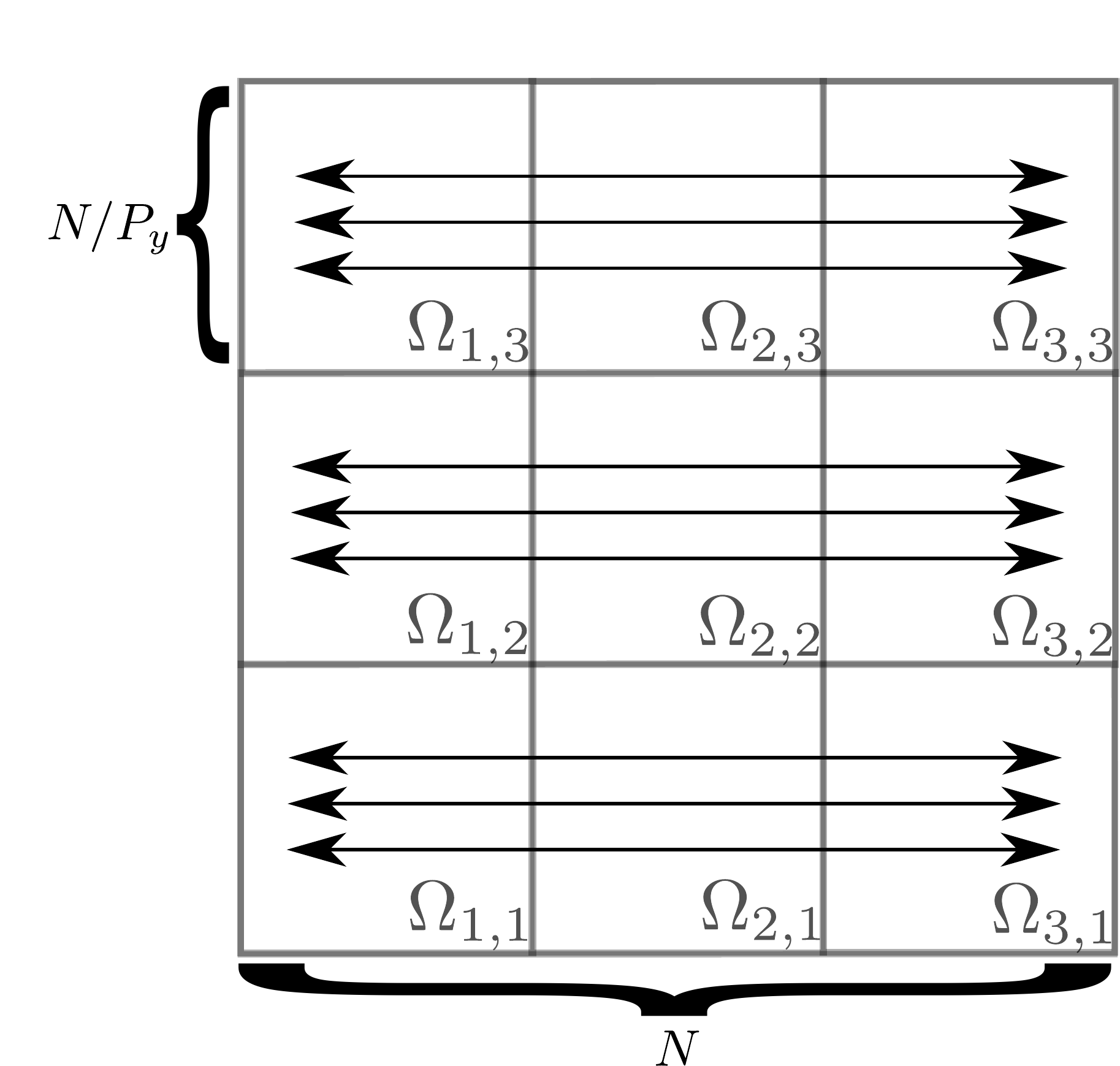} 
      \label{fig:LinearSystemXDirection}}
    \qquad 
    \subfigure[]{\includegraphics[width=0.44\textwidth]{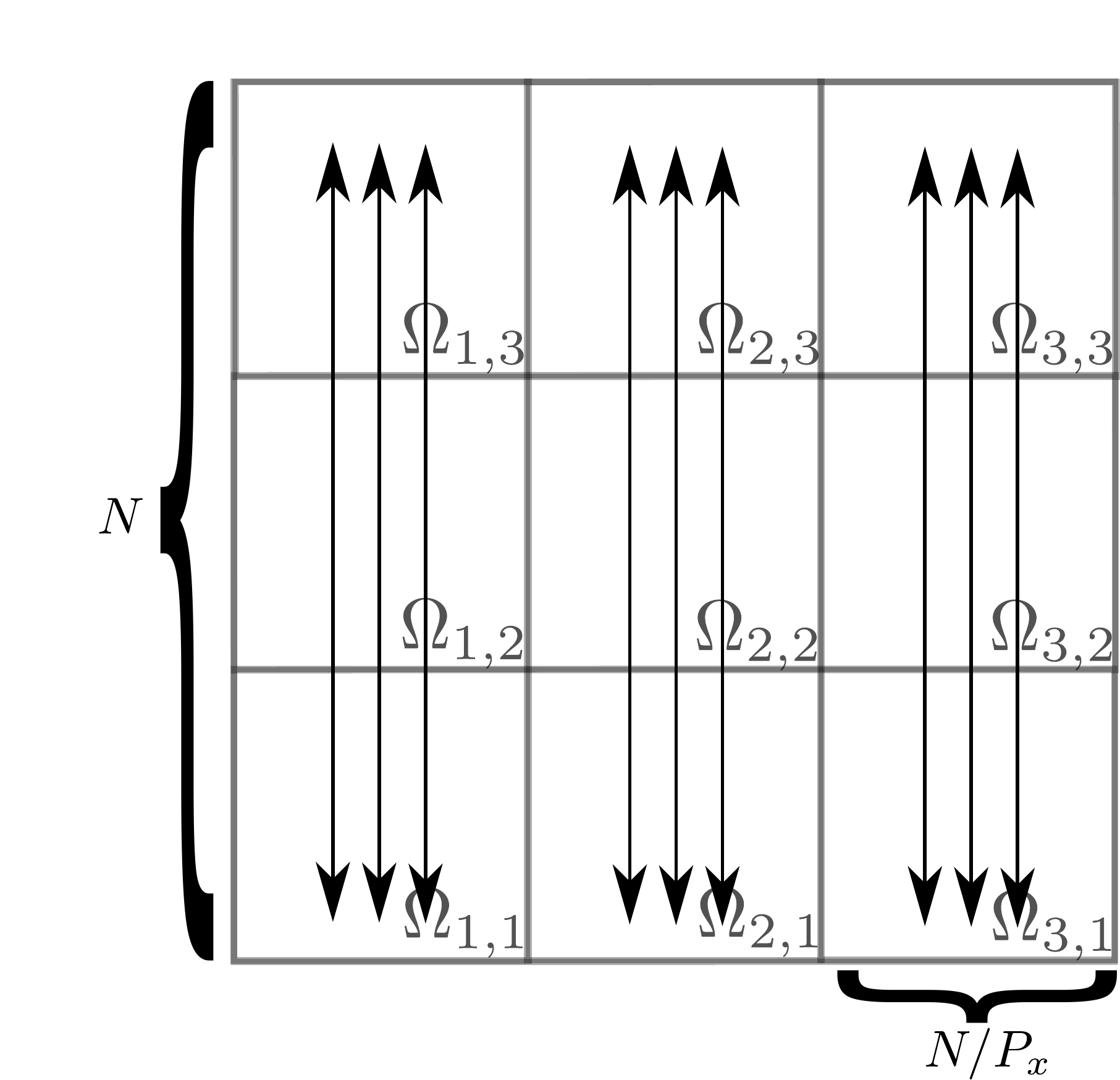} 
      \label{fig:LinearSystemYDirection}}
    \caption{\subref{fig:LinearSystemXDirection} Coupling direction for
      linear systems \eqref{eq:umacstep1} and
      \eqref{eq:PenaltyStepPart2}. \subref{fig:LinearSystemYDirection}
      Coupling direction for linear systems \eqref{eq:umacstep2} and
      \eqref{eq:PenaltyStepPart1}.  Each processing node participates in
      solving $N/P_{x,y}$ tridiagonal systems and requires communication
      in the direction of the arrows.}
    \label{fig:TriDiagSystems} 
  \end{center}
\end{figure}

Because the processing node $(\ell,m)$ contains only fluid data
residing in subdomain $\Omega_{\ell,m}$, these tridiagonal linear
systems divide naturally between nodes.  Each node solves those
linear systems for which it has the corresponding data $b_{i,j} \in
\Omega_{\ell,m}$ as illustrated in Figure~\ref{fig:TriDiagSystems}.
For example, when solving \eqref{eq:TriDiagX} along the $x$-direction,
each processing node solves $N/P_y$ linear systems and the total work is
spread over $P_x$ nodes.  Similarly, when solving the corresponding
systems along the $y$-direction, each processing node solves $N/P_x$
systems spread over $P_y$ nodes.

Each periodic tridiagonal system is solved directly using a
Schur-complement technique~\cite[sec.~14.2.1]{Saad1996}. This is
achieved by rewriting the linear equations as a block-structured system
where the interfaces between blocks correspond to those for the
subdomains.  To illustrate, let us consider an example with $P=2$
processors only, for which the periodic tridiagonal
system
\newcommand{\colorhline}{\arrayrulecolor{lightgray}\hline}
\newcommand{\colorvline}{\color{lightgray}\vrule}
\begin{gather*}
  \left[ \begin{array}{cccccccccc}
      a_1 & b_1 &        &         &         &         &         &         &        & c_1 \\
      c_2 & a_2 & b_2    &         &         &         &         &         &        &     \\
          &     & \ddots &         &         &         &         &         &        &     \\
          &     &        & \ddots  &         &         &         &         &        &     \\ 
          &     &        & c_{M-1} & a_{M-1} & b_{M-1} &         &         &        &     \\\colorhline 
          &     &        &         & c_M     & a_M     & b_M     &         &        &     \\ 
          &     &        &         &         & c_{M+1} & a_{M+1} & b_{M+1} &        &     \\ 
          &     &        &         &         &         &         & \ddots  &        &     \\ 
          &     &        &         &         &         &         &         & \ddots &     \\ 
      b_N &     &        &         &         &         &         &         & c_N    & a_N
    \end{array} \right]
  \left[ \begin{array}{c}
      y_1 \\
      x_2 \\
      \vdots \\
      \vdots \\
      x_{M-1} \\\colorhline
      y_2 \\
      x_{M+1} \\
      \vdots \\
      \vdots \\
      x_N 
    \end{array} \right]
  =
  \left[ \begin{array}{c}
      g_1 \\
      f_2 \\
      \vdots \\
      \vdots \\
      f_{M-1} \\\colorhline
      g_2 \\
      f_{M+1} \\
      \vdots \\
      \vdots \\
      f_N
    \end{array} \right]
\end{gather*}
arises from a single row of unknowns in
Figure~\ref{fig:TriDiagSystems}\subref{fig:LinearSystemXDirection} (or a
single column in
Figure~\ref{fig:TriDiagSystems}\subref{fig:LinearSystemYDirection}).  In
this example, the indices $M-1$ and $M$ refer to the subdomain boundary
points (denoted with a vertical line in the matrix above) so that the
data ($y_1$, $x_2$, \dots, $x_{M-1}$, $g_1$, $f_2$, \dots, $f_{M-1}$)
reside on processor~1 and ($y_2$, $x_{M+1}$, \dots, $x_{N}$, $g_2$,
$f_{M+1}$, \dots, $f_{N}$) on processor~2.  To isolate the coupling
between subdomains, the rows in the matrix are reordered to shift the
unknowns at periodic subdomain boundaries ($y_1$ and $y_2$) to the last
two rows, and then the columns are reordered to keep the diagonal
entries on the main diagonal.  This yields the equivalent linear system
\begin{gather*}
  \left[ \begin{array}{cccc!{\colorvline}cccc!{\colorvline}cc}
      a_2 & b_2    &         &         &         &         &        &         & c_2    & \\
          & \ddots &         &         &         &         &        &         &        & \\
          &        & \ddots  &         &         &         &        &         &        & \\
          &        & c_{M-1} & a_{M-1} &         &         &        &         &        & b_{M-1} \\\colorhline 
          &        &         &         & a_{M+1} & b_{M+1} &        &         &        & c_{M+1} \\ 
          &        &         &         &         & \ddots  &        &         &        & \\ 
          &        &         &         &         &         & \ddots &         &        & \\ 
          &        &         &         &         &         & c_N    & a_N     & b_N    & \\\colorhline 
      b_1 &        &         &         &         &         &        & c_1     & a_1    & \\
          &        &         & c_M     & b_M     &         &        &         &        & a_M 
    \end{array} \right]
  \left[ \begin{array}{c}
      x_2    \\
      \vdots \\
      \vdots \\
      x_{M-1}\\\colorhline
      x_{M+1}\\
      \vdots \\
      \vdots \\
      x_N    \\\colorhline
      y_1    \\
      y_2
    \end{array} \right]
  =
  \left[ \begin{array}{c}
      f_2    \\
      \vdots \\
      \vdots \\
      f_{M-1}\\\colorhline
      f_{M+1}\\
      \vdots \\
      \vdots \\
      f_N    \\\colorhline
      g_1    \\
      g_2    
    \end{array} \right],
\end{gather*}
which has the block structure
\begin{gather*}
  \left[ \begin{array}{ccc}
      \bs{B}_1 &     & \bs{E}_1  \\
      & \bs{B}_2 & \bs{E}_2  \\
      \bs{F}_1 & \bs{F}_2 & \bs{C} 
    \end{array} \right]
  \left[ \begin{array}{c}
      \bs{x}_1 \\
      \bs{x}_2 \\
      \bs{y}
    \end{array} \right]
  =
  \left[ \begin{array}{c}
      \bs{f}_1 \\
      \bs{f}_2 \\
      \bs{g}
    \end{array} \right].
\end{gather*}
In the more general situation with $P$ subdomains, the block structure
becomes
\begin{gather*}
  \left[ \begin{array}{ccccc}
      \bs{B}_1 &          &        &          & \bs{E}_1  \\
               & \bs{B}_2 &        &          & \bs{E}_2  \\
               &          & \ddots &          & \vdots  \\
               &          &        & \bs{B}_P &  \bs{E}_P  \\
      \bs{F}_1 & \bs{F}_2 & \cdots & \bs{F}_P & \bs{C}
    \end{array} \right]
  \left[ \begin{array}{c}
      \bs{x}_1 \\
      \bs{x}_2 \\
      \vdots \\
      \bs{x}_P \\
      \bs{y}
    \end{array} \right]
  =
  \left[ \begin{array}{c}
      \bs{f}_1 \\
      \bs{f}_2 \\
      \vdots \\
      \bs{f}_P \\
      \bs{g}
    \end{array} \right],
\end{gather*}
or more compactly
\begin{gather}
  \label{eq:BlockSchurSystem} 
  \left[ \begin{array}{ccc}
      \bs{B} & \bs{E}  \\
      \bs{F} & \bs{C}  \\
    \end{array} \right]
  \left[ \begin{array}{c}
      \bs{x} \\
      \bs{y}
    \end{array} \right]
  =
  \left[ \begin{array}{c}
      \bs{f} \\
      \bs{g}
    \end{array} \right],
\end{gather}
where $\bs{C} \in \R^{P \times P}$, $\bs{B} \in \R^{(N-P) \times (N-P)}$, 
$\bs{E} \in \R^{(N-P) \times P}$, and $\bs{F} \in \R^{P \times (N-P)}$.
Here, $\bs{x}$ and $\bs{f}$ denote the data located in the interior of a
subdomain while $\bs{y}$ and $\bs{g}$ denote the data residing on the
interface between subdomains. 
Next, we use the LU factorization to
rewrite the block matrix from \eqref{eq:BlockSchurSystem} as
\begin{gather*}
  \left[ \begin{array}{ccc}
      \bs{B} & \bs{E}  \\
      \bs{F} & \bs{C}  \\
    \end{array} \right]
  =
  \left[ \begin{array}{ccc}
      \bs{I} & \bs{0}  \\
      \bs{F}\bs{B}^{-1} & \bs{I}
    \end{array} \right]
  \left[ \begin{array}{ccc}
      \bs{B} & \bs{E}  \\
      \bs{0} & \bs{S} 
    \end{array} \right],
\end{gather*}
where $\bs{S} = \bs{C} - \bs{F} \bs{B}^{-1} \bs{E}$ is the Schur
complement.  Using this factorized form, we can decompose the block
system into the following three smaller problems:
\begin{align}
  \label{eq:LocalTriSystem}
  \bs{B} \bs{f}^\mystar &= \bs{f}, \\
  \label{eq:SchurSystem}
  \bs{S} \bs{y} &= \bs{g} - \bs{F} \bs{f}^\mystar, \\
  \label{eq:SchurCorrection}
  \bs{B} \bs{x} &= \bs{B} \bs{f}^\mystar - \bs{E} \bs{y}.
\end{align}

Based on this decomposition, we can now summarize the solution procedure
as follows:
\begin{itemize}
\item \emph{Local tridiagonal solver:} Each processor solves a 
  local non-periodic tridiagonal system
  \begin{gather*}
    \bs{B}_p \bs{f}^\mystar_p = \bs{f}_p, 
  \end{gather*}
  which can be solved efficiently using Thomas's algorithm.  The
  matrices $\bs{B}_p$ are the non-periodic tridiagonal blocks in the
  block diagonal matrix $\bs{B}$.
\item \emph{Gather data to master node:} Each processor sends three
  scalar values to the master node corresponding to the first and last
  entries of the vector $\bs{f}^\mystar_p$, as well as the scalar
  $g_p$. Because $\bs{F}_p$ is sparse, only a few values are required to
  construct the right hand side of the Schur complement system.
\item \emph{Solve Schur complement system:} On the master node, solve
  the reduced $P \times P$ Schur complement
  system~\eqref{eq:SchurSystem}.  Based on the sparsity patterns of
  $\bs{F}$ and $\bs{E}$, the Schur complement matrix $\bs{S}$ is
  periodic and tridiagonal and therefore can be inverted efficiently
  using Thomas's algorithm.
\item \emph{Scatter data from master node:} The master node scatters two
  scalar values from $\bs{y}$ to each processor.  Because of the
  sparsity of $\bs{B}^{-1}\bs{E}$, only a few values of $\bs{y}$ are
  required in the next step. Therefore, the $p$th processor only
  requires the entries of $\bs{y}$ numbered $p$ and $\text{mod}(p+1,P)$.
\item \emph{Correct local solution:} Each processor corrects its local
  solution 
  \begin{gather*}
    \bs{x}_p = \bs{f}^\mystar_p - \bs{B}^{-1}_p \bs{E}_p \bs{y},
  \end{gather*}
  using the local values $\bs{f}^\mystar_p$ computed in the first step. 
\end{itemize}
The tridiagonal systems above can be parallelized very efficiently.  As
already indicated earlier, this procedure only requires two collective
communications -- scatter and gather -- and because global communication
only occurs along one spatial direction the communication overhead
increases only marginally with the number of processors.  A further cost
savings derives from the fact that the tridiagonal systems do not change
from one time step to the next, and so the matrices $\bs{S}$ and
$\bs{B}^{-1}\bs{E}$ can be precomputed.

The only potential bottleneck in this procedure is in solving the
reduced Schur complement system \eqref{eq:SchurSystem}.  
Since the reduced system is solved only on
the master node, clock cycles on the remaining idle nodes are wasted
at this time.  Furthermore, this wasted time increases as the number of 
processors increase since the Schur complement system grows with $P$.
Fortunately, the IB algorithm never solves just a single tridiagonal
system.  For example, when solving \eqref{eq:TriDiagX} in the $x$-direction,
$P=P_x$ processing nodes work together to solve $N/P_y$ tridiagonal systems.
Therefore, the $N/P_y$ systems when solved together require solving 
$N/P_y$ different Schur complement systems. This workload can be 
spread out evenly between the $P$ processors thereby keeping all 
processors occupied.

\section{Numerical Results}
\label{sec:Results}

\changed{ To test the accuracy of our algorithm, we consider two model
  problems.  The first is an idealized one-dimensional elliptical
  membrane with zero thickness that is immersed in a 2D fluid and
  undergoes a damped periodic motion. Here, the immersed boundary exerts
  a singular force on the fluid and results in a pressure jump across
  the membrane that reduces the method's order of accuracy.  The second
  model problem is a generalization of the first in which we consider a
  thick immersed boundary, made up of multiple fiber layers in which the
  elastic stiffness is reduced smoothly to a value of zero at the edges.
  By providing the immersed boundary in this example with a physical
  thickness, the external force is no longer singular, which then leads
  to higher-order convergence rates.  }

%%%%%%%%%%%%%%%%%%%%%%%%%%%%%%%%%%%%%%%%%%%%%%%%%%%%%%%%%%%%%%%%%%%%%%%%%%%%%
\subsection{Thin Ellipse}
\label{sec:ThinEllipseProblem}

\begin{figure}[!tbp]
  \begin{center}
    \subfigure[]{\includegraphics[width=0.40\textwidth]{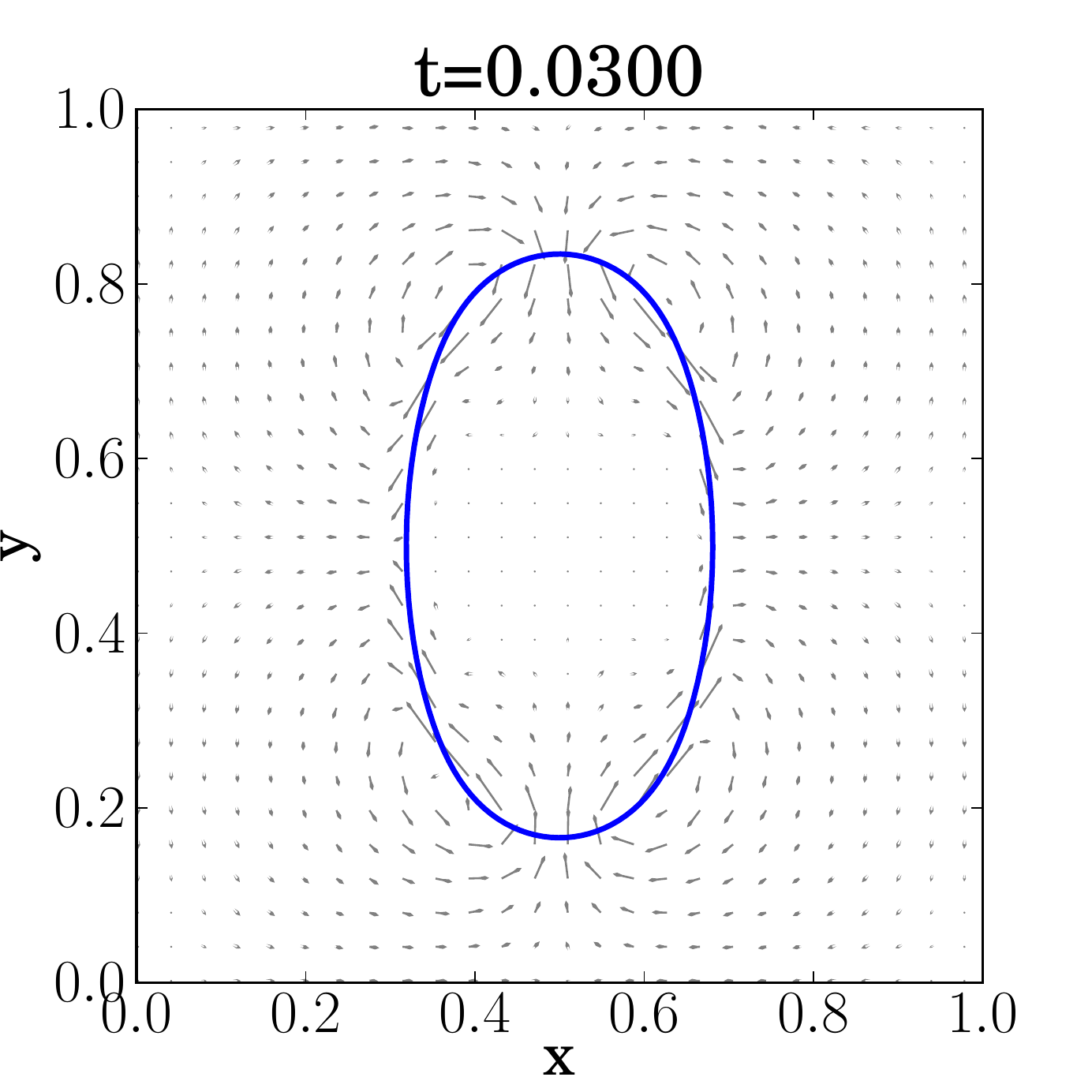} 
      \label{fig:ThinEllipseSim1}}
    \qquad 
    \subfigure[]{\includegraphics[width=0.40\textwidth]{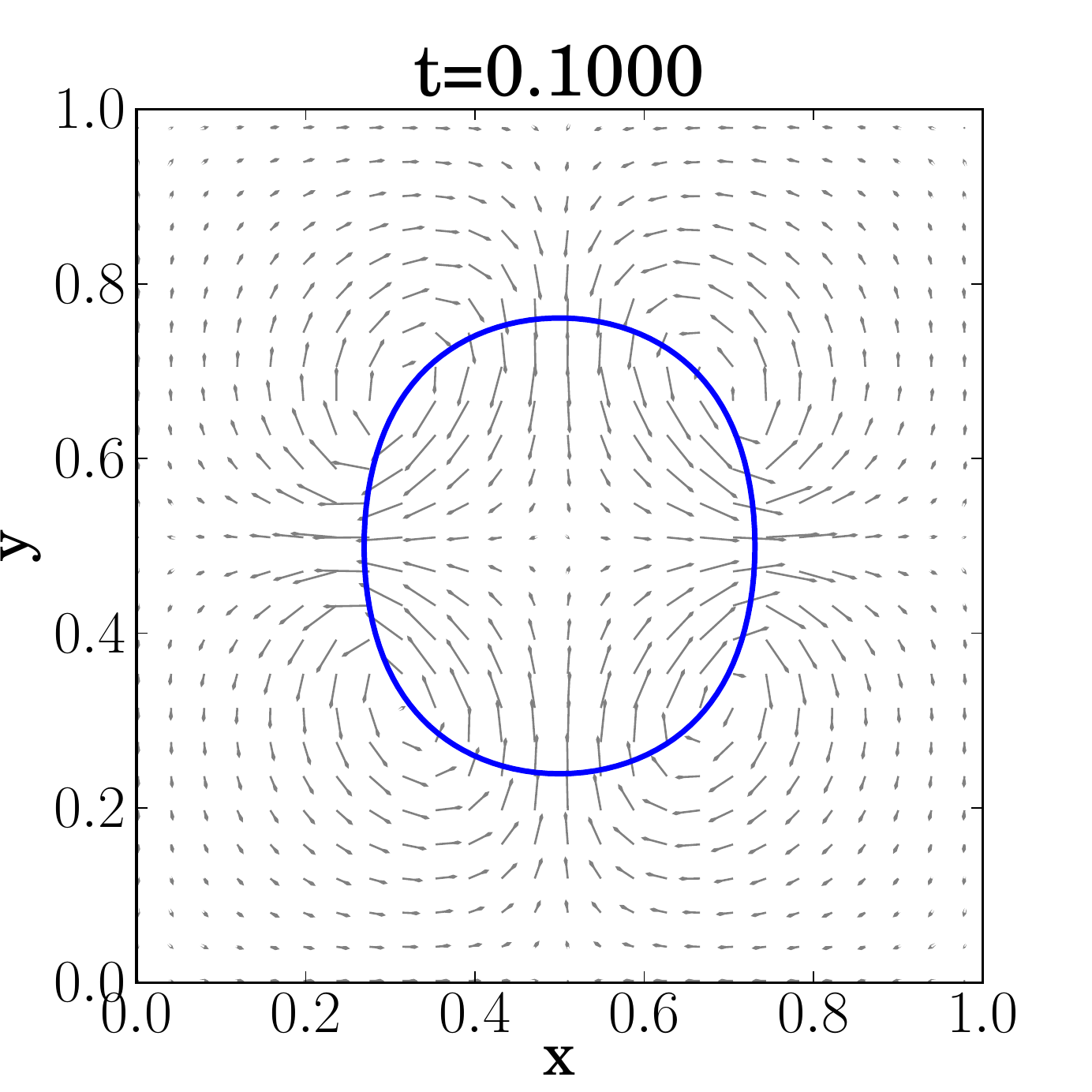} 
      \label{fig:ThinEllipseSim2}}
    \qquad 
    \subfigure[]{\includegraphics[width=0.40\textwidth]{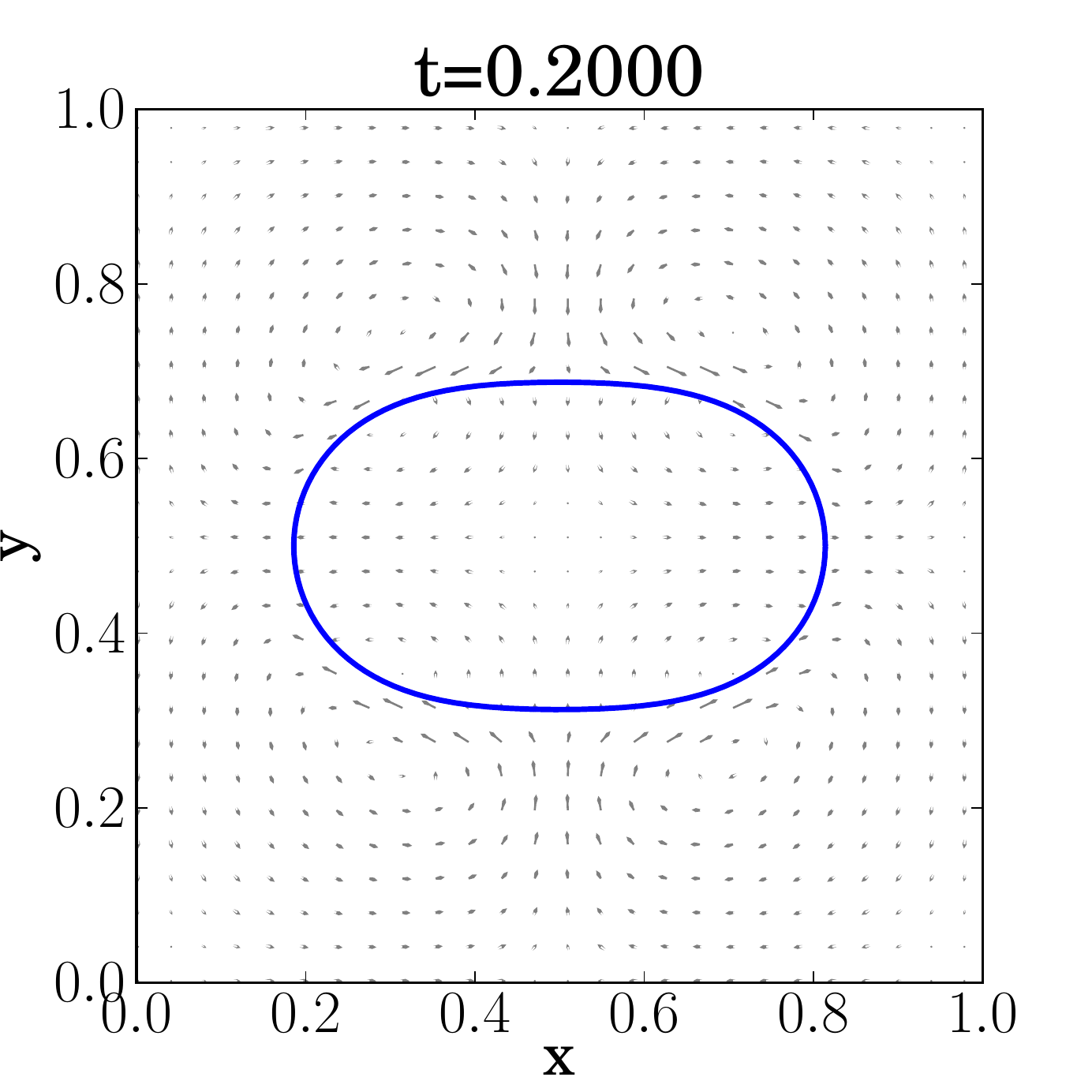} 
      \label{fig:ThinEllipseSim3}}
    \qquad 
    \subfigure[]{\includegraphics[width=0.40\textwidth]{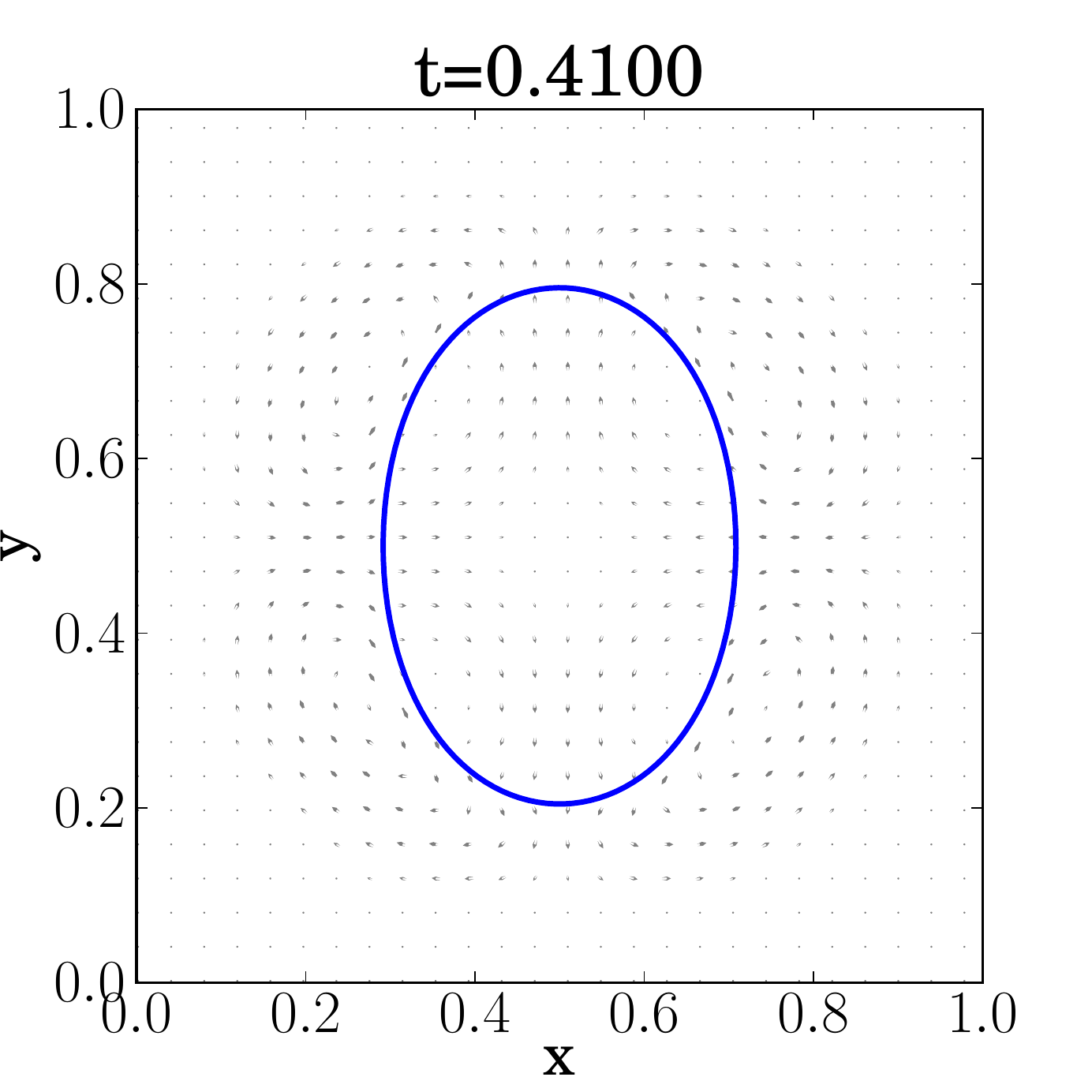} 
      \label{fig:ThinEllipseSim4}}
    \caption{Snapshots of a thin oscillating ellipse using the GM-IB
      method, with parameters $\sigma=1$, $N=256$ and $\dt=0.04/512$.} 
    \label{fig:ThinEllipseSim} 
  \end{center}
\end{figure}

For our first 2D model problem, the initial configuration is an
elliptical membrane with semi-axes $r_1$ and $r_2$, parameterized by
\begin{gather*}
  \bs{X}(s,0) = \left( \half + r_1 \cos(2 \pi s) ,~
    \half + r_2  \sin(2 \pi s) \right),
\end{gather*}
with $s\in[0,1]$.  The ellipse is placed in a unit square containing
fluid that is initially stationary with $\bs{u}(\bs{x},0)=0$.  We see
from the solution snapshots in Figure~\ref{fig:ThinEllipseSim} that the
elastic membrane undergoes a damped periodic motion, oscillating back
and forth between elliptical shapes having a semi-major axis aligned
with the $x$- and $y$-directions.  The amplitude of the oscillations
decreases over time, and the membrane tends ultimately toward a circular
equilibrium state with radius approximately equal to $\sqrt{r_1 r_2}$
(which has the same area as the initial ellipse).

For this problem, we actually computed results using two immersed boundary
algorithms corresponding to different fluid solvers.  The first
algorithm, denoted GM-IB, is the same one described in
section~\ref{sec:algorithm} that uses Guermond and Minev's fluid
solver. The second algorithm, denoted BCM-IB, is identical to the first
except that the fluid solver is replaced with the second-order projection
method described by Brown, Cortez, and Minion~\cite{Brown2001}.  We take
values of the parameters from Griffith~\cite{Griffith2012}, who used
$\mu=0.01$, $\rho=1$, $r_1=\frac{5}{28}$, $r_2=\frac{7}{20}$ and
$N_s=\frac{19}{4}N$.  We then compare our numerical results for
different choices of the membrane elastic stiffness ($\sigma$) and
spatial discretization ($N$).  Unless stated otherwise, the time step is
chosen so that the simulation is stable on the finest spatial grid (with
$N=512$).  This is a conservative choice for the time step that attempts
to avoid any unreasonable accumulation of errors in time, but it also
provides limited information regarding the time step restrictions for
the two methods.  \changed{However, we observe in practice that there is
  little difference between the time step restrictions for the GM-IB and
  BCM-IB algorithms, although GM-IB does have a slightly stricter time
  step restriction than BCM-IB.}

Because the fluid contained within the immersed boundary cannot escape,
the area of the oscillating ellipse should remain constant in time.
However, many other IB computations for this thin ellipse problem
exhibit poor volume conservation which manifests itself as an apparent
``leakage'' of fluid out of the immersed boundary.  The source of this
volume conservation error is numerical error in the discrete
divergence-free condition for the interpolated velocity field located on
immersed boundary points, which can be non-zero even when the fluid
solver guarantees that the velocity is discretely divergence-free on the
Eulerian fluid grid~\cite{NewrenPhdthesis2007,PeskinPrintz1993}.
Griffith~\cite{Griffith2012} observed that volume conservation can be
improved by using a pressure-increment fluid solver instead of a
pressure-free solver, and furthermore that fluid solvers based on a
staggered grid tended to perform better than those using a collocated
grid.  We have employed both of these ideas in our proposed method and
so we expect to see significant improvement in volume conservation
relative to other IB methods.

\begin{figure}[!tbp]
  \begin{center}
    \subfigure[]{\includegraphics[width=0.45\textwidth]{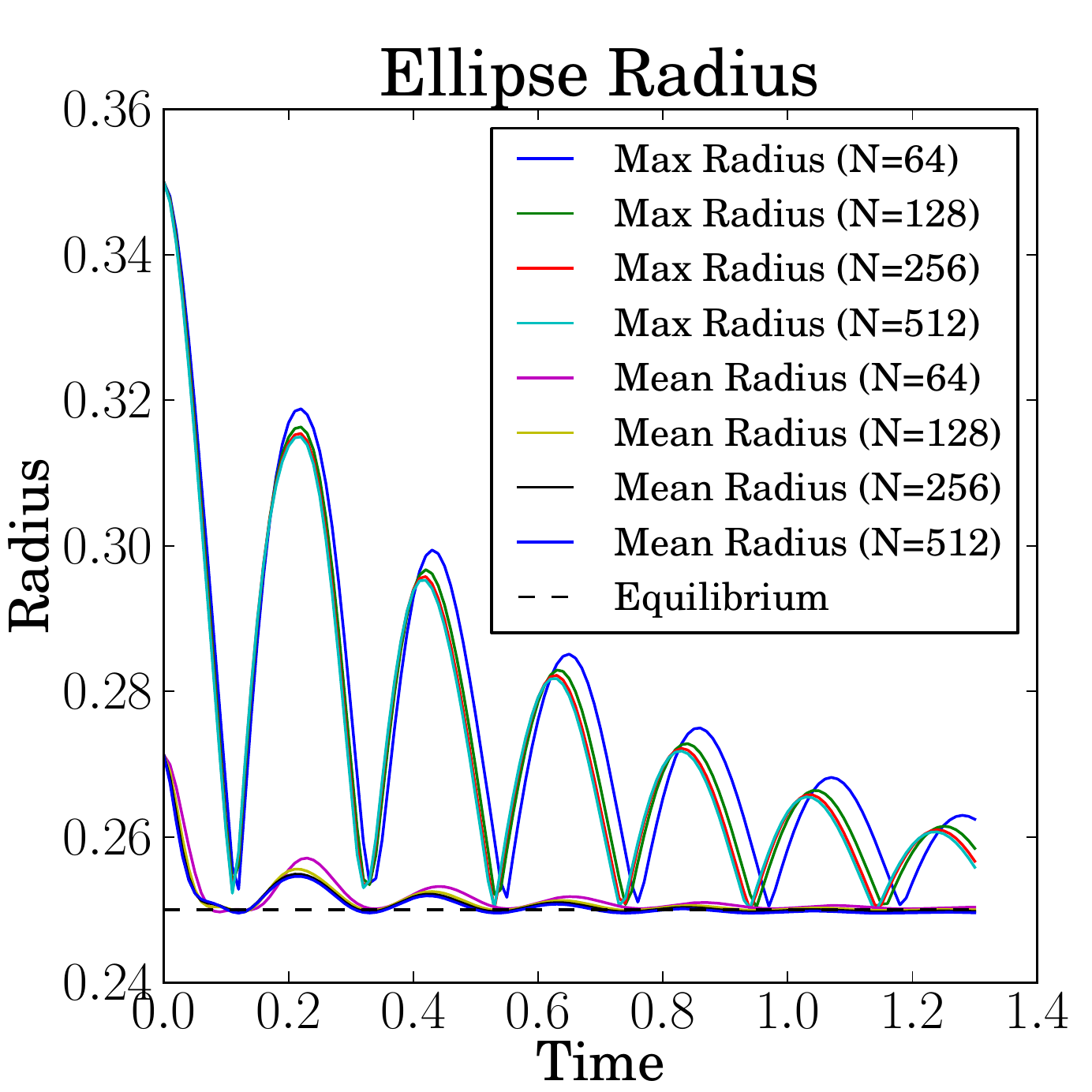} 
      \label{fig:ThinEllipseRadius}}
    \qquad 
    \subfigure[]{\includegraphics[width=0.45\textwidth]{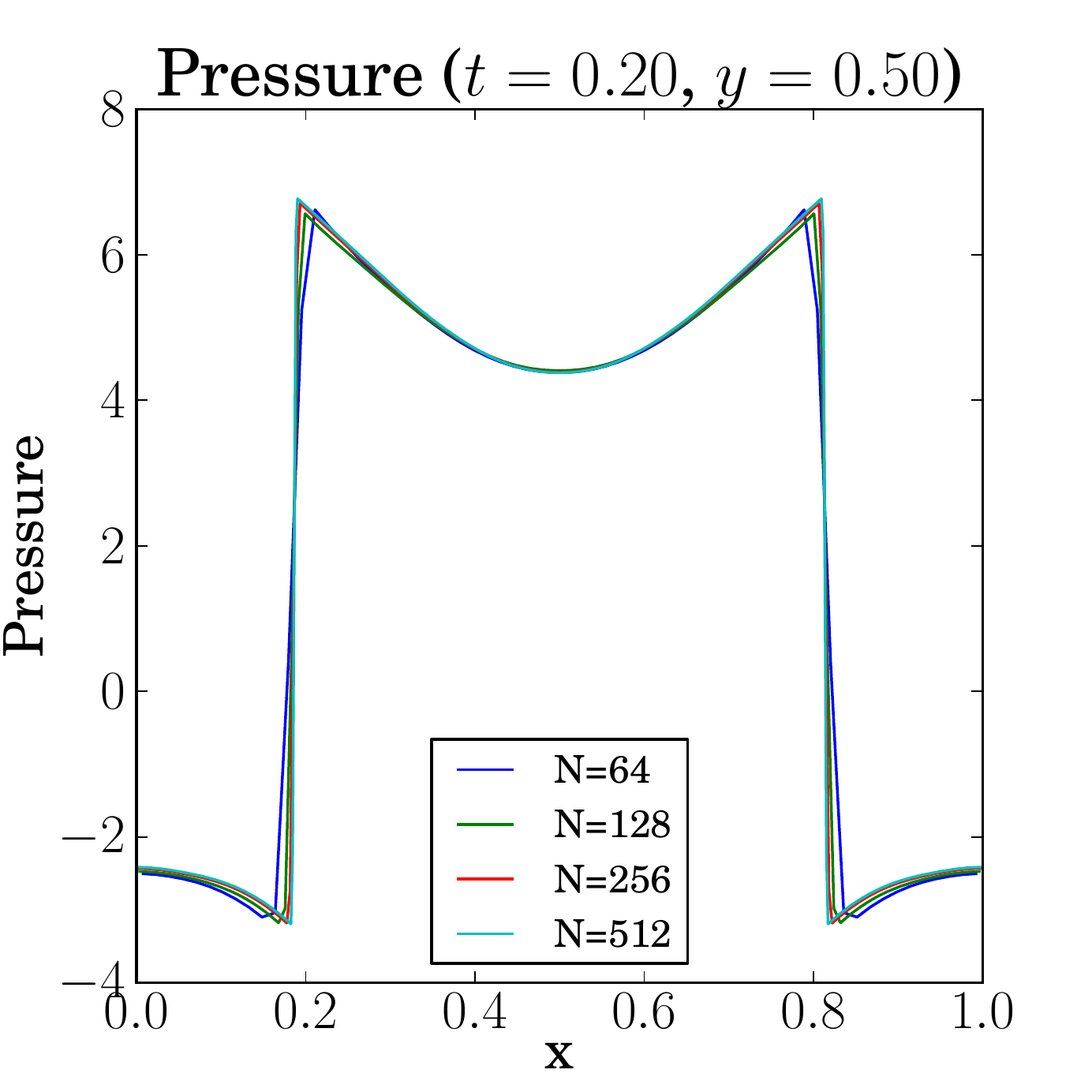} 
      \label{fig:ThinEllipsePressure}}
    \caption{Results for the thin ellipse problem using the GM-IB method
      with $\sigma=1$ and $\dt=0.04/512$. \subref{fig:ThinEllipseRadius}
      Maximum and mean radii. \subref{fig:ThinEllipsePressure} Pressure
      slices across the $x$--axis with $y=0.5$ and $t=0.2$.}
    \label{fig:ThinEllipseRadiusPressure} 
  \end{center}
\end{figure}

We begin by plotting the maximum and mean radii of the ellipse versus
time in
Figure~\ref{fig:ThinEllipseRadiusPressure}\subref{fig:ThinEllipseRadius},
from which it is clear that the immersed boundary converges to a
circular steady state having radius $\sqrt{r_1 r_2}=\frac{1}{4}$.  The
BCM-IB results are indistinguishable from those using GM-IB, and so only
the latter are depicted in this figure.  The low rate of volume loss
observed in both algorithms is consistent with the numerical experiments
of Griffith~\cite{Griffith2012}.  Owing to the relatively high Reynolds
number for this flow ($\Reynolds \approx 150$) there exists a noticeable
error in the oscillation frequency for coarser discretizations, although
we note that this error is much smaller for lower $\Reynolds$ flows.  We
suspect that this frequency error could be reduced significantly by
employing higher-order approximations in the nonlinear advection term
and the IB evolution equation~\eqref{eq:membrane}, such as has been done
by Griffith~\cite{Griffith2007}.  Finally, we note that
Figure~\ref{fig:ThinEllipseRadiusPressure}\subref{fig:ThinEllipsePressure}
shows that the GM-IB algorithm captures the discontinuity in pressure
without any visible oscillations.

We next estimate the error and convergence rate for both algorithms.
Because the thin ellipse problem is characterized by a singular IB
force, there is a discontinuity in velocity derivatives and
pressure and so our numerical scheme is limited to first order accuracy.
We note that improvements in the convergence rate could be achieved by
explicitly incorporating these discontinuities into the difference
scheme, for example as is done in the immersed interface method
\cite{Lee2003,Leveque1997}.  

When reporting the error in a discrete variable $q_N$ that is
approximated on a grid at refinement level $N$, we use the notation
\begin{gather}
  \myerror{q}{N} = \| q_N - q_{\text{exact}} \|_2.  
  \label{eq:EstError}
\end{gather}
Because the exact solution for the thin ellipse problem is not known, we
estimate $q_\text{exact}$ by using the approximate solution on the
finest mesh corresponding to $N_f=512$ \changed{with the BCM-IB algorithm}, and then take
$q_{\text{exact}}=\mathcal{I}^{N_f\to N} q_{N_f}$, where
$\mathcal{I}^{N_f \to N}$ is an operator that interpolates the finest
mesh solution $q_{N_f}$ onto the current coarse mesh with $N$ points.
We use the discrete $\ell^2$ norm to estimate errors, which is
calculated for an Eulerian quantity such as the pressure using
\begin{gather}
  \|p _{i,j}\|_2 = \brac{ h^2 \sum_{i,j} |p_{i,j}|^2 }^{1/2},
  \label{eq:ScalarDiscreteL2}
\end{gather}
and similarly for a Lagrangian quantity such as the IB position using  
\begin{gather}
  \| \bs{X}_k\|_2 = \brac{ h_s \sum_k |\bs{X}_k|^2 }^{1/2},
  % \mbox{~~or~~} \| \bs{X} _{k,\ell}\|_2 = \brac{ h_s h_r \sum_{k,\ell} 
  %   |\bs{X}_{k,\ell}|^2 }^{1/2}, 
  \label{eq:VectorDiscreteL2}
\end{gather}
where $|\cdot|$ represents the absolute value in the first formula and
the Euclidean distance in the second.  The convergence rate can then be
estimated using solutions $q_N$, $q_{2N}$ and $q_{4N}$ on successively
finer grids as
\begin{gather}
  \myrate{q}{N} = \log_2 \brac{ \frac{\| q_N -
      \mathcal{I}^{2N\to N} q_{2N} \|_2}{\| q_{2N} -
      \mathcal{I}^{4N\to 2N} q_{4N} \|_2} }.
  \label{eq:EstConvergence}
\end{gather}

A summary of convergence rates and errors is given in
Tables~\ref{Table:ThinEllipse_ConvergenceRate}
and~\ref{Table:ThinEllipse_Error} for both the GM-IB and BCM-IB
algorithms, taking different values of the elastic stiffness parameter
$\sigma$.  The error in all cases is measured at a time three-quarters
through the ellipse's first oscillation, when the membrane is roughly
circular in shape.  Table~\ref{Table:ThinEllipse_ConvergenceRate}
clearly shows that the two algorithms exhibit similar convergence rates
for all state variables.  First-order convergence is seen in both the
fluid velocity and membrane position, while the pressure shows the
expected reduction in accuracy to $\order{h^{1/2}}$ owing to the
pressure discontinuity.  The errors in
Table~\ref{Table:ThinEllipse_Error} show that GM-IB and BCM-IB are
virtually indistinguishable from each other except for the error in the
divergence-free condition, $\myerror{\nabla \cdot \bs{u}}{N}$, where the
BCM-IB algorithm appears to enforce the incompressibility constraint
better than GM-IB.  Because Guermond and Minev's fluid solver does not
project the velocity field onto the space of divergence-free velocity
fields (even approximately), it is not surprising that BCM-IB performs
better in this regard.  We remark that the magnitude of the fluid
variables increases with the stiffness $\sigma$, so that the error
increases as well (since ${\cal E}$ is defined as an absolute error
measure); however, the relative error and convergence rates remain
comparable as $\sigma$ varies over several orders of magnitude.

\begin{table}[htbp]\centering\small
  \caption{Estimated $\ell^2$ convergence rates for the thin
    ellipse problem with three different parameter sets:
    ($\sigma=0.1$, $t=1.06$, $\dt=0.08/512$), ($\sigma=1$, $t=0.31$, 
    $\dt=0.04/512$), ($\sigma=10$, $t=0.0975$, $\dt=0.01/512$).}  
  %\ra{1.6}
  \begin{tabular}{ll cccccccccc}\toprule
    & & \multicolumn{2}{c}{$\myrate{\bs{u}}{N}$}  & \multicolumn{2}{c}{$\myrate{p}{N}$} & \multicolumn{2}{c}{$\myrate{\bs{X}}{N}$} \\
    \cmidrule(r){3-4} \cmidrule(r){5-6}  \cmidrule(r){7-8}  
    $\sigma$ & $N$ & GM & BCM & GM & BCM & GM & BCM  \\
    \midrule
    \multirow{2}{*}{$0.1$} & $64$ & $1.02$ & $1.02$ & $0.55$ & $0.55$ & $1.46$ & $1.46$  \\
                          & $128$ & $1.05$ & $1.06$ & $0.53$ & $0.53$ & $1.28$ & $1.29$ \\
    \midrule
    \multirow{2}{*}{$1$} & $64$ & $1.48$ & $1.51$ & $0.72$ & $0.73$ & $1.34$ & $1.35$  \\
                        & $128$ & $0.96$ & $1.03$ & $0.57$ & $0.58$ & $1.31$ & $1.37$  \\
    \midrule
    \multirow{2}{*}{$10$} & $64$ & $1.27$ & $1.33$ & $0.88$ & $0.84$ & $1.35$ & $1.33$  \\
                         & $128$ & $0.89$ & $1.03$ & $0.68$ & $0.82$ & $1.32$ & $1.71$  \\
    \bottomrule
  \end{tabular}
  \label{Table:ThinEllipse_ConvergenceRate}
\end{table}

\begin{table}[htbp]\centering\small
  \caption{Estimated $\ell^2$ errors the thin ellipse problem with three
    different parameter sets: ($\sigma=0.1$, $t=1.06$,
    $\dt=0.08/512$), ($\sigma=1$, $t=0.31$, $\dt=0.04/512$),
    ($\sigma=10$, $t=0.0975$, $\dt=0.01/512$).}   
  %\ra{1.6}
  \begin{tabular}{ll cccccccccc}\toprule
    & & \multicolumn{2}{c}{$\myerror{\bs{u}}{N}$} & \multicolumn{2}{c}{$\myerror{p}{N}$} & \multicolumn{2}{c}{$\myerror{\bs{X}}{N}$} & \multicolumn{2}{c}{$\myerror{\nabla \cdot \bs{u}}{N}$} \\
    \cmidrule(r){3-4} \cmidrule(r){5-6}  \cmidrule(r){7-8}  \cmidrule(r){9-10}  
    $\sigma$ & $N$ & GM & BCM & GM & BCM & GM & BCM & GM & BCM  \\
    \midrule
    \multirow{4}{*}{$0.1$} & $64$ & $7.41\EE{-3}$ & $7.44\EE{-3}$ & $4.13\EE{-2}$ & $4.13\EE{-2}$ & $2.81\EE{-4}$ & $2.82\EE{-4}$ & $4.77\EE{-3}$ & $3.67\EE{-16}$ \\
                          & $128$ & $3.10\EE{-3}$ & $3.16\EE{-3}$ & $2.36\EE{-2}$ & $2.36\EE{-2}$ & $6.69\EE{-5}$ & $6.75\EE{-5}$ & $1.05\EE{-2}$ & $7.22\EE{-16}$ \\
                          & $256$ & $9.64\EE{-4}$ & $1.03\EE{-3}$ & $1.03\EE{-2}$ & $1.04\EE{-2}$ & $1.36\EE{-5}$ & $1.39\EE{-5}$ & $1.86\EE{-2}$ & $1.41\EE{-15}$ \\
                          & $512$ & $1.91\EE{-4}$ & --            & $8.84\EE{-4}$ & --            & $1.80\EE{-6}$ & --            & $2.91\EE{-2}$ & $2.83\EE{-15}$ \\
    \midrule
    \multirow{4}{*}{$1$} & $64$ & $5.61\EE{-2}$ & $5.69\EE{-2}$ & $4.56\EE{-1}$ & $4.57\EE{-1}$ & $4.32\EE{-4}$ & $4.36\EE{-4}$ & $1.04\EE{-1}$ & $1.86\EE{-15}$ \\
                        & $128$ & $1.88\EE{-2}$ & $1.98\EE{-2}$ & $2.56\EE{-1}$ & $2.57\EE{-1}$ & $1.06\EE{-4}$ & $1.09\EE{-4}$ & $2.50\EE{-1}$ & $3.60\EE{-15}$ \\
                        & $256$ & $6.32\EE{-3}$ & $6.65\EE{-3}$ & $1.13\EE{-1}$ & $1.11\EE{-1}$ & $2.08\EE{-5}$ & $2.15\EE{-5}$ & $4.55\EE{-1}$ & $7.04\EE{-15}$ \\
                        & $512$ & $5.46\EE{-3}$ & --            & $4.77\EE{-2}$ & --            & $9.21\EE{-6}$ & --            & $7.23\EE{-1}$ & $1.40\EE{-14}$ \\
    \midrule
    \multirow{4}{*}{$10$} & $64$ & $3.37\EE{-1}$ & $3.38\EE{-1}$ & $5.88\EE{+0}$ & $5.89\EE{+0}$ & $6.33\EE{-4}$ & $6.38\EE{-4}$ & $6.12\EE{-1}$ & $7.01\EE{-15}$ \\
                         & $128$ & $1.61\EE{-1}$ & $1.62\EE{-1}$ & $3.12\EE{+0}$ & $3.04\EE{+0}$ & $1.56\EE{-4}$ & $1.57\EE{-4}$ & $2.12\EE{+0}$ & $1.40\EE{-14}$ \\
                         & $256$ & $7.06\EE{-2}$ & $5.48\EE{-2}$ & $1.42\EE{+0}$ & $1.20\EE{+0}$ & $3.08\EE{-5}$ & $2.60\EE{-5}$ & $3.94\EE{+0}$ & $2.72\EE{-14}$ \\
                         & $512$ & $6.24\EE{-2}$ & --            & $1.12\EE{+0}$ & --            & $2.01\EE{-5}$ & --            & $6.34\EE{+0}$ & $5.39\EE{-14}$ \\
    \bottomrule
  \end{tabular}
  \label{Table:ThinEllipse_Error}
\end{table}

\changed{ Lastly, we examine the issue of volume conservation by
  considering the volume (area) of the membrane for the GM-IB and BCM-IB
  methods as the solution approaches steady-state. Ideally, the membrane
  area should remain constant in time with a value of $\pi r_1 r_2$
  because the fluid contained inside the immersed boundary cannot
  escape. For a membrane with an elastic stiffness of $\sigma=1$, the
  volume conservation is illustrated in
  Table~\ref{Table:ThinEllipse_AreaError}.  For both numerical schemes,
  the loss of enclosed volume is less than one percent by the time the
  solution attains a quasi-steady state near $t=4$.  The same is true of
  the corresponding simulations using $\sigma=10$ and $\sigma=0.1$,
  where the quasi-steady state is reached at $t=2$ and $t=4$
  respectively.  
  
  When comparing methods, we observe
  that BCM-IB conserves volume better than GM-IB.  However, 
  as observed in Table~\ref{Table:ThinEllipse_Error} (see $\myerror{\bs{X}}{N}$), 
  the difference in volume conservation has negligible impact on the solution's
  accuracy. 
  It is only when approaching the stability boundaries (in
  terms of the allowable time step) that the difference in volume
  conservation becomes noticeable.  
  Lastly, when reducing the time step, the volume
  conservation in the GM-IB algorithm improves noticeably, which is not surprising since the
  GM fluid solver introduces an $\order{\dt}$ perturbation to the
  incompressibility constraint~\eqref{eq:PerturbIncompressible}. Furthermore,
  from Table~\ref{Table:ThinEllipse_AreaErrorRate}, we see that leakage rate 
  of the membrane is not affected by the time step and is nearly
  identical to BCM-IB.  }

\begin{table}[htbp]\centering\small
  \caption{ The loss of enclosed volume (relative error) of the immersed boundary at time $t=4$ when $\sigma=1$. }     
  %\ra{1.6}
  \begin{tabular}{l ccccc c}\toprule
     &\multicolumn{4}{c}{GM-IB} & BCM-IB \\
     \cmidrule(r){2-5} \cmidrule(r){6-6}
     N & $\dt=\frac{0.04}{512}$ & $\dt=\frac{0.02}{512}$ & $\dt=\frac{0.01}{512}$ & $\dt=\frac{0.005}{512}$ & $\dt=\frac{0.04}{512}$  \\
    \midrule
    $64$  & $2.21\EE{-3}$ & $1.63\EE{-3}$ & $1.45\EE{-3}$ & $1.38\EE{-3}$ & $1.35\EE{-3}$ \\
    $128$ & $2.74\EE{-3}$ & $1.61\EE{-3}$ & $1.21\EE{-3}$ & $1.07\EE{-3}$ & $6.42\EE{-4}$ \\
    $256$ & $3.31\EE{-3}$ & $1.55\EE{-3}$ & $9.36\EE{-4}$ & $7.12\EE{-4}$ & $3.32\EE{-4}$ \\
    $512$ & $4.25\EE{-3}$ & $1.65\EE{-3}$ & $7.89\EE{-4}$ & $4.81\EE{-4}$ & $1.57\EE{-4}$ \\
    \bottomrule
  \end{tabular}
  \label{Table:ThinEllipse_AreaError}
\end{table}

\begin{table}[htbp]\centering\small
  \caption{ The temporal leakage rate of the membrane (relative error) over the
    time interval $t\in[2,4]$ when $\sigma=1$ which is obtained using linear least squares fit. }      
  %\ra{1.6}
  \begin{tabular}{l ccccc c}\toprule
     &\multicolumn{4}{c}{GM-IB} & BCM-IB \\
     \cmidrule(r){2-5} \cmidrule(r){6-6}
     N & $\dt=\frac{0.04}{512}$ & $\dt=\frac{0.02}{512}$ & $\dt=\frac{0.01}{512}$ & $\dt=\frac{0.005}{512}$ & $\dt=\frac{0.04}{512}$  \\
    \midrule
    $64$  & $1.03\EE{-3}$ & $1.04\EE{-3}$ & $1.05\EE{-3}$ & $1.05\EE{-3}$ & $1.05\EE{-3}$ \\
    $128$ & $6.35\EE{-4}$ & $6.39\EE{-4}$ & $6.41\EE{-4}$ & $6.41\EE{-4}$ & $6.41\EE{-4}$ \\
    $256$ & $3.29\EE{-4}$ & $3.31\EE{-4}$ & $3.32\EE{-4}$ & $3.32\EE{-4}$ & $3.32\EE{-4}$ \\
    $512$ & $1.57\EE{-4}$ & $1.57\EE{-4}$ & $1.57\EE{-4}$ & $1.57\EE{-4}$ & $1.57\EE{-4}$ \\
    \bottomrule
  \end{tabular}
  \label{Table:ThinEllipse_AreaErrorRate}
\end{table}

%%%%%%%%%%%%%%%%%%%%%%%%%%%%%%%%%%%%%%%%%%%%%%%%%%%%%%%%%%%%%%%%%%%%%%%%%%%%%
\subsection{Thick Elliptical Shell}
\label{sec:ThickEllipseProblem}

\begin{figure}[!tbp]
  \begin{center}
    \subfigure[]{\includegraphics[width=0.40\textwidth]{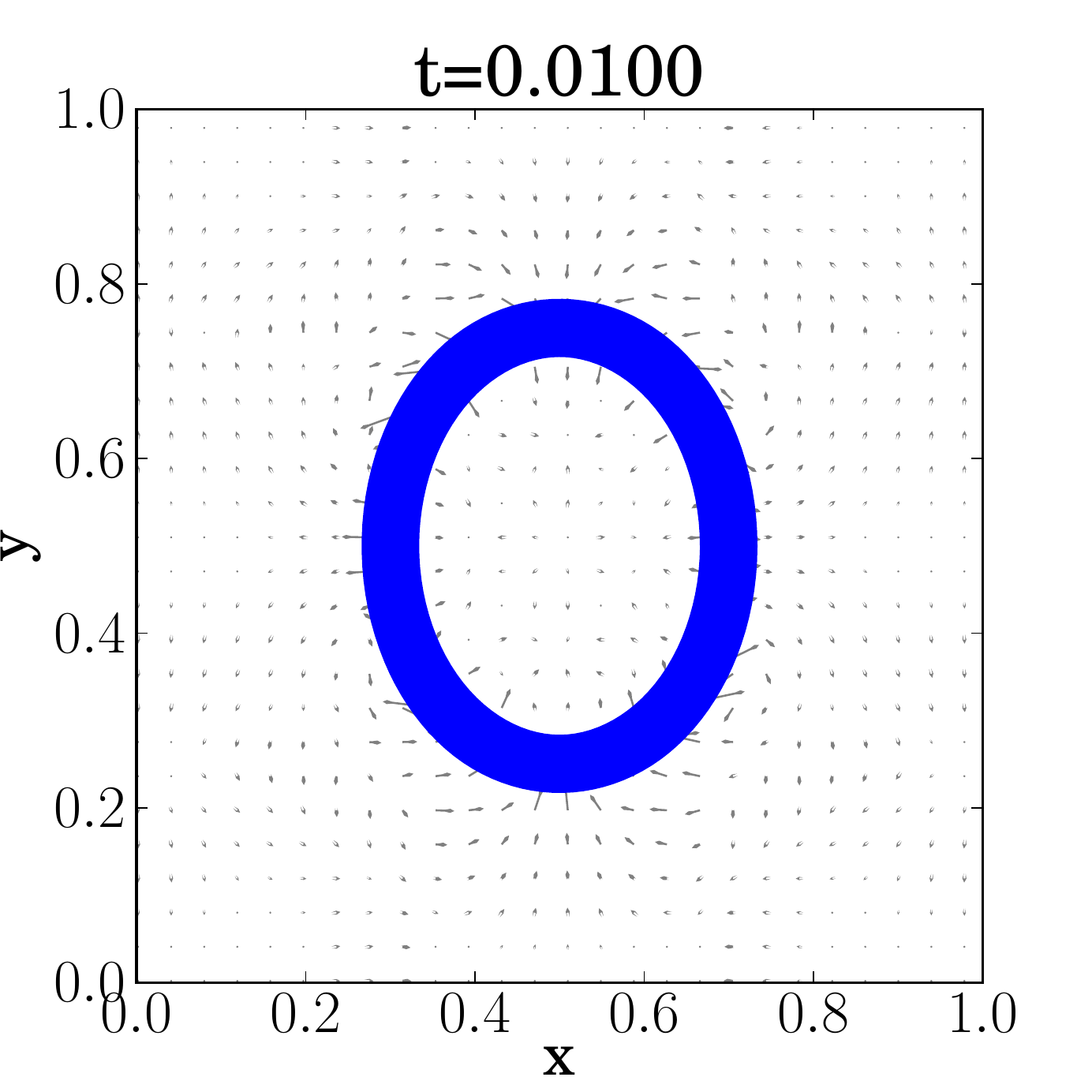} 
      \label{fig:ThickEllipseSim1}}
    \qquad 
    \subfigure[]{\includegraphics[width=0.40\textwidth]{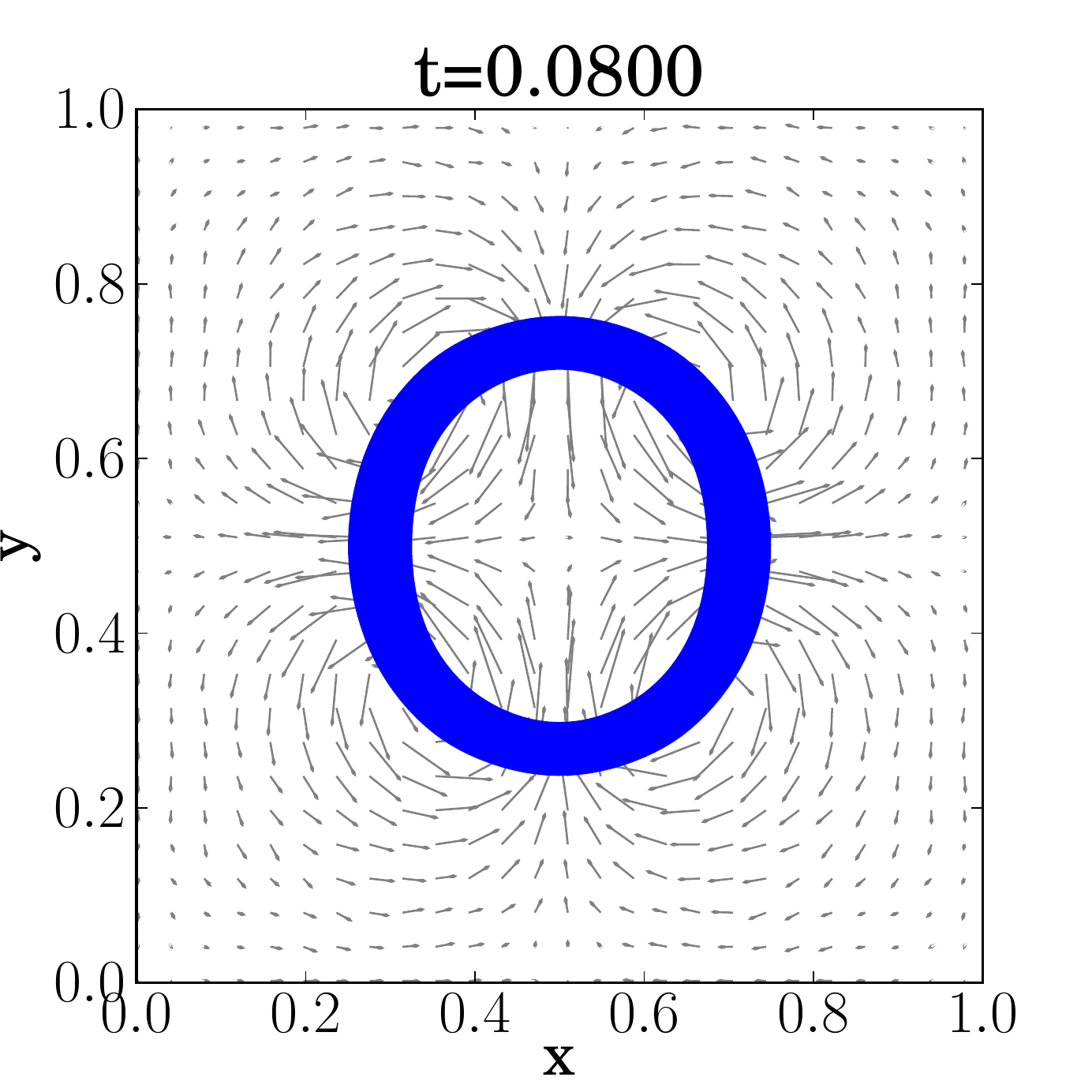} 
      \label{fig:ThickEllipseSim2}}
    \qquad 
    \subfigure[]{\includegraphics[width=0.40\textwidth]{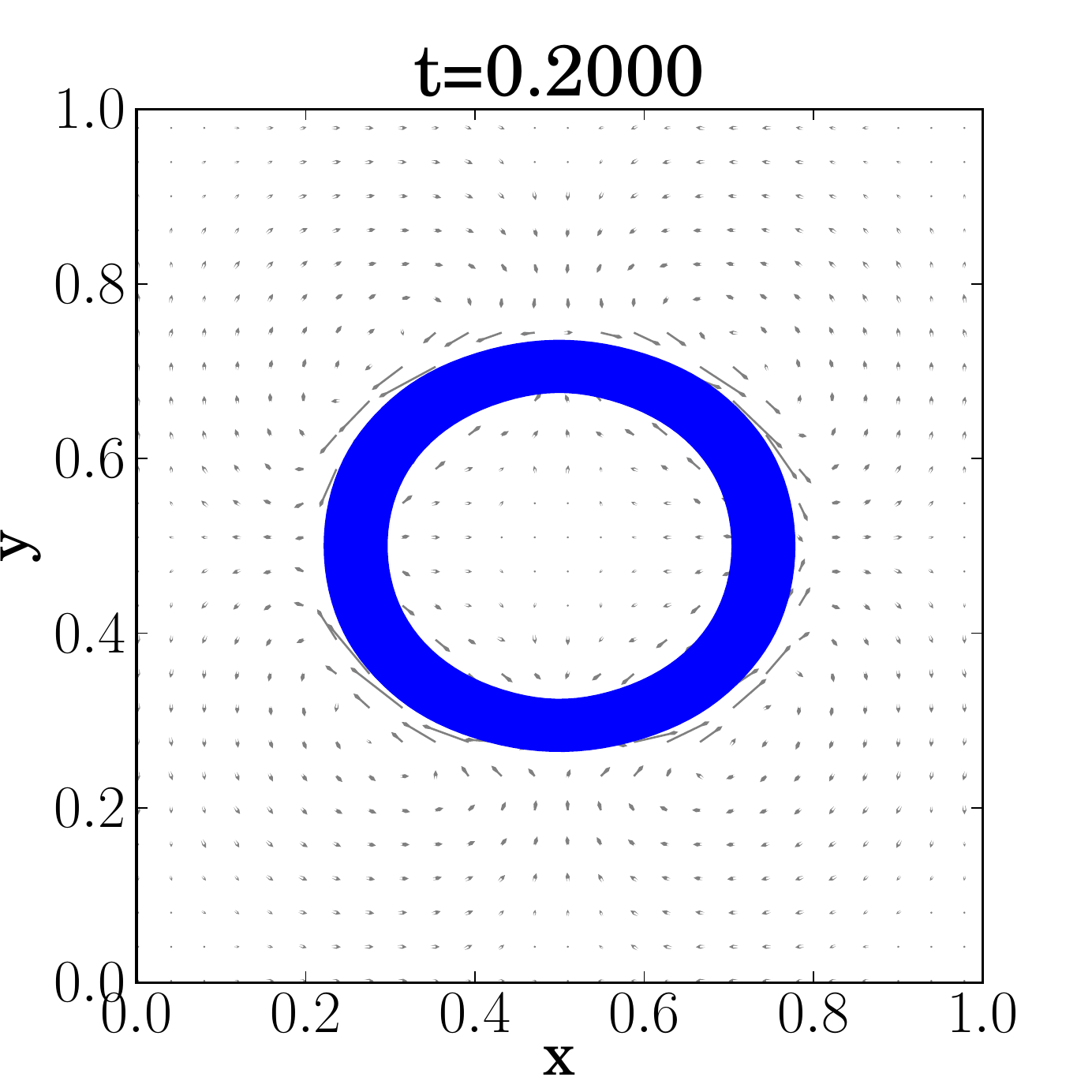} 
      \label{fig:ThickEllipseSim3}}
    \qquad 
    \subfigure[]{\includegraphics[width=0.40\textwidth]{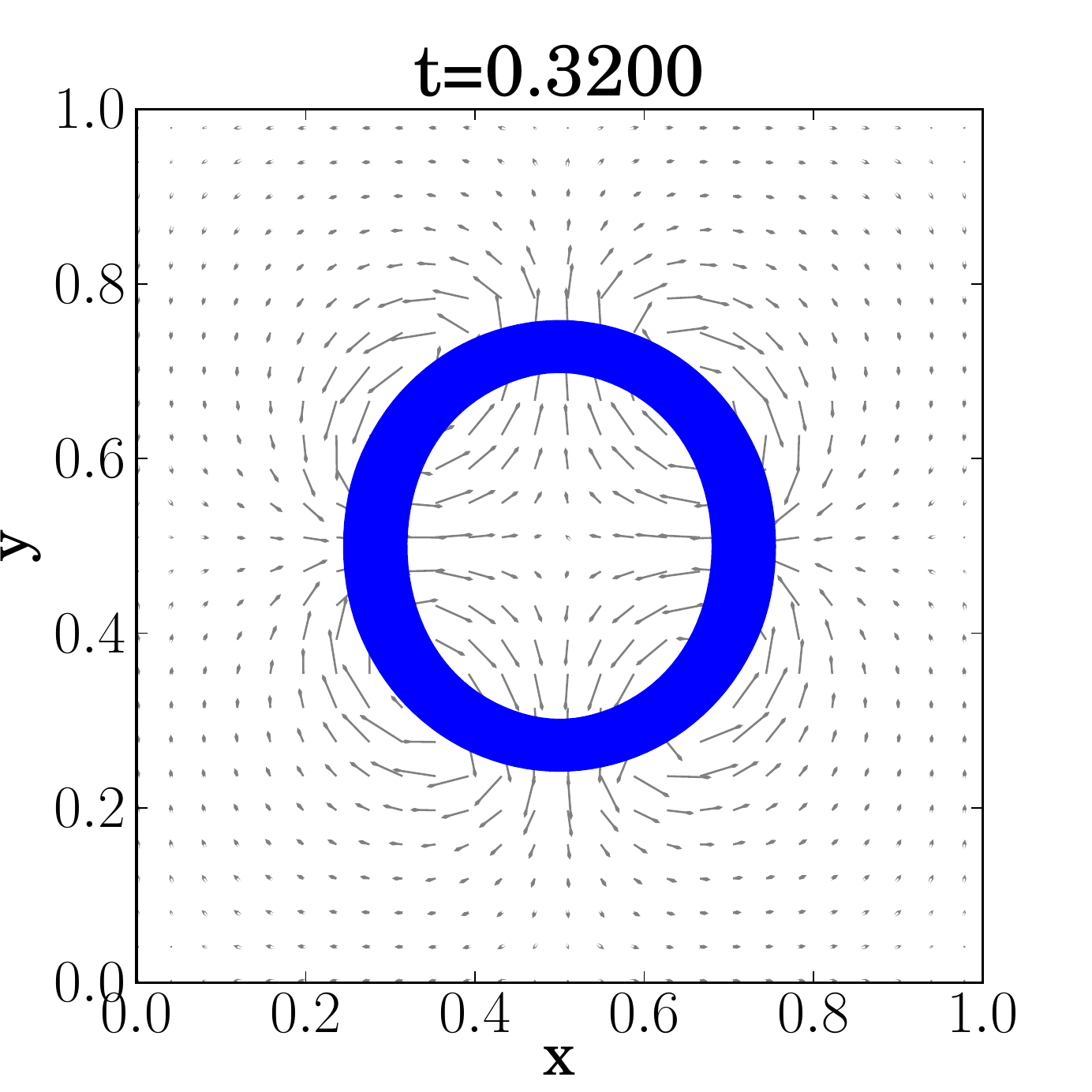} 
      \label{fig:ThickEllipseSim4}}
    \caption{Snapshots of a thick oscillating ellipse using GM-IB
      method, with parameters $\mu=0.005$, $N=256$ and $\dt=0.08/512$.}
    \label{fig:ThickEllipseSim} 
  \end{center}
\end{figure}

Our second test problem involves the thick elastic shell pictured in
Figure~\ref{fig:ThickEllipseSim} that has been studied before by
Griffith and Peskin~\cite{Griffith2005}.  This is a natural
generalization of the thin ellipse problem, wherein the shell is treated
using a nested sequence of elliptical immersed fibers.  The purpose of
this example is not only to illustrate the application of our algorithm
to more general solid elastic structures, but also to illustrate the
genuine second-order accuracy of our numerical method for problems that
are sufficiently smooth.

To this end, we take an elliptical elastic shell with thickness
$\gamma$ using two independent Lagrangian parameters $s,r\in [0,1]$ and
specify the initial configuration by
\begin{gather*}
  \bs{X}(s,r,0) = \left( \half + (r_1 + \gamma (r-1/2))
    \cos(2 \pi s) ,~ \half  + (r_2 + \gamma (r-1/2))  \sin(2 \pi
    s) \right).
\end{gather*}
The shell is composed of circumferential fibers having an elastic
stiffness that varies in the radial direction according to
\begin{gather*}
  \sigma(r) = 1 - \cos(2 \pi r ) .
\end{gather*}
Because the elastic stiffness drops to zero at the inner and outer edges
of the shell, the corresponding Eulerian force $\bs{f}$ is a continuous
function of $\bs{x}$; this should be contrasted with the ``thin
ellipse'' example in which the fluid force is singular, since it
consists of a 1D delta distribution in the tangential direction along
the membrane.  As a result, we expect in this example to observe higher
order convergence because the solution does not contain the
discontinuities in pressure and velocity derivatives that were present
in the thin ellipse problem.  Unless otherwise indicated, we take the
parameter values $\rho=1$, $r_1=0.2$, $r_2=0.25$, $\gamma=0.0625$,
$N_s=(75/16)N$, $N_r=(3/8)N$ and $\dt=0.08/512$ that are consistent with
the computations in~\cite{Griffith2005}.

The dynamics of the thick ellipse problem illustrated in
Figure~\ref{fig:ThickEllipseSim} are qualitatively similar to those in
the previous section, in that the elastic shell undergoes a damped
oscillation.  In Table~\ref{Table:ThickEllipse_ConvergenceRate}, we
present the $\ell^2$ convergence rates in the solution for different
values of fluid viscosity $\mu$.  We also include the corresponding
results computed by Griffith and Peskin~\cite{Griffith2005} and observe
that the GM-IB, BCM-IB, and Griffith-Peskin algorithms all exhibit
remarkably similar convergence rates.  The $\ell^2$ errors for the GM-IB
and BCM-IB methods are almost identical to Griffith and Peskin's, and so
we have not reported them for this example.  

It is only when the viscosity is taken very small ($\mu=0.0005$) that
the Griffith-Peskin algorithm begins to demonstrate superior results.
Because this improvement corresponds to a higher Reynolds number, we
attribute it to differences in the treatment of the nonlinear advection
term and the IB evolution equation.  Indeed, Griffith and Peskin
approximate the nonlinear advection term using a high-order Godunov
method \cite{Colella1990,Minion1996} and integrate the IB equation
\eqref{eq:membrane} using a strong stability preserving Runge-Kutta
method \cite{Gottlieb2001}.  \changed{We have made no attempt to
  incorporate these modifications into our algorithm because the time
  integration they used for their Godunov method requires solution of a
  Poisson problem, whereas a Runge-Kutta time integration would require
  an additional velocity interpolation step that reduces computational
  efficiency.}  Recall that one of our primary aims is precisely to
avoid the pressure Poisson solves required in so many other IB methods.
With this in mind, we have restricted our attention in this paper to
lower Reynolds number flows corresponding roughly to
$\Reynolds\lessapprox 1000$.

\begin{table}[htbp]\centering\small
  \caption{Estimated $\ell^2$ convergence rates $\myrate{q}{128}$ for
    the thick ellipse problem at time $t=0.4$. For comparison,
    Griffith's results~\cite{Griffith2005} are reported in the final
    row. Since Griffith reports the component-wise convergence rate of
    the velocity field, we approximate  $\myrate{\bs{u}}{128} \approx
    \max(\myrate{u}{128},~\myrate{v}{128})$.}    
  %\ra{1.6}
  \begin{tabular}{l cccc cccc cccc}\toprule
    & \multicolumn{3}{c}{$\mu=0.05$} & \multicolumn{3}{c}{$\mu=0.01$}  & \multicolumn{3}{c}{$\mu=0.005$} \\
    \cmidrule(r){2-4} \cmidrule(r){5-7}  \cmidrule(r){8-10}
      &$\bs{u}$ & $p$ & $\bs{X}$ &$\bs{u}$ & $p$ & $\bs{X}$ &$\bs{u}$ & $p$ & $\bs{X}$\\
     GM-IB & $2.10$ &  $1.88$ & $1.69$ & $2.12$ & $1.88$ & $1.76$ & $2.11$ & $1.88$ & $1.99$  \\
     BCM-IB & $2.11$ & $1.88$ & $1.69$ & $2.09$ & $1.87$ & $1.74$ & $2.09$ & $1.87$ & $1.99$ \\
     Griffith \cite{Griffith2005}& $2.16^*$ & $1.89$ & $1.98$ & -- & -- & -- & $2.20^*$ & $1.86$ & $1.74$  \\
    \bottomrule
  \end{tabular}
  \label{Table:ThickEllipse_ConvergenceRate}
\end{table}

Lastly, we investigate the accuracy with which our discrete solution
satisfies the discrete divergence-free condition for a variety of time
steps and spatial discretizations.  Our aim in this instance is to
determine how well the fluid solver of Guermond and Minev approximates
the incompressibility constraint, which is related to the volume
conservation issue discussed in the thin ellipse example.
Table~\ref{Table:ThinEllipse_DivergenceError} lists values of the error
in the discrete divergence of velocity, $\myerror{\nabla \cdot
  \bs{u}}{N}$, measured at time $t=0.4$ and estimated using
equation~\eqref{eq:EstError}.  Observe that $\myerror{\nabla \cdot
  \bs{u}}{N}$ increases slightly as the spatial discretization is
refined, but decreases when a smaller time step is used. This last
result is to be expected because Guermond and Minev use a $\order{\dt}$
perturbation of the incompressibility
constraint~\eqref{eq:PerturbIncompressible}.

\begin{table}[htbp]\centering\small
  \caption{Error in the divergence-free condition $\myerror{\nabla \cdot 
      \bs{u}}{N}$ for the thick ellipse problem using the GM-IB method
    and $\mu=0.01$.}     
  %\ra{1.6}
  \begin{tabular}{l cccccc}\toprule
     & $\dt=\frac{0.08}{512}$ & $\dt=\frac{0.04}{512}$ & $\dt=\frac{0.02}{512}$ & $\dt=\frac{0.01}{512}$ & $\dt=\frac{0.005}{512}$ & $\dt=\frac{0.0025}{512}$  \\
    \midrule
    $N=64$  & $6.34\EE{-3}$ & $2.37\EE{-3}$ & $8.93\EE{-4}$ & $3.20\EE{-4}$ & $1.08\EE{-4}$ & $3.42\EE{-5}$ \\
    $N=128$ & $8.94\EE{-3}$ & $3.68\EE{-3}$ & $1.49\EE{-3}$ & $5.75\EE{-4}$ & $2.11\EE{-4}$ & $7.36\EE{-5}$ \\
    $N=256$ & $1.00\EE{-2}$ & $4.19\EE{-3}$ & $1.73\EE{-3}$ & $6.79\EE{-4}$ & $2.55\EE{-4}$ & $9.06\EE{-5}$ \\
    $N=512$ & $1.04\EE{-2}$ & $4.35\EE{-3}$ & $1.80\EE{-3}$ & $7.11\EE{-4}$ & $2.68\EE{-4}$ & $9.59\EE{-5}$ \\
    \bottomrule
  \end{tabular}
  \label{Table:ThinEllipse_DivergenceError}
\end{table}

\section{\changed{Parallel Performance Results}}
\label{sec:Performance}

We now focus on comparing the parallel performance of our algorithm
(GM-IB) with an analogous projection-based scheme (BCM-IB).  We begin by
comparing the performance difference between solving the pure Poisson
problem that plays a central role in projection schemes, versus Guermond
and Minev's directional-split counterpart.  This captures the major
differences between the fluid solvers used in the corresponding IB
algorithms.  We then follow by performing weak and strong scalability
tests for the full immersed boundary problem.

%%%%%%%%%%%%%%%%%%%%%%%%%%%%%%%%%%%%%%%%%%%%%%%%%%%%%%%%%%%%%%%%%%%%%%%%%%%%%
\subsection{Comparison with Poisson Solvers}
\label{sec:PoissonComparison}

In this section, we compare the performance of several Poisson solvers
with our tridiagonal solver described in section~\ref{sec:linearsolver}.
Since any standard projection scheme requires solving a Poisson problem,
this performance study encapsulates the major differences between
Guermond and Minev's fluid solver and other projection-based approaches
(for example, that of Brown-Cortez-Minion).  To illustrate the
comparison, we consider the problem
\begin{gather}
\left\{
\begin{array}{c l}      
    \ctsop{A} \psi = f(\bs{x}) & \mbox{~in } \Omega = [0,1]^d,\\
    \psi \mbox{~is periodic } &\mbox{~on } \partial\Omega,
\end{array}\right.
   \label{eqn:PoissonGMProblem}
\end{gather}
where 
\begin{gather*}
f(\bs{x}) = \left\{
\begin{array}{c l}      
    \sin( 2\pi x ) \cos( 2\pi y ) & \mbox{~when } d=2,\\
    \sin( 2\pi x ) \cos( 2\pi y ) \cos( 2\pi z ) & \mbox{~when } d=3,\\
\end{array}\right.
\end{gather*}
and $\ctsop{A}$ is either the Laplacian operator ($\ctsop{A} =
\Laplacian$) or the directional-split operator $\ctsop{A} =
(1-\partial_{xx})(1-\partial_{yy})$ (when $d=2$).

When solving the Poisson problem, we compare with two other solvers: one
based on FFTs and the other on multigrid.  For both of these solvers we
discretize the problem using a second-order finite difference scheme.
In the FFT-based solver, the difference scheme is rewritten in terms of
the Fourier coefficients and solved using the real-to-complex and
complex-to-real transformations found in FFTW~\cite{FFTW}.  For the
multigrid solver, we use a highly scalable multigrid preconditioner
(PFMG) implemented in Hypre~\cite{Hypre} that is used with a conjugate
gradient solver.  When performing the comparison for the
directional-split problem, the discrete system decouples into a set of
one-dimensional tridiagonal systems that we solve using the techniques
described in section~\ref{sec:linearsolver}.  The major differences here
occur in terms of the domain partitioning, which for the FFT-based
solver involves a slab decomposition, whereas the multigrid and
directional-split solvers use square-like subdomains.

Throughout our performance study, times are collected using MPI and the
best result of multiple runs is reported.  All simulations are performed
using the Bugaboo cluster managed by WestGrid~\cite{Bugaboo}, a member
of the high-performance computing consortium Compute Canada.  This
cluster consists of 12-core blades, each containing two Intel Xeon X5650
6-core processors (2.66 GHz) that are connected by Infiniband using a
288-port QLogic switch.

First, we evaluate the strong scaling property of each solver by
running a sequence of simulations in which the problem size is held
fixed as the number of processors increases.  For the three-dimensional
computations, the problems are solved on $N = 128$ and $N = 256$
grids, which are common resolutions used in 3D IB calculations.  For the
two-dimensional computations, the problems are solved on grids that are
larger than usual ($N = 2048$ and $4096$).  The strong
scaling results are given in
Tables~\ref{Table:StrongScaling:PoissonProblem2d}
and~\ref{Table:StrongScaling:PoissonProblem3d}, which 
report the execution time $T_P$ and parallel efficiency 
\begin{gather*}
  E_P = \frac{T_1}{P T_P},  
\end{gather*}
for $P$ processors. The parallel efficiency
quantifies how well the processors are utilized throughout a
computation, where a value of $E_P=1$ corresponds to the ideal case
and smaller values indicate a reduced parallel efficiency.  Note that
a reduction in efficiency is expected since the serial computation involves
no Schur complement systems but instead computes the 
tridiagonal systems directly.

In all parallel computations, we observe that the directional-split
solver is strongly scalable ($E_P > 0.8$) and outperforms both Poisson
solvers by a significant margin.  Indeed, when comparing the
directional-split solver to multigrid, there is an order of magnitude
difference in execution time.  For all multigrid computations, the
conjugate gradient solver required $6$ iterations which makes the
directional-split solver a factor of $2$ to $5$ times faster than a
single multigrid iteration. When comparing the directional-split solver
to the FFT-based solver, the difference in execution times is much
smaller, particularly when using fewer processors. However, as the
processor count increases, we still see a two-fold or greater
performance improvement when using the directional-split solver.

Besides the performance improvements, the Guermond and Minev fluid
solver has a few additional advantages over FFT-based fluid
solvers. First of all, FFT-based solvers are restricted to periodic
boundaries while the directional-split solver has no such
restriction. For example, the Guermond-Minev fluid solver can
inexpensively compute driven-cavity
and (periodic) channel flows without the use of
immersed boundaries~\cite{Guermond2011-2}.  Secondly, the slab decomposition used by many FFT
libraries (such as FFTW) can lead to serious load-balancing issues in
the immersed boundary context as indicated by Yau~\cite{Yau2002}. Note
that this could be mitigated somewhat in 3D simulations by moving to a
pencil decomposition that consequently allows for more processors to be
used in the computation~\cite{Pippig2013}.

\begin{table}[htbp]\centering
  \caption{ Execution time $T_P$ and parallel efficiency $E_P$ for
    the 2D problem~\eqref{eqn:PoissonGMProblem} on an $N^2$ grid
    with $P$ processors.}    
  %\ra{1.6}
  \begin{tabular}{ll cccccc}\toprule
    & & \multicolumn{2}{c}{Multigrid} & \multicolumn{2}{c}{FFT} & \multicolumn{2}{c}{Directional-Split}  \\
    \cmidrule(r){3-4} \cmidrule(r){5-6}  \cmidrule(r){7-8} 
                             & $P$ & $T_P$            & $E_P$  & $T_P$          & $E_P$  & $T_P$          & $E_P$   \\
    \midrule
    \multirow{6}{*}{$N=2048$}& $1$   & $4.26\EE{+0}$  & $--$   & $3.10\EE{-1}$  & $--$   & $2.77\EE{-1}$  & $--$  \\
                             & $8$   & $6.66\EE{-1}$  & $0.80$ & $7.93\EE{-2}$  & $0.49$ & $3.77\EE{-2}$  & $0.92$  \\
                             & $16$  & $3.07\EE{-1}$  & $0.87$ & $3.57\EE{-2}$  & $0.54$ & $1.82\EE{-2}$  & $0.95$  \\
                             & $32$  & $1.72\EE{-1}$  & $0.77$ & $2.50\EE{-2}$  & $0.39$ & $8.78\EE{-3}$  & $0.99$  \\
                             & $64$  & $8.30\EE{-2}$  & $0.80$ & $1.43\EE{-2}$  & $0.34$ & $4.40\EE{-3}$  & $0.99$  \\
                             & $128$ & $4.33\EE{-2}$  & $0.77$ & $1.20\EE{-2}$  & $0.20$ & $2.45\EE{-3}$  & $0.88$  \\
    \midrule
    \multirow{7}{*}{$N=4096$}& $1$   & $1.38\EE{+1}$  & $--$   & $1.33\EE{+0}$  & $--$   & $1.09\EE{+0}$  & $--$  \\
                             & $8$   & $3.18\EE{+0}$  & $0.54$ & $3.58\EE{-1}$  & $0.47$ & $1.54\EE{-1}$  & $0.88$  \\
                             & $16$  & $1.83\EE{+0}$  & $0.47$ & $2.15\EE{-1}$  & $0.39$ & $7.86\EE{-2}$  & $0.87$  \\
                             & $32$  & $8.92\EE{-1}$  & $0.48$ & $1.17\EE{-1}$  & $0.36$ & $4.06\EE{-2}$  & $0.84$  \\
                             & $64$  & $4.84\EE{-1}$  & $0.45$ & $6.36\EE{-2}$  & $0.33$ & $2.03\EE{-2}$  & $0.84$  \\
                             & $128$ & $2.23\EE{-1}$  & $0.48$ & $3.84\EE{-2}$  & $0.27$ & $9.62\EE{-3}$  & $0.84$  \\
                             & $256$ & $1.05\EE{-1}$  & $0.51$ & $2.57\EE{-2}$  & $0.20$ & $5.22\EE{-3}$  & $0.81$  \\
    \bottomrule
  \end{tabular}
  \label{Table:StrongScaling:PoissonProblem2d}
\end{table}

\begin{table}[htbp]\centering
  \caption{ Execution time $T_P$ and parallel efficiency $E_P$ for
    the 3D problem~\eqref{eqn:PoissonGMProblem} on an $N^3$ grid
    with $P$ processors.}    
  %\ra{1.6}
  \begin{tabular}{ll cccccc}\toprule
    & & \multicolumn{2}{c}{Multigrid} & \multicolumn{2}{c}{FFT} & \multicolumn{2}{c}{Directional-Split}  \\
    \cmidrule(r){3-4} \cmidrule(r){5-6}  \cmidrule(r){7-8} 
                             & $P$ & $T_P$            & $E_P$  & $T_P$          & $E_P$  & $T_P$          & $E_P$   \\
    \midrule
    \multirow{6}{*}{$N=128$} & $1$   & $2.77\EE{+0}$  & $--$   & $1.60\EE{-1}$  & $--$   & $2.05\EE{-1}$  & $--$  \\
                             & $8$   & $4.68\EE{-1}$  & $0.74$ & $3.16\EE{-2}$  & $0.63$ & $2.65\EE{-2}$  & $0.97$  \\
                             & $16$  & $2.38\EE{-1}$  & $0.73$ & $1.88\EE{-2}$  & $0.53$ & $1.35\EE{-2}$  & $0.95$  \\
                             & $32$  & $1.40\EE{-1}$  & $0.62$ & $1.21\EE{-2}$  & $0.42$ & $6.75\EE{-3}$  & $0.95$  \\
                             & $64$  & $7.48\EE{-2}$  & $0.58$ & $6.67\EE{-3}$  & $0.38$ & $3.50\EE{-3}$  & $0.92$  \\
                             & $128$ & $5.67\EE{-2}$  & $0.38$ & $4.57\EE{-3}$  & $0.27$ & $1.88\EE{-3}$  & $0.85$  \\
    \midrule
    \multirow{7}{*}{$N=256$} & $1$   & $2.16\EE{+1}$  & $--$   & $1.43\EE{+0}$  & $--$   & $1.71\EE{+0}$  & $--$  \\
                             & $8$   & $3.58\EE{+0}$  & $0.75$ & $2.68\EE{-1}$  & $0.67$ & $2.26\EE{-1}$  & $0.95$  \\
                             & $16$  & $1.93\EE{+0}$  & $0.70$ & $1.53\EE{-1}$  & $0.58$ & $1.12\EE{-1}$  & $0.95$  \\
                             & $32$  & $1.15\EE{+0}$  & $0.59$ & $8.63\EE{-2}$  & $0.52$ & $5.51\EE{-2}$  & $0.97$  \\
                             & $64$  & $6.77\EE{-1}$  & $0.50$ & $5.48\EE{-2}$  & $0.41$ & $2.80\EE{-2}$  & $0.95$  \\
                             & $128$ & $3.97\EE{-1}$  & $0.42$ & $3.43\EE{-2}$  & $0.32$ & $1.52\EE{-2}$  & $0.88$  \\
                             & $256$ & $2.26\EE{-1}$  & $0.37$ & $2.35\EE{-2}$  & $0.24$ & $7.46\EE{-3}$  & $0.90$  \\
    \bottomrule
  \end{tabular}
  \label{Table:StrongScaling:PoissonProblem3d}
\end{table}

\begin{figure}[!tbp]
  \begin{center}
    \subfigure[]{\includegraphics[width=0.46\textwidth]{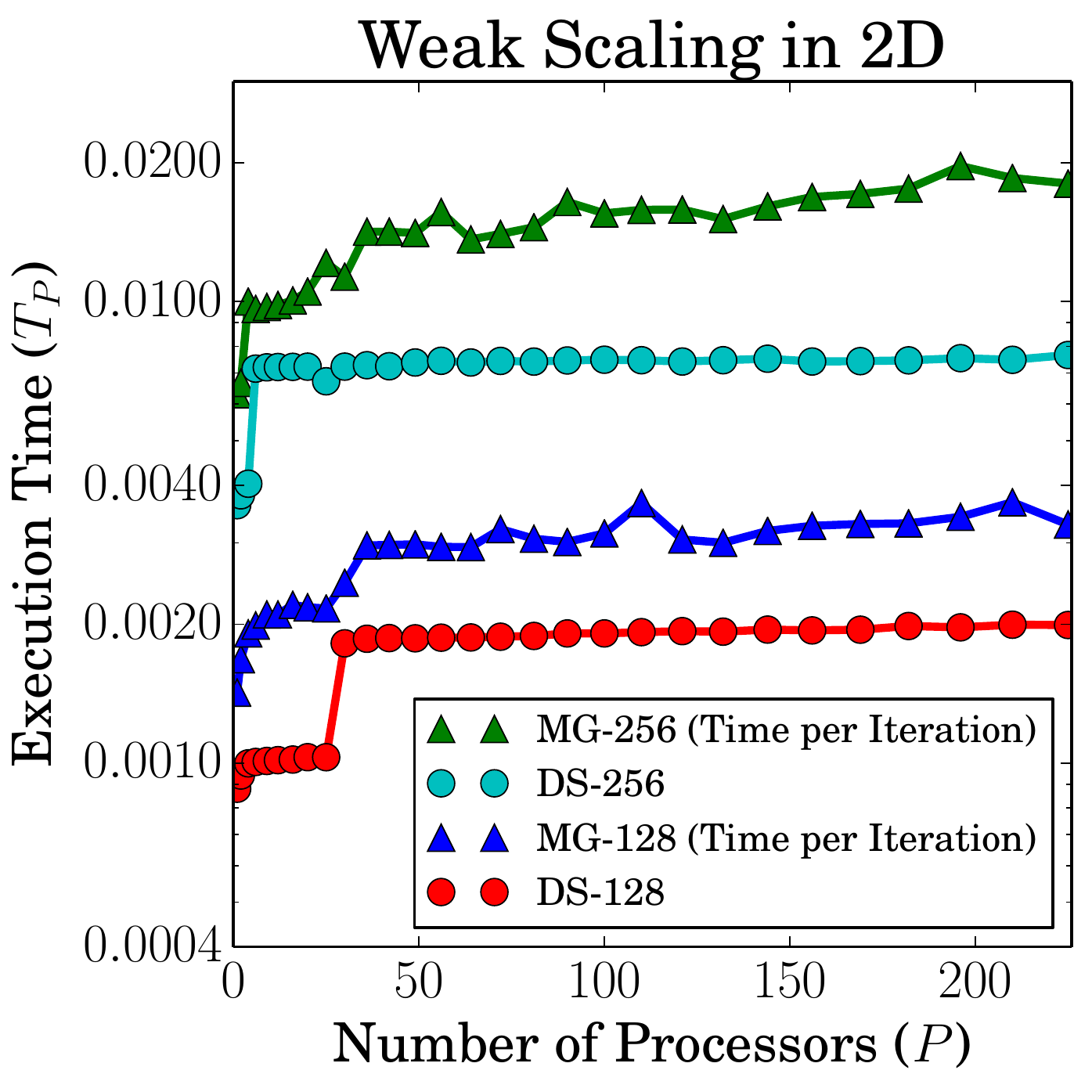} 
      \label{fig:WeakScaling:PoissonProblem:2d}}
    \qquad 
    \subfigure[]{\includegraphics[width=0.46\textwidth]{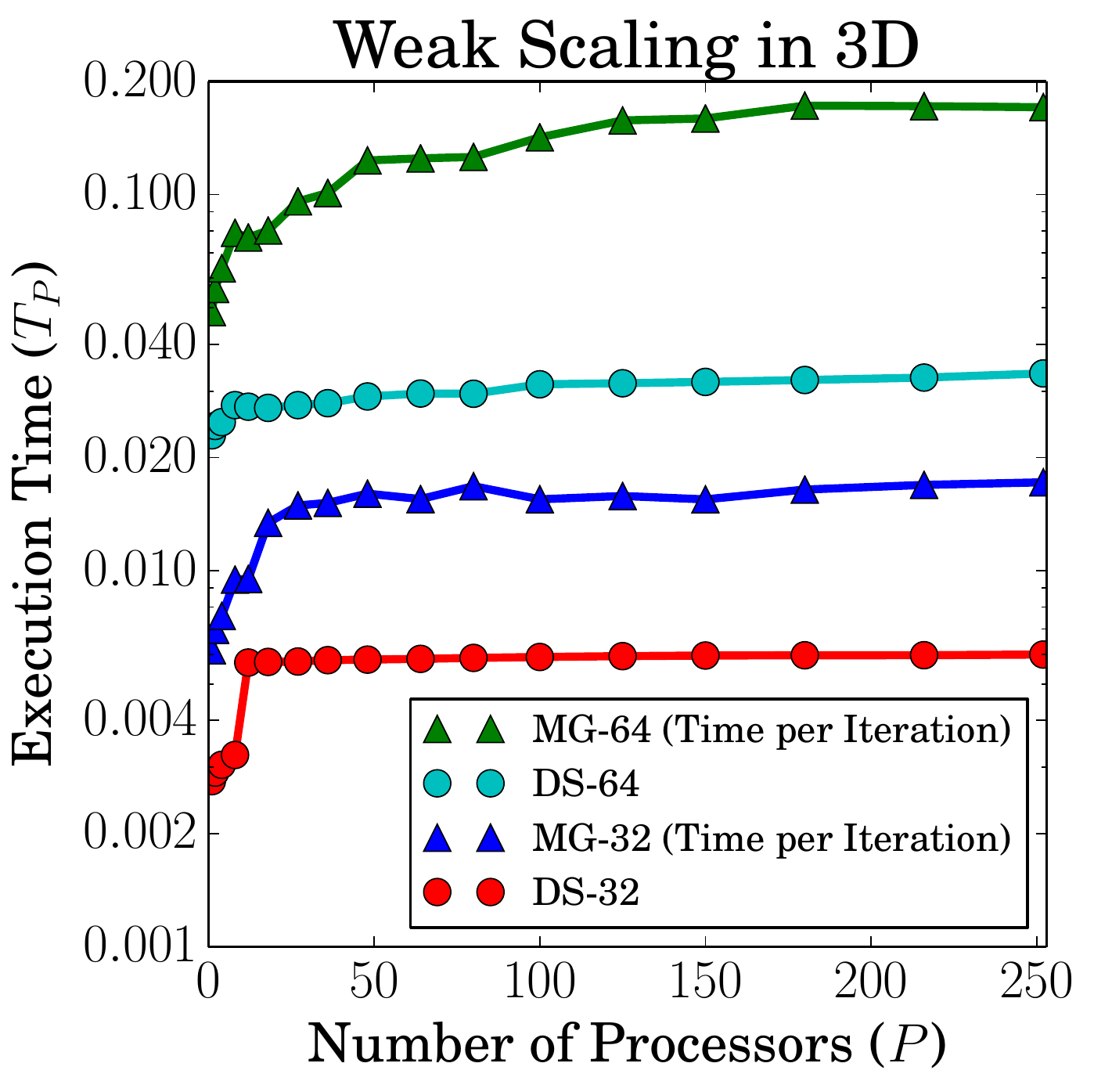} 
      \label{fig:WeakScaling:PoissonProblem:3d}}
    \caption{ Weak scaling of the multigrid and directional-split
      solvers when approximating problem~\eqref{eqn:PoissonGMProblem} in
      2D and 3D.  For the 2D computations, MG-128 and MG-256 denote the
      execution time of a single multigrid iterations using local
      $n=128$ and $256$ grids. Likewise, DS-128 and DS-256 denote the
      execution times for completely solving the directional-split
      problem. For 3D computations, we use a local $n=32$ and $n=64$
      grids where the solver is specified using the same 2D naming
      convention.}
        \label{fig:WeakScaling:PoissonProblem}
  \end{center}
\end{figure}

Next we report the weak scaling results for the multigrid solver and the
directional-split solver as shown in
Figure~\ref{fig:WeakScaling:PoissonProblem}.  In each set of
computations, the local problem size $n^d$ (where $N=n\cdot P_x$) is
held fixed as the number of processors is increased.  In the ideal case,
the execution time should stay constant so that the workload
per processor does not change as the number of nodes increase.  For 2D
problems ($d=2$), the local grid resolution on each subdomain is either
$n = 128$ or 256, while for 3D problems ($d=3$) we use either $n=32$ or
$64$.

As expected~\cite{Baker2012,Guermond2011}, both solvers are weakly
scalable since the execution time stays essentially constant as the
problem size and number of processors increase. For the $n=128$
simulations, the execution time jumps suddenly at $P=25$, which 
is due to increased communication costs occurring between blades
inside the same chassis. Since the blades in a chassis are connected through the same
switch where multiple cores share the same connection, and since the work load per
processor is so small, a noticeable jump appears as a result 
of resource contention.

When comparing the execution time between solvers, the directional-split
solver is around $1.5$ to $5$ times faster than a single multigrid
iteration. Since the multigrid solver requires $5$ to $6$ iterations of
conjugate gradient, this results in an order of magnitude difference in
the total execution time. Of course, this difference would be reduced by
using a better initial guess in the multigrid solver which would in turn
require fewer iterations.

As can be seen from this comparison, the directional-split solver
outperformed the Poisson solvers in all non-serial computations. The
precise difference depends on the problem size $N$, the number of
processors $P$, and the hardware configuration of the
cluster. Furthermore, when solving the Poisson problem, we used the two
highly optimized libraries Hypre~\cite{Hypre} and
FFTW~\cite{FFTW}. Therefore, we would expect to see even greater
performance differences if we were to optimize the
directional-split solver to the same degree as these other solvers.

%%%%%%%%%%%%%%%%%%%%%%%%%%%%%%%%%%%%%%%%%%%%%%%%%%%%%%%%%%%%%%%%%%%%%%%%%%%%%
\subsection{Multiple Thin Ellipses in 2D}
\label{sec:TilingThinEllipsesProblem}

The next example is designed to explore in more detail the parallel
performance of GM-IB and BCM-IB (with Hypre) by computing a variation of
the thin ellipse problem from section~\ref{sec:ThinEllipseProblem}.
Because our 2D computations are performed on a doubly-periodic fluid
domain, the thin ellipse geometry is actually equivalent to an infinite
array of identical elliptical membranes.  This periodicity in the
solution provides a simple mechanism for increasing the computational
complexity of a simulation by explicitly adding multiple periodic copies
while technically solving a problem with a solution that is identical to
that for a single membrane.  Each copy of the original domain (see
section~\ref{sec:ThinEllipseProblem}) may then be handled by a different
processing node, which allows us to explore the parallel performance in
an idealized geometry.

\begin{figure}[!tbp]
  \begin{center}
    \subfigure[]{\includegraphics[width=0.40\textwidth]{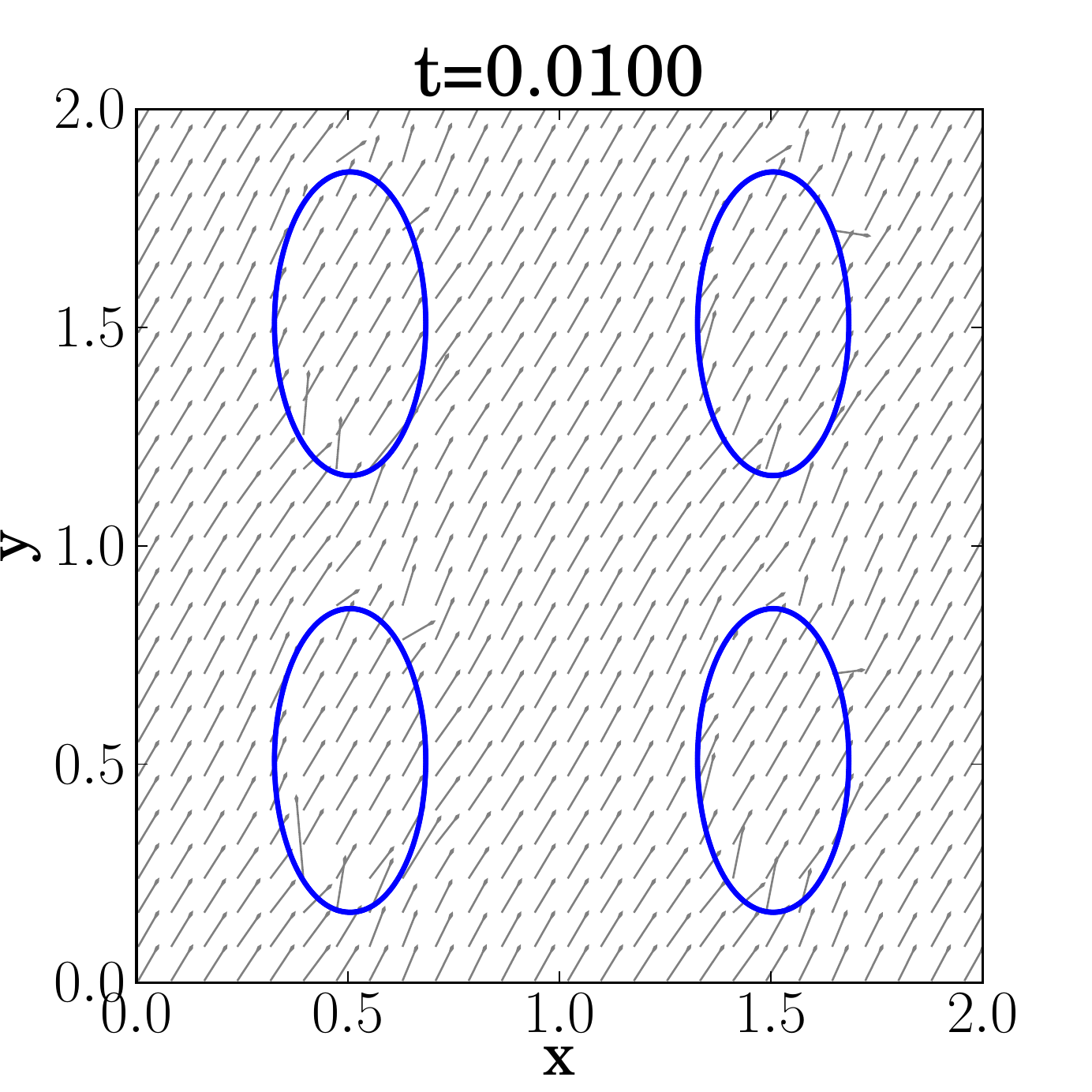} 
      \label{fig:MultipleThinEllipseSim1}}
    \qquad 
    \subfigure[]{\includegraphics[width=0.40\textwidth]{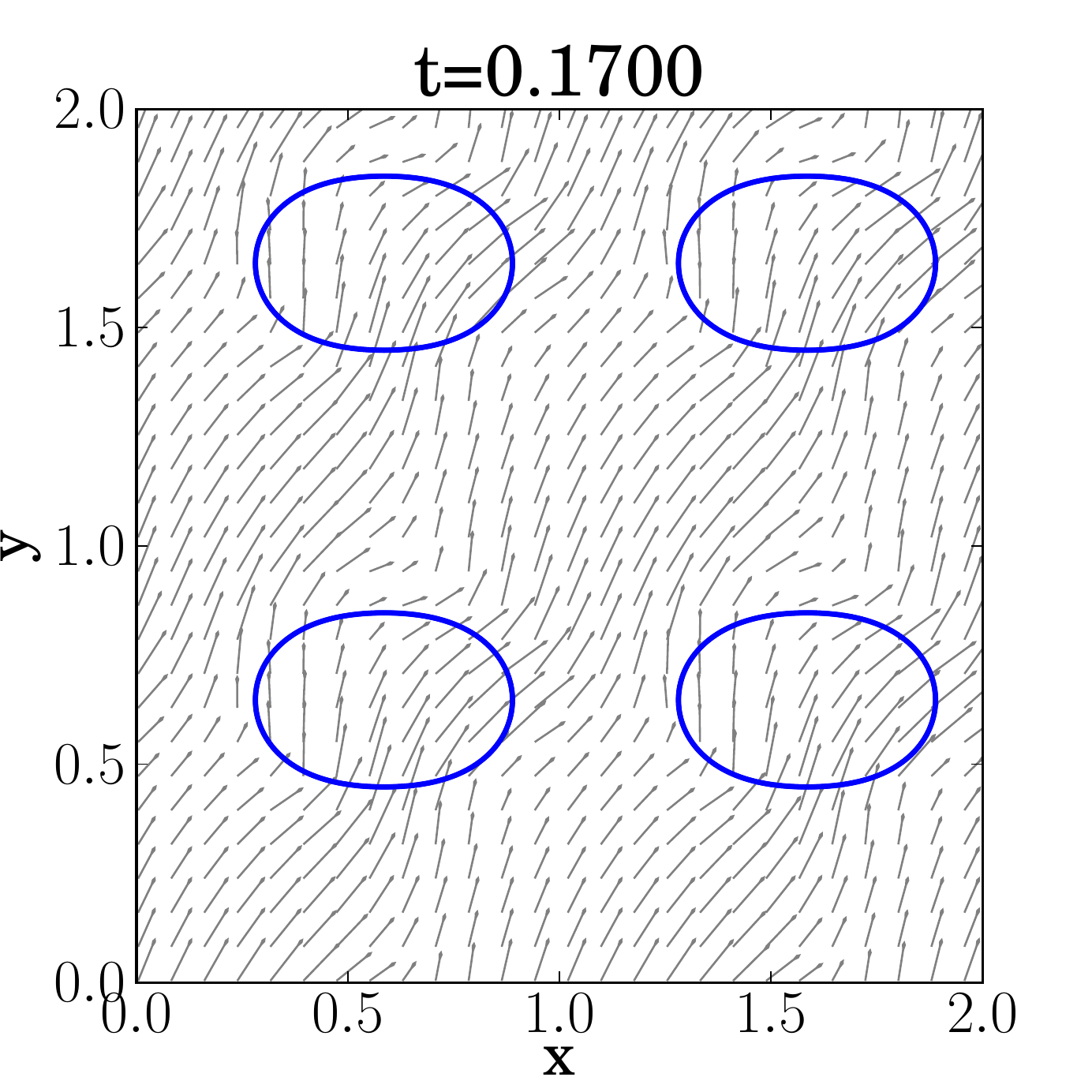} 
      \label{fig:MultipleThinEllipseSim2}}
    \qquad 
    \subfigure[]{\includegraphics[width=0.40\textwidth]{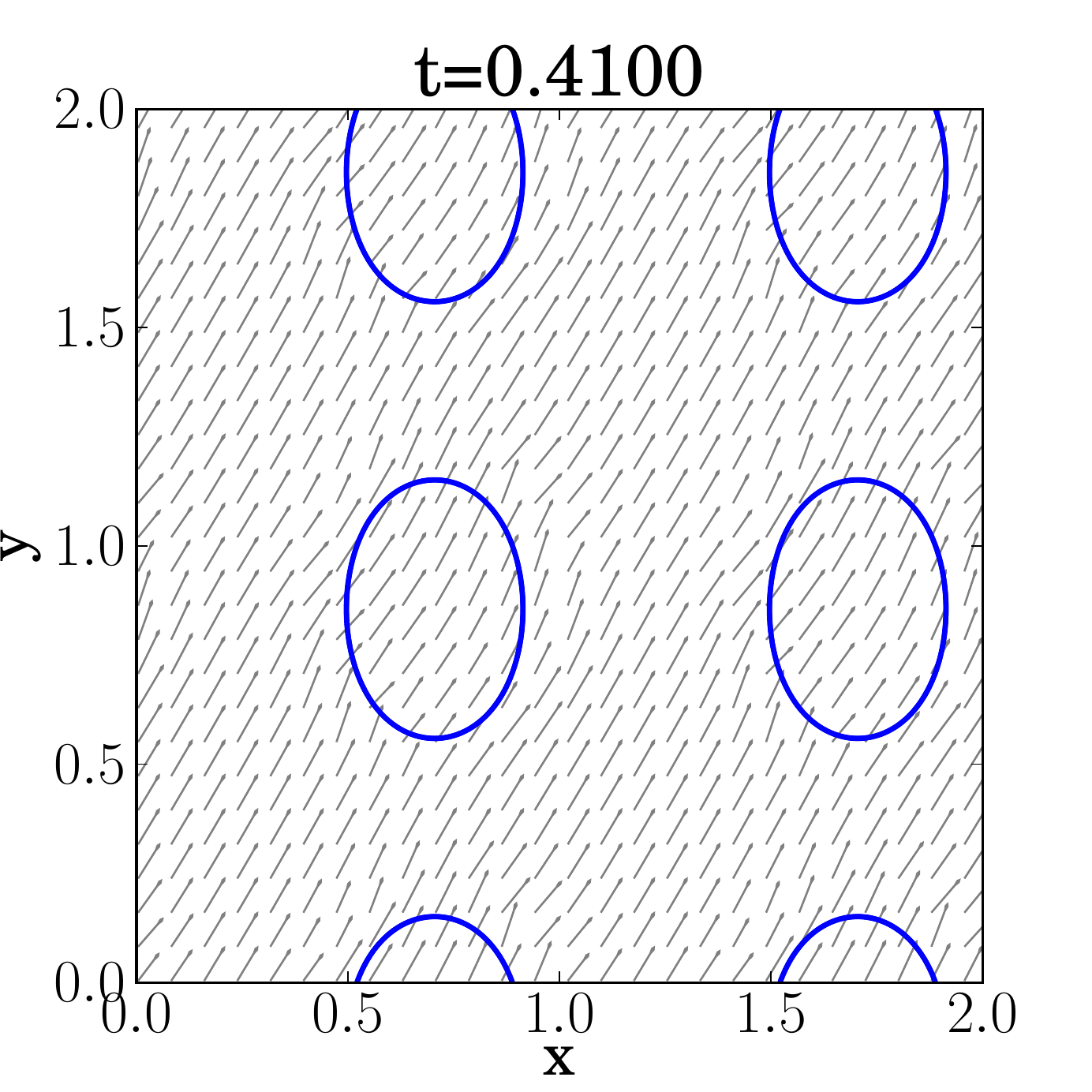} 
      \label{fig:MultipleThinEllipseSim3}}
    \qquad 
    \subfigure[]{\includegraphics[width=0.40\textwidth]{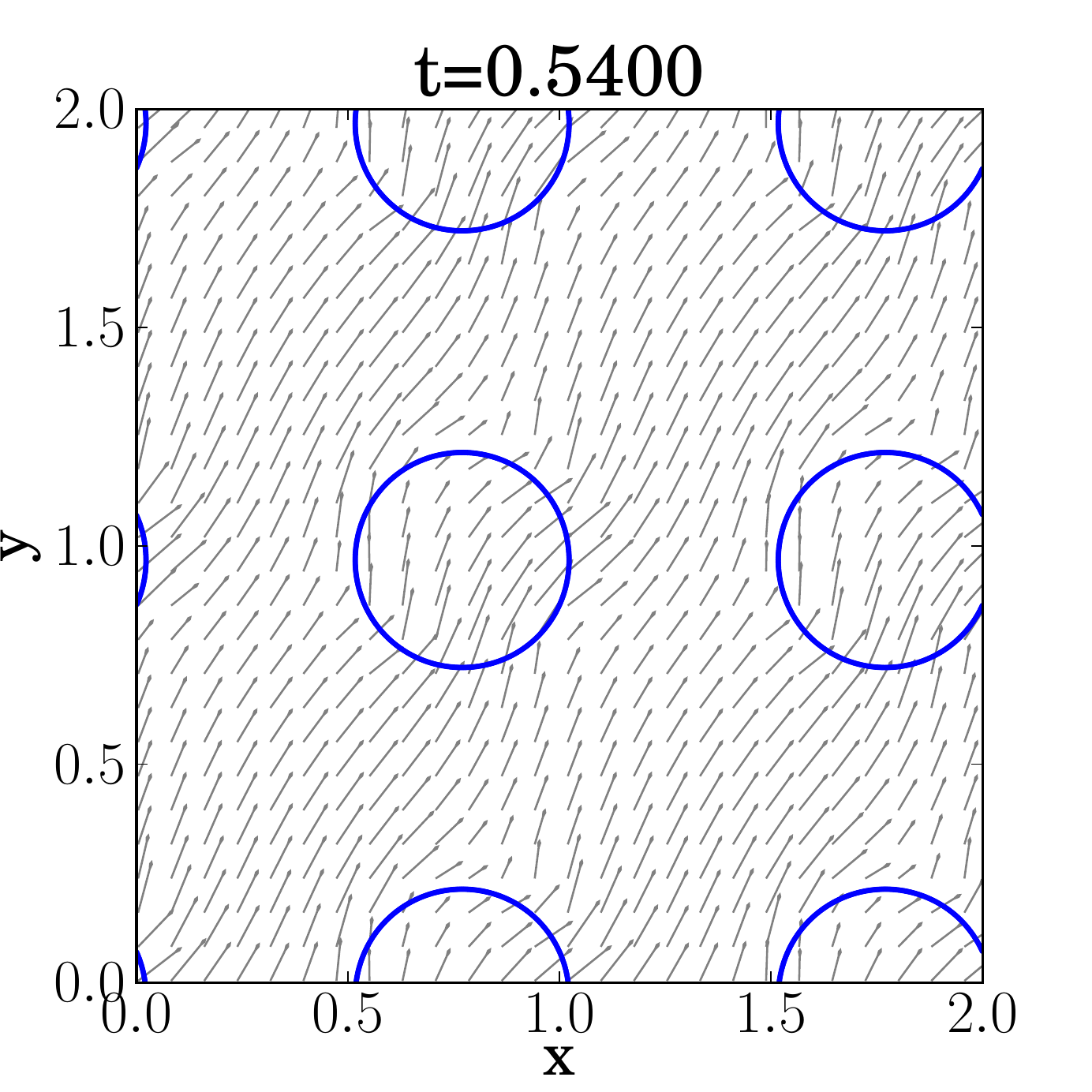} 
      \label{fig:MultipleThinEllipseSim4}}
    \caption{Simulation of a $2\times 2$ array of thin ellipses.}
    \label{fig:MultipleThinEllipseSim} 
  \end{center}
\end{figure}

Suppose that we would like to perform a parallel simulation using
$P=P_x\cdot P_y$ processing nodes.  On such a cluster, we can simulate a
rectangular $P_x\times P_y$ array of identical ellipses, situated on the
fluid domain $\Omega=[0, P_x] \times [0, P_y]$.  We subdivide the domain
into equal partitions so that each processor handles the unit-square
subdomain $\Omega_{\ell,m}=[\ell-1, \ell] \times [m-1, m]$, for
$\ell=1,2,\ldots, P_x$ and $m=1,2,\ldots P_y$.  If we denote by
$(x_{\ell,m}, y_{\ell,m})$ the centroid of $\Omega_{\ell,m}$, then each
such subdomain contains a single ellipse having the initial
configuration
\begin{gather*}
  \bs{X}_{\ell,m}(s,0) = \left( x_{\ell,m} + r_1 \cos(2 \pi s), \;
    y_{\ell,m} + r_2  \sin(2 \pi s) \right), 
\end{gather*}
where $s\in[0,1]$ is the same Lagrangian parameter as before.  In order
to make the flow slightly more interesting, and to test the ability of
our parallel algorithm to handle immersed boundaries that move between
processing nodes, we impose a constant background fluid velocity field
$\bs{u}(\bs{x},0) = \half\brac{1, \sqrt{3}}$ instead of the zero initial
velocity used in section~\ref{sec:ThinEllipseProblem}.  Snapshots of the
solution for a $2\times 2$ array of ellipses are given in
Figure~\ref{fig:MultipleThinEllipseSim}.

To investigate the parallel performance of the GM-IB and BCM-IB
algorithms, we simulate different-sized arrays of thin ellipses
corresponding to values of $P_x$ and $P_y$ in the range $[1,16]$. For
each simulation, we use parameters $\mu=0.01$, $\rho=1$, $\sigma=1$,
$r_1=\frac{5}{28}$, $r_2=\frac{7}{20}$, $h_s=\frac{4}{19}h$ and
$\dt=0.01 h$, and we compute up to time $t=1.00$ using two values of the
fluid mesh width $h=\frac{1}{128}$ and $\frac{1}{256}$.  In the case of
perfect parallel scaling the execution time should remain constant
between simulations, because of our problem constructon in which
doubling the problem size also double the number of nodes
$P$. Therefore, the problem represents a weak scalability test for our
algorithm in which the workload per processor node remains constant as
the number of nodes increase.

\begin{figure}[!tbp]
  \begin{center}
    \includegraphics[width=0.6\textwidth]{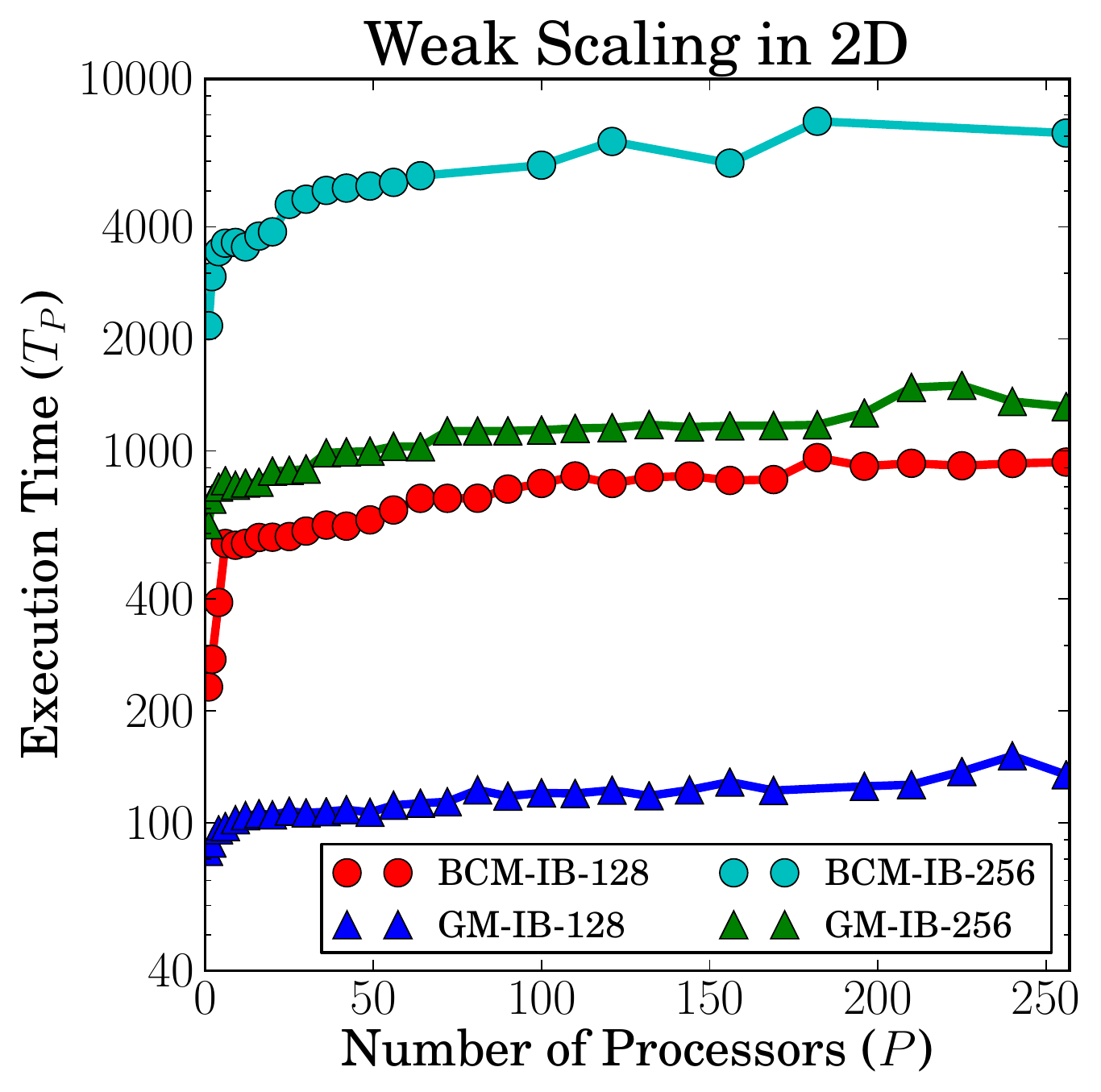} 
    \caption{ Execution time (in seconds) for the multiple thin ellipse
      problem using the BCM-IB and GM-IB algorithms ($P$ ellipses, $P$
      processors, local grids with $n=128$ and $256$).}
        \label{fig:WeakScaling:IBProblem}
  \end{center}
\end{figure}

The execution times for various array sizes ($P_x$, $P_y$) are
summarized in Figure~\ref{fig:WeakScaling:IBProblem} for both IB
solvers. The execution time remains roughly constant in both cases,
which indicates that the GM-IB and BCM-IB implementations are
essentially weakly scalable.  Notice that there is a slight degradation
in performance as $P$ increases; for example, on the $h=\frac{1}{128}$
grid the GM-IB execution time increases by roughly $20\%$ between $P=64$
to $P=254$, which is minimal considering the large variation in problem
size.

When comparing solvers, GM-IB outperforms BCM-IB by more than a factor
of 5 in execution time. This difference in performance is largely due to
the efficiency of the linear solvers. Here, the BCM-IB solver uses
conjugate gradient with a multigrid preconditioner (PFMG) implemented
within Hypre~\cite{Hypre}, where the initial guess is set to the
solution from the previous time step. For this particular problem, the
multigrid solver typically requires $2$ iterations to solve the momentum
equations and $5$ iterations for the projection step. Naturally, if
either iteration count could be reduced, the performance of the BCM-IB
solver would improve significantly. However, since an iteration of multigrid 
is substantially slower than the directional-split solver 
(see section~\ref{sec:PoissonComparison}), GM-IB  
would continue to outperform BCM-IB.

%%%%%%%%%%%%%%%%%%%%%%%%%%%%%%%%%%%%%%%%%%%%%%%%%%%%%%%%%%%%%%%%%%%%%%%%%%%%%
\subsection{Cylindrical Shell in 3D}
\label{sec:CylinderSimulation}

For our final test case, we consider a three-dimensional example in
which the immersed boundary is a cylindrical elastic shell.  The
cylinder initially has an elliptical cross-section with semi-axes
$r_1$ and $r_2$ that is parameterized by
\begin{gather*}
  \bs{X}(s,r,0) = \left( r ,~ \half + r_1 \cos(2 \pi s) ,~
    \half + r_2  \sin(2 \pi s) \right),
\end{gather*}
using the two Lagrangian parameters $s,r\in[0,1]$.  The force density
is 
\begin{gather*}
  \bs{\mathcal{F}}[\bs{X}(s,r,t)] = 
    \sigma_s \pdd{\bs{X}}{s} +
    \sigma_r \pd{}{r}\brac{\pd{\bs{X}}{r} \brac{ 1 -
        \frac{L}{\left|\pd{\bs{X}}{r}\right|} }},
\end{gather*}
which corresponds to an elastic shell made up of an interwoven mesh of
one-dimensional elastic fibers.  The $s$ parameterization identifies
individual fibers running around the elliptical cross-section of the
cylinder, each having zero resting length and elastic stiffness
$\sigma_s$.  On the other hand, the $r$ parameterization describes
fibers running axially along the length of the cylinder, each having a
non-zero resting-length $L$ and stiffness $\sigma_r$.  Since the domain
is periodic in all directions, the ends of the cylinder are connected to
their periodic copies so that there are no ``cuts'' along the
fibers. This problem is essentially equivalent to the two-dimensional
thin ellipse problem considered in section~\ref{sec:ThinEllipseProblem},
with the only difference being that the 2D problem does not have any
fibers running along the non-existent third dimension.  The 2D thin
ellipse and 3D cylinder problems are only strictly equivalent when
$\sigma_r=0$.  However, we take $\sigma_r=\sigma_s=1$ and $L=1$ in order
to maintain the integrity of the elastic shell and to avoid any drifting
of elliptical cross-sections in the $x$-direction.  The elastic shell is
discretized using equally-spaced values of the Lagrangian parameters $s$
and $r$.  In the simulations that follow, we use the parameter values
$\mu=0.01$, $\rho=1$, $r_1=\frac{5}{28}$, $r_2=\frac{7}{20}$,
$N_s=\frac{19}{4}N$, $N_r=3 N$ and $\dt=0.04/N$ where $N=128$ and
$N=256$.

\begin{sidewaysfigure}[!tbp]
  \begin{center}
    \subfigure[]{\includegraphics[width=0.38\textwidth]{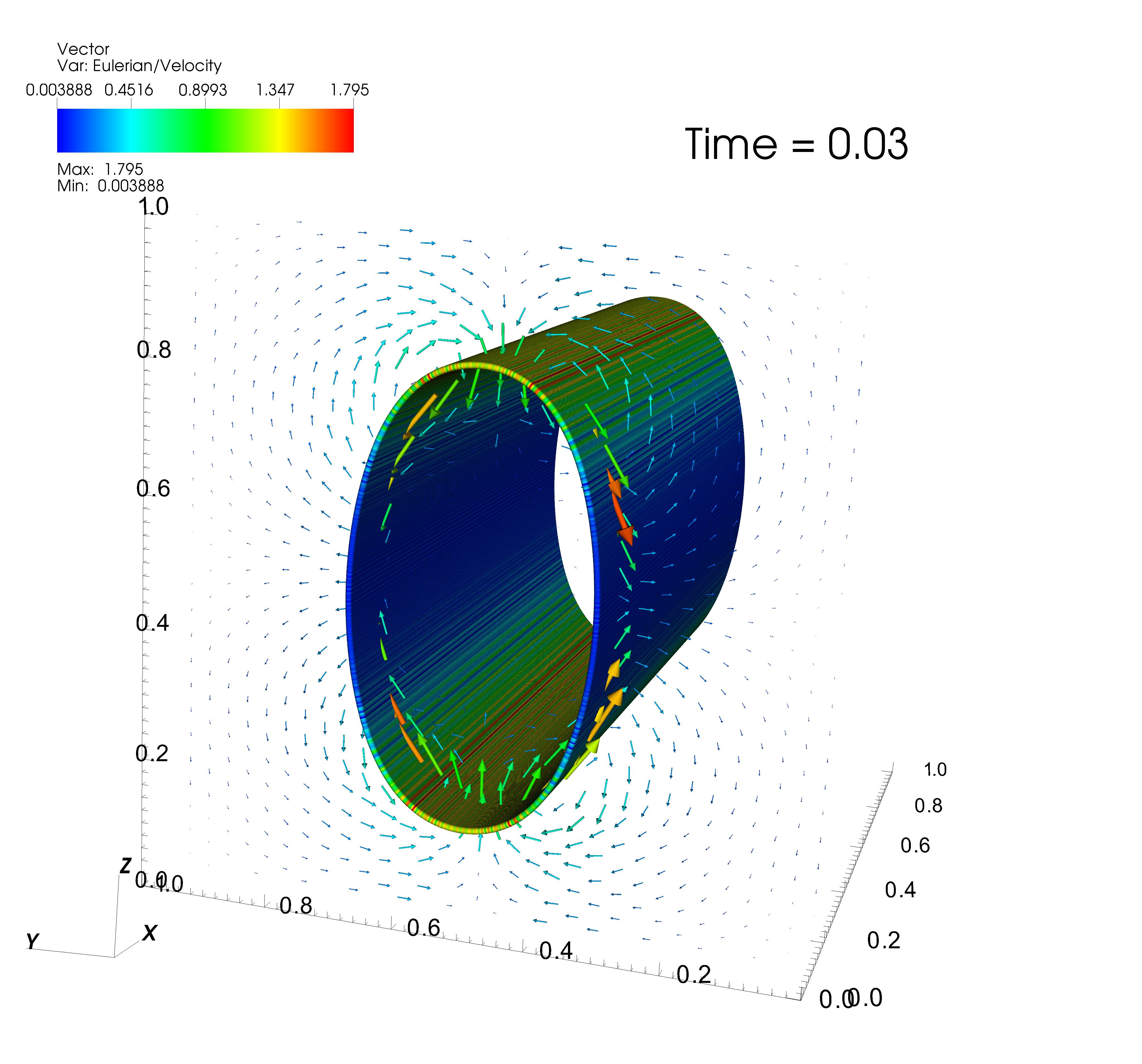}
      \label{fig:CylinderSim1}}
    \qquad
    \subfigure[]{\includegraphics[width=0.38\textwidth]{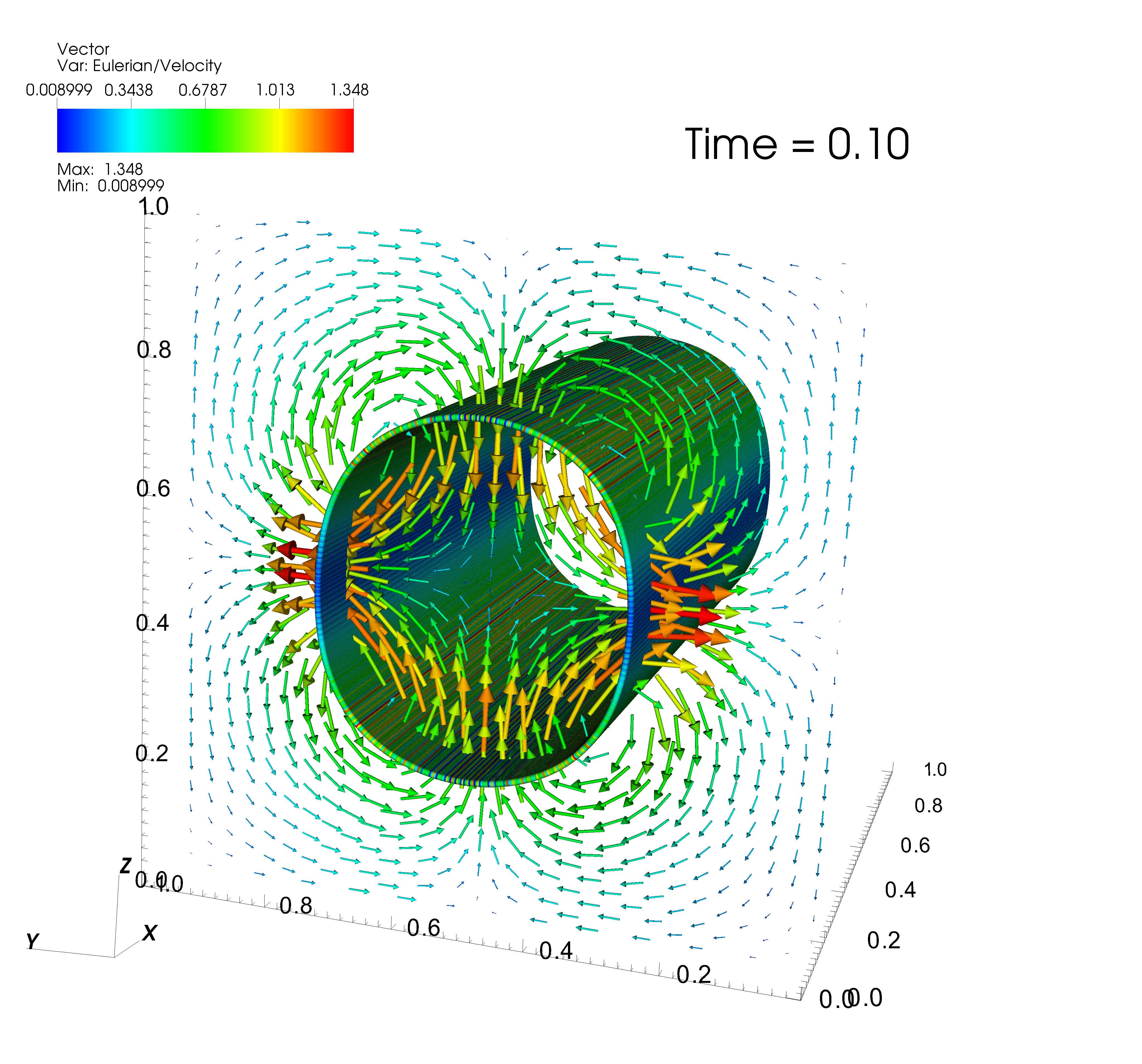}
      \label{fig:CylinderSim2}}
    \qquad
    \subfigure[]{\includegraphics[width=0.38\textwidth]{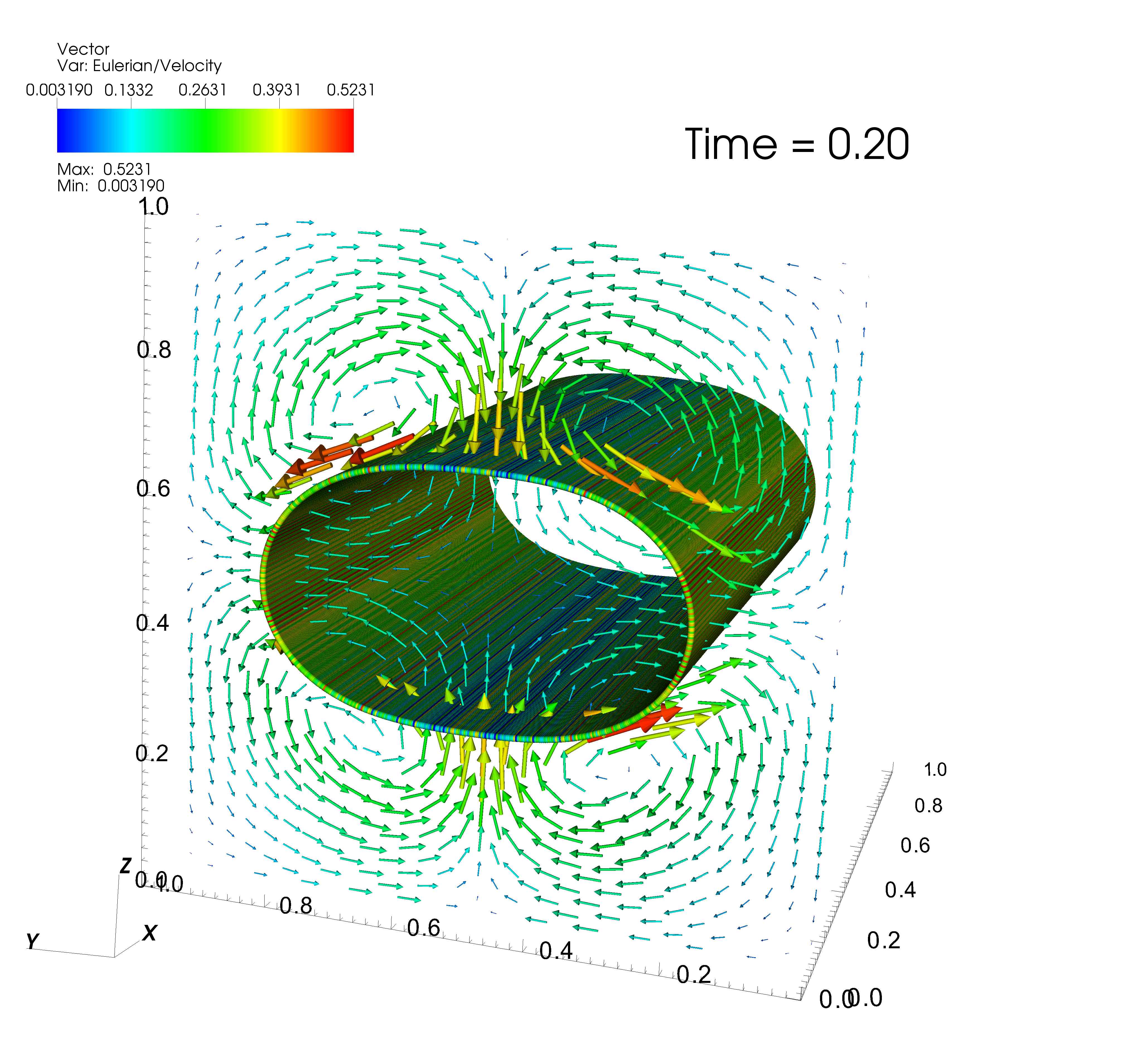}
      \label{fig:CylinderSim3}}
    \qquad
    \subfigure[]{\includegraphics[width=0.38\textwidth]{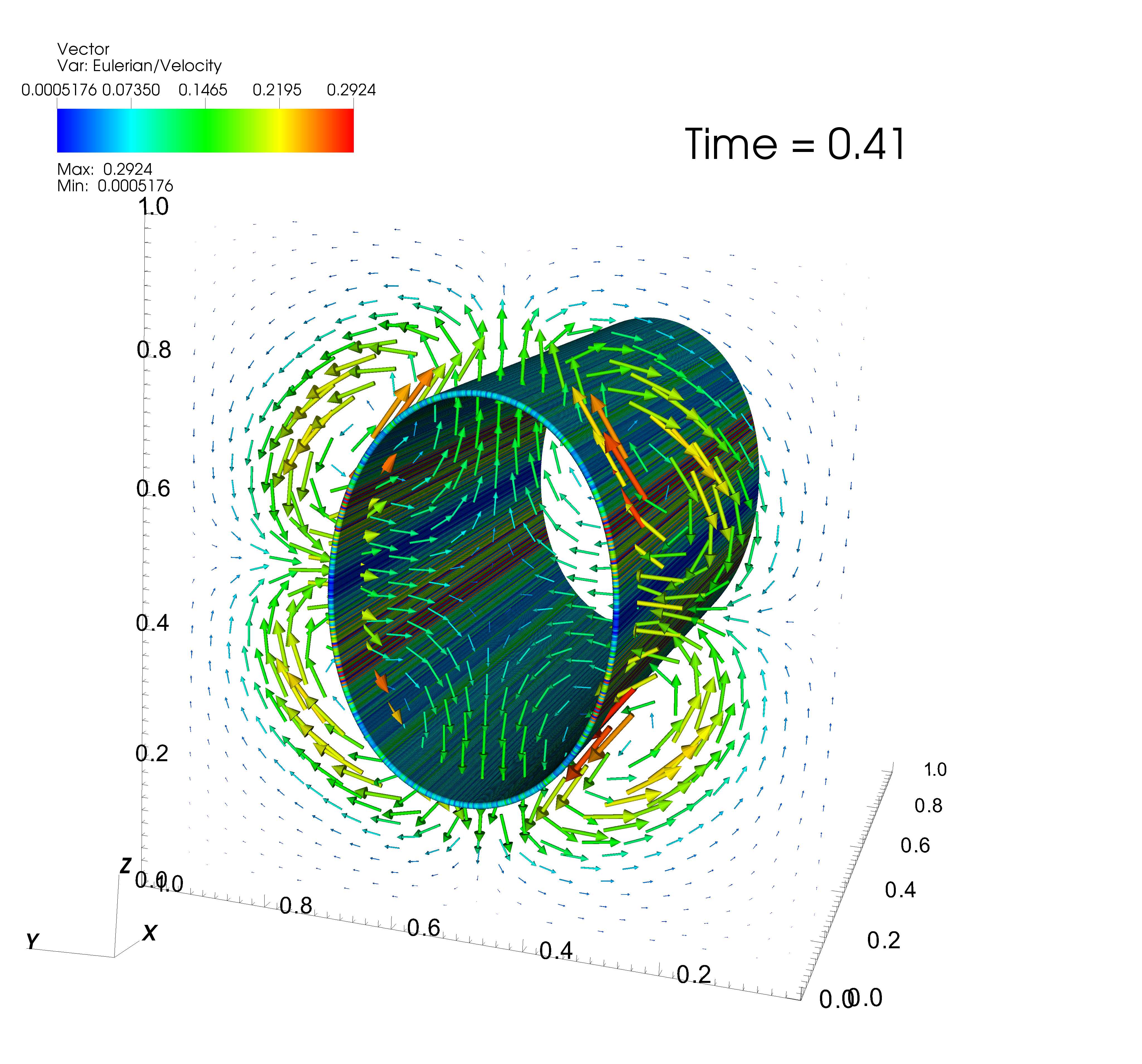}
      \label{fig:CylinderSim4}}
    \caption{Snapshots of an oscillating 3D cylindrical shell ($N=128$)
      that is initially stretched outward along the $x$--direction.}
    \label{fig:CylinderSim}
  \end{center}
\end{sidewaysfigure}

The solution dynamics are illustrated by the snapshots pictured in
Figure~\ref{fig:CylinderSim}, and we observe that the 3D elastic shell
oscillates at roughly the same frequency as the 2D ellipse shown in
Figure~\ref{fig:ThinEllipseSim}.  Although the geometry of this problem
may seem somewhat of a special case because of the alignment of axial
fibers along the $x$-coordinate direction, this feature has no
noticeable impact on parallel performance measurements.  Indeed, the
reason that fiber alignment doesn't affect communication cost is because
all IB points and force connections residing in the ghost region are
communicated regardless of whether or not they actually cross subdomain
boundaries.

In Table~\ref{Table:CylinderScaling}, we present measurements of
execution time and efficiency that illustrate the parallel scaling over
the first $100$ time steps with the number of processors $P$ varying
between 1 and 128. In all runs, the domain is partitioned evenly between
the $P$ processing nodes using rectangular boxes. When the IB points are
evenly distributed between all domain partitions, we observe good parallel
efficiency. On average, we obtain a speedup factor of $1.85$ when doubling the
number of processors for this particular problem.

For larger runs, the parallel efficiency does deteriorate as the local
subdomain shrinks in size. For example, when subdividing a $N=128$
grid to a $(P_x,P_y,P_z) = (32,2,2)$ array of processors, the local grid
size is $4 \times 64 \times 64$. Therefore, every processor has to
communicate all data within its subdomain to neighbouring processors,
since the entire domain overlaps with ghost regions. For this reason,
the GM-IB algorithm performs remarkably well, given the circumstances.

Lastly, in Table~\ref{Table:CylinderScaling}, we show the execution time
for situations where the immersed boundary is not evenly distributed
between processors. Since no load balancing strategy is incorporated in
our implementation, the parallel efficiency drops as the computational
work becomes more unevenly divided.  Here, the total execution time
becomes increasingly dominated by the IB portion of the calculation as
the workload becomes more unbalanced.  The parallel efficiency in this
situation could be improved by partitioning the fluid domain in a
dynamic manner, such as is done by IBAMR using SAMRAI
in~\cite{Griffith2010}.

\begin{table}[htbp]\centering\small
  \caption{Execution time (in seconds) and efficiency for the
    3D cylindrical shell problem for a fixed problem size
    while varying the number of processing nodes $P$.}
  %\ra{1.6}
  \begin{tabular}{cc cc cc}\toprule
      & & \multicolumn{2}{c}{$N=128$} & \multicolumn{2}{c}{$N=256$} \\
    \cmidrule(r){3-4} \cmidrule(r){5-6}
    $P$  & ($P_x,P_y,P_z$)& Wall Time  & Efficiency & Wall Time  & Efficiency \\
    \midrule
    $1$   &  ($1,1,1$)  & $4.01\EE{+2}$ & $1.00$ & $2.60\EE{+3}$ & $1.00$ \\
    $2$   &  ($2,1,1$)  & $2.11\EE{+2}$ & $0.93$ & $1.43\EE{+3}$ & $0.91$ \\
    $4$   &  ($4,1,1$)  & $1.18\EE{+2}$ & $0.81$ & $7.06\EE{+2}$ & $0.92$ \\
    $8$   &  ($2,2,2$)  & $6.07\EE{+1}$ & $0.82$ & $3.52\EE{+2}$ & $0.92$ \\
    $16$  &  ($4,2,2$)  & $3.18\EE{+1}$ & $0.82$ & $1.83\EE{+2}$ & $0.89$ \\
    $32$  &  ($8,2,2$)  & $1.65\EE{+1}$ & $0.79$ & $9.64\EE{+1}$ & $0.84$ \\
    $64$  & ($16,2,2$)  & $9.42\EE{+0}$ & $0.70$ & $5.39\EE{+1}$ & $0.75$ \\
    $128$ & ($32,2,2$)  & $7.76\EE{+0}$ & $0.55$ & $3.02\EE{+1}$ & $0.67$ \\
    \midrule
    $32$  &  ($4,4,2$)  & $2.53\EE{+1}$ & $0.52$ & $1.30\EE{+2}$ & $0.62$ \\
    $64$  &  ($4,4,4$)  & $1.42\EE{+1}$ & $0.46$ & $7.13\EE{+1}$ & $0.57$ \\
    $128$ &  ($8,4,4$)  & $7.76\EE{+0}$ & $0.42$ & $3.72\EE{+1}$ & $0.55$ \\
    \bottomrule
  \end{tabular}
  \label{Table:CylinderScaling}
\end{table}

\section{Conclusions}
\label{sec:Conclusions}

We have developed a new algorithm for the immersed boundary
problem on distributed-memory parallel computers that is based on the
pseudo-compressibility method of Guermond and Minev for solving the
incompressible Navier-Stokes equations.  The fundamental advantage of
this fluid solver is the direction-splitting strategy applied to the
incompressibility constraint, which reduces to solving a series of
tridiagonal linear systems with an extremely efficient parallel
implementation.  

Numerical computations demonstrate the ability of our method to simulate
a wide range of immersed boundary problems that includes not only 2D
flows containing isolated fibers and thick membranes constructed of
multiple nested fibers, but also 3D flows containing immersed elastic
surfaces.  The strong and weak scalability of our algorithm is
demonstrated in tests with up to \changed{256} distributed processors,
where excellent speedups are observed. \changed{Furthermore, comparisons
  against FFT-based (FFTW~\cite{FFTW}) and multigrid
  (Hypre~\cite{Hypre}) solvers shows substantial performance
  improvements when using the Guermond and Minev solver.}  We observe
that since our implementation does not apply any load balancing
strategy, some degradation in the parallel efficiency is observed in
immersed boundary portion of the computation when the elastic membrane
is not equally divided between processors.

We believe that our computational approach is a promising one for
solving fluid-structure interaction problems in which the solid elastic
component takes up a large portion of the fluid domain, such as occurs
with dense particle suspensions~\cite{TornbergShelley2004} or very
complex elastic structures that are distributed throughout the fluid.
These are problems where local adaptive mesh refinement is less likely
to offer any advantage because of the need to use a nearly-uniform fine
mesh over the entire domain in order to resolve the immersed boundary.
It is for this class of problems that we expect our approach to offer
significant advantages over methods such as that of Griffith et
al.~\cite{Griffith2007}.

We plan in future to implement modifications to our algorithm that will
improve the parallel scaling, and particularly on improving memory
access patterns for the Lagrangian portion of the calculation related to
force spreading and velocity interpolation.  We will also investigate
code optimizations that aim to reduce cache misses and exploit on-chip
parallelism. \changed{Lastly, work has started on applying this algorithm to
study spherical membrane dynamics, particle sedimentation, and 
fiber suspensions.}

\bibliography{Paper}
\bibliographystyle{abbrv}

\end{document}